\documentclass[prd,twocolumn,nofootinbib,superscriptaddress,floatfix,preprintnumbers]{revtex4-2}
\usepackage{amsmath,amssymb,amsfonts,bm,graphicx,hyperref,xcolor,mathtools}
\usepackage[normalem]{ulem}
\hypersetup{colorlinks=true, linkcolor=blue, citecolor=blue, urlcolor=blue}

\newcommand{\meson}{meson}

\usepackage{mathtools,slashed,soul,xcolor}
\usepackage[english]{babel}
\usepackage{comment}

\begin{document}

\preprint{CERN-TH-2026-012}

\title{Enhanced Cosmic-Ray Antinuclei Fluxes with Dark Matter Annihilation into SUEPs}

\author{Mattia Di Mauro}\email{dimauro.mattia@gmail.com}
\affiliation{Istituto Nazionale di Fisica Nucleare, Sezione di Torino, Via P. Giuria 1, 10125 Torino, Italy}

\author{Caleb Gemmell}\email{cgemmell2@wisc.edu}
\affiliation{Department of Physics, University of Wisconsin, Madison, WI, USA}
\affiliation{Wisconsin IceCube Particle Astrophysics Center, University of Wisconsin, Madison, WI, USA}
\affiliation{Department of Physics, University of Toronto, Toronto, Ontario, Canada M5S 1A7}

\author{Austin Batz}\email{abatz@uoregon.edu}
\affiliation{Institute for Fundamental Science and Department of Physics, University of Oregon, Eugene, OR 97403, USA}
\affiliation{Theoretical Physics Group, Lawrence Berkeley National Laboratory, Berkeley, CA 94720, USA}

\author{David Curtin}\email{d.curtin@utoronto.ca}
\affiliation{Department of Physics, University of Toronto, Toronto, Ontario, Canada M5S 1A7}

\author{Fiorenza Donato}\email{fiorenza.donato@unito.it}
\affiliation{Istituto Nazionale di Fisica Nucleare, Sezione di Torino, Via P. Giuria 1, 10125 Torino, Italy}
\affiliation{Department of Physics, University of Torino, via P. Giuria, 1, 10125 Torino, Italy}

\author{Nicolao Fornengo}\email{nicolao.fornengo@unito.it}
\affiliation{Istituto Nazionale di Fisica Nucleare, Sezione di Torino, Via P. Giuria 1, 10125 Torino, Italy}
\affiliation{Department of Physics, University of Torino, via P. Giuria, 1, 10125 Torino, Italy}

\author{Graham D. Kribs}\email{kribs@uoregon.edu}
\affiliation{Institute for Fundamental Science and Department of Physics, University of Oregon, Eugene, OR 97403, USA}
\affiliation{Theoretical Physics Department, CERN, 1211 Geneva 23, Switzerland}

\date{\today}

\begin{abstract}
Standard-Model (SM) hadronic parton showers initiated by secondary cosmic-ray  production or dark matter (DM) annihilations robustly predict very low antinuclei yields and a strong additional suppression for heavier antinuclei. We show that an important exception can arise if DM annihilates into a confining dark sector that produces Soft Unclustered Energy Patterns (SUEPs). The hallmark of SUEPs is the emission of very large multiplicities of soft dark {\meson}s ($\pi_D$), which can overcome the usual phase-space suppression of antinuclei formation in parton showers, provided that the dark {\meson}s decay promptly into SM quarks, i.e.\ within a SM hadronization length.
We study several benchmark realizations and find that for DM masses $m_{\rm DM}\sim\mathcal{O}(10~\mathrm{TeV})$, dark {\meson} masses $m_{\pi_D} \sim 400~\mathrm{GeV}$, $\pi_D$ dominantly decaying to $t\bar t$, and a SUEP temperature $T_{\rm SUEP}\simeq 0.1\,m_{\pi D}$, DM annihilation into SUEPs can yield tens of antideuterons and a few antihelium--3 events at \textsf{AMS-02} at kinetic energies of $\mathcal{O}(\mathrm{GeV}$/n) and a few antideuterons and antihelium-3 events in \textsf{GAPS} at energies below 0.5 GeV/n. A future confirmation of an antinuclei signal by the \textsf{AMS-02} or \textsf{GAPS} experiments could provide hints for hidden confining dynamics and would significantly constrain the relevant SUEP parameters.
\end{abstract}

\maketitle

\section{Introduction}
\label{sec:intro}

The nature of dark matter (DM) remains elusive despite sustained theoretical progress and a broad experimental campaign. If DM is made of particles, its existence points to physics beyond the Standard Model (BSM), since no known Standard Model (SM) particle can account for it.
A realistic DM particle must be stable on cosmological timescales, electrically neutral, very weakly coupled to visible matter, non-relativistic by matter--radiation equality with a sufficiently small free-streaming length,
and have self-interactions compatible with astrophysical bounds (e.g., the Bullet Cluster)~\cite{Clowe:2006eq}. The underlying BSM framework must also relate early-Universe dynamics to the present-day abundance, which is determined with percent-level precision from CMB anisotropies as $\Omega_{\rm DM}h^2 \simeq 0.120$~\cite{Planck:2018vyg}. Among the many possibilities, Weakly Interacting Massive Particles (WIMPs) remain particularly well motivated~\cite{Lee:1977ua,1978ApJ...223.1015G}. The most pursued detection strategies are direct detection~\cite{Schumann:2019eaa}, collider searches~\cite{Boveia:2018yeb}, and indirect detection~\cite{Gaskins:2016cha} (see \cite{Cirelli:2024ssz} for an exhaustive recent review). Current direct-detection limits from LZ and XENONnT probe spin-independent WIMP--nucleon cross sections down to $\mathcal{O}(10^{-47}\text{--}10^{-48})\,\mathrm{cm}^2$ for weak-scale masses~\cite{LZ:2023,Aprile:2023XENONnT,LZ:2024zvo}, severely restricting the viable parameter space of most WIMP models (see, e.g.,~\cite{Arcadi:2017kky,Arcadi:2019lka,DiMauro:2023tho,Arcadi:2024ukq,DiMauro:2025jia}). In this regard, secluded DM models have been proposed as scenarios in which dark matter retains a thermal history while still evading stringent laboratory constraints~\cite{Pospelov:2007mp,DiMauro:2025jsb,DiMauro:2025uxt}.

Indirect probes look for excesses in cosmic messengers such as positrons, antiprotons, antinuclei, $\gamma$ rays, and neutrinos~\cite{Gaskins:2016cha,Fermi-LAT:2016afa, Cirelli:2024ssz}. However, the fluxes of these cosmic particles are often dominated by conventional astrophysical sources, making it harder to isolate any potential DM contribution (see, e.g.,~\cite{DiMauro:2015jxa,DiMauro:2015tfa,Fermi-LAT:2016afa,Gaskins:2016cha,Cuoco:2019kuu,Genolini:2021doh,Calore:2022stf,DiMauro:2021qcf,McDaniel:2023bju,Balan:2023lwg,Orusa:2024ewq}).

Cosmic antinuclei generated from DM annihilation or decay provide a particularly sensitive probe. Antinuclei formation proceeds by coalescence: antinucleons produced nearby in phase space (small relative momentum and spatial separation) can bind into antideuterons ($\overline{\rm D}$) and antihelium (in particular ${}^3\overline{\rm He}$). $\overline{\rm D}$~\cite{Donato:1999gy} and, to a lesser extent, ${}^3\overline{\rm He}$~\cite{Cirelli:2014qia,Carlson:2014ssa} are attractive because expected secondary backgrounds fall steeply at kinetic energies per nucleon $K\lesssim1\,\mathrm{GeV}/\mathrm{n}$. This suppression is mainly a consequence of baryon--number conservation: producing an antideuteron in a $p\!-\!p$ collision requires simultaneously producing at least four additional baryons. As a result, the minimum energy of the incoming cosmic proton is of order $17\,m_p$.
WIMPs, which do not possess baryon number and travel at non-relativistic speed, can produce, upon annihilation in the Galactic halo, a $\overline{\rm D}$ flux around $K=(0.1-1)\,\mathrm{GeV}/\mathrm{n}$ that exceeds the secondary component by at least an order of magnitude (see, e.g.,~\cite{Donato:1999gy,Ibarra:2012cc,Fornengo:2013osa,Herms:2016vop,Korsmeier:2017xzj,DeLaTorreLuque:2024htu,Heisig:2024jkk}). Consequently, even a handful of low-energy $\overline{\rm D}$ events would constitute a compelling hint of DM~\cite{vonDoetinchem:2020vbj}.

No conclusive $\overline{\rm D}$ detection has yet been observed. The most stringent constraint to date is from \textsf{BESS}, which sets an upper limit of $6.7\times10^{-5}\,\mathrm{(m^2\,s\,sr\,GeV/n)^{-1}}$ in the range $K=(0.163-1.100)\,\mathrm{GeV}/\mathrm{n}$~\cite{PhysRevLett.132.131001}. Current and upcoming experiments --- \textsf{AMS-02} running on the International Space Station~\cite{2008ICRC....4..765C} and the balloon-borne \textsf{GAPS} mission~\cite{Aramaki:2015laa}, which has just completed its first flight in Antarctica --- are poised to substantially sharpen sensitivities, targeting flux levels down to $\sim(0.8-2)\times10^{-6}\,\mathrm{(m^2\,s\,sr\,GeV/n)^{-1}}$ for $K<1\,\mathrm{GeV}/\mathrm{n}$~\cite{vonDoetinchem:2020vbj}. Cosmic antideuterons and antihelium nuclei therefore represent potentially the cleanest indirect DM detection signature due to the highly suppressed astrophysical background.

Coalescence kinematics and SM hadronization strongly suppress the formation probability of antinuclei with each additional antinucleon.
For illustration, using the \textsc{Pythia~8} Monte Carlo code and applying the simple coalescence setup used in this paper, for a $50\,\mathrm{GeV}$ DM particle annihilating to $b\bar b$, the following ratios are found \cite{DiMauro:2025vxp}:
\begin{equation}
\label{eq:ratios}
\overline{p} : \overline{\mathrm{D}} : {}^3\overline{\mathrm{He}} \;\sim\; 1 : 1.4\times10^{-4} : 3.4\times10^{-8}~.
\end{equation}
This means that the formation probability drops drastically, by about a factor of $10^4$, for each additional antinucleon in the nucleus. Moreover, standard DM models able to produce observable fluxes of $\overline{\rm D}$, ${}^3\overline{\rm He}$, and ${}^4\overline{\rm He}$ would also predict antiproton fluxes from DM that should already have been detected by \textsf{AMS-02}, on top of the known secondary astrophysical component~\cite{Korsmeier:2017xzj}.

A mechanism that has been proposed to enhance the production of ${}^3\overline{\rm He}$ with respect to $\overline{\rm D}$ is that a fraction of the antihelium yield could originate from weakly decaying $b$-baryons produced in $b\bar b$ final states, in particular $\bar{\Lambda}^0_b$, which can efficiently generate multi-antinucleon systems with small relative momenta and hence enhance coalescence~\cite{Winkler:2020ltd}. However, Ref.~\cite{DiMauro:2025vxp} demonstrated that this channel is subdominant with respect to prompt production and therefore not effective in producing a significant enhancement of the antinuclei fluxes.

Various BSM scenarios able to produce enhanced multiplicities of antinuclei have been explored,
such as DM particles with (anti)baryon number~\cite{Heeck:2019ego}, antimatter stars~\cite{Poulin:2018wzu,Bykov:2023nnr}, energetic injections of SM antiquarks~\cite{Fedderke:2024hfy}, or a confining dark sector~\cite{DeLaTorreLuque:2024htu}. Such dark sectors, or hidden valleys~\cite{Strassler:2006im}, would need to produce final states that deviate from expected QCD-like behavior to overcome the standard phase-space suppression that results in the $10^{-4}$ factors shown in Eq.~(\ref{eq:ratios}) for WIMP annihilation within a purely standard QCD context. Ref.~\cite{DeLaTorreLuque:2024htu} achieved this goal with a simple model that includes a sequential chain of decays resulting in a large multiplicity of dark pions in the final state, which then decay to SM particles.

Phenomenologically similar, but theoretically distinct and well motivated, are the \emph{Soft Unclustered Energy Patterns} (SUEPs)~\cite{Strassler:2008bv,Knapen:2016hky}, a specific instance of confining dark sectors that, for a given portal coupling to the SM, contain showers of dark mesons that can collectively decay into $\mathcal{O}(10^2\text{--}10^3)$ SM quarks. Provided the soft mesons are produced and decay to SM quarks within a SM hadronization length, SUEP production can thus boost the $\overline{\mathrm{He}}/\overline{p}$ and $\overline{\mathrm D}/\overline{p}$ ratios by orders of magnitude relative to conventional WIMP annihilations.
These predictions could potentially reconcile the reports of $\mathcal{O}(10)$ candidate events consistent with ${}^3\overline{\rm He}$ and ${}^4\overline{\rm He}$, along with a few candidates compatible with $\overline{\rm D}$~\cite{Tingcern2016,Miapp2022DbarHebar,Miapp2022Dbar}.

In this work, we propose a concrete dark sector model in which DM annihilates into one or more SUEPs and investigate the regions of parameter space that could generate an enhancement of antinuclei production with respect to the antiproton flux. This is done on an event-by-event basis, utilising the \textsc{Pythia~8} Monte Carlo event generator \cite{Bierlich:2022pfr} and a state-of-the-art model for the coalescence of final-state antinucleons taken from Ref.~\cite{DiMauro:2024kml}. Additionally, we consider the effect of various standard portals mediating the decay of the dark {\meson}s. In particular, we show that the $\overline{\mathrm{He}}/\overline{p}$ ratio from SUEP-like theories can be boosted by several orders of magnitude relative to conventional QCD-like expectations, and $\overline{\mathrm{D}}/\overline{p}$ by at least two orders of magnitude. Assuming a $\mathcal{O}(10)\,\mathrm{TeV}$ mass DM candidate coupled to these SUEP states, we then compute the antiproton, antideuteron, and antihelium-3 source spectra in the Galaxy for various DM masses and annihilation cross sections. We propagate the source spectra to the Earth and verify that observable $\overline{\rm D}$ and ${}^3\overline{\mathrm{He}}$ fluxes can arise in realistic regions of the SUEP parameter space while keeping the antiproton flux well below the \textsf{AMS-02} data.

The paper is structured as follows. In Sec.~\ref{sec:model} we outline the benchmark dark matter and SUEP models and their implementation to generate spectra of antinucleons. The coalescence model used to generate antinuclei spectra is discussed in Sec.~\ref{sec:coalescence}. Sec.~\ref{sec:propagation} highlights the propagation of antimatter in the Galaxy. In Sec.~\ref{sec:results} we present our results, including the predicted fluxes of antinuclei at \textsf{AMS-02} and \textsf{GAPS}, before concluding in Sec.~\ref{sec:conclusion}.

\section{Dark Matter Model}
\label{sec:model}

\subsection{Confining Dark Sectors and SUEPs}
\label{sec:SUEPs}

Confining dark sectors are a class of BSM theories that include an array of new fields transforming under a new dark confining force, typically with gauge group $SU(N_D)$~\cite{Kang:2008ea,Baumgart:2009tn,Bai_2014,Renner:2018fhh}. The dark hadrons in the confined phase of the theory can be SM singlets, allowing these models to evade current collider constraints and potentially provide DM candidates~\cite{Hur:2007uz,Kribs:2009fy,Bai:2010qg,Buckley:2012ky,Antipin:2014qva,Appelquist:2015yfa,Antipin:2015xia,Mitridate:2017oky,Beauchesne:2018myj,Francis:2018xjd,Contino:2020god,Asadi:2021yml,Asadi:2021pwo,Cline:2021itd,Asadi:2024bbq,Asadi:2024tpu,Fleming:2024flc}.
Additionally, confining dark sectors can appear in theories of neutral naturalness~\cite{Chacko:2005pe,Craig:2015pha,Burdman:2006tz}, which can address the little hierarchy problem.
Many of these theoretical studies of confining dark sectors generally assume the theory behaves like a variant of QCD\@, i.e. a theory with light and/or heavy dark quarks with a running coupling that is weak at high energies and becomes strong only near the dark confinement scale.

A more exotic possibility is that there are a sufficient number of dark quarks to cause dark sector to be quasi-conformal over a large range of energies. These theories are characterized by a large 't~Hooft coupling
\begin{equation}
  \lambda \equiv g_D^2 N_D \gg 1 \, ,
\end{equation}
with $g_D$ and $N_D$ the gauge coupling and number of dark-sector colors, respectively.

When an external process, such as DM annihilation, transfers energy into the dark sector well within the quasi-conformal regime, it will undergo a dark-sector parton shower.
Unlike QCD, since $\lambda$ is large throughout the shower, parton splittings are unsuppressed and a large multiplicity of soft dark quarks and gluons is emitted at wide angles. Thus, rather than the jet-like behaviour observed in QCD, these soft dark partons spread into a quasi-isotropic angular distribution. When the partons eventually bind into hadrons, those hadrons are likewise soft, unclustered, and high in multiplicity.
This is the soft unclustered energy pattern (SUEP) that can arise from these quasi-conformal dark sectors, which is our main interest in this paper.
A SUEP leads to distinct signatures that have been searched for at colliders~\cite{Knapen:2016hky,Barron:2021SUEP,CMS:2024nca,Curtin:2025ngf}. We also point out that a similar behavior can be achieved by models with a large number of cascade decays~\cite{Elor:2015bho,DeLaTorreLuque:2024htu} or extra spatial dimensions~\cite{Costantino:2020msc,Cesarotti:2020uod}.

\subsection{Modeling a SUEP with a quasi-conformal dark sector}

DM annihilation into a SUEP could arise in a variety of ways.
One possibility is that DM annihilates into one or more portal fields $S$ that couple DM to the fields in the quasi-conformal sector. This allows DM annihilation to generate a SUEP while the portal field and the DM remain neutral under the strongly coupled quasi-conformal sector. The DM annihilation rate is thus independent of the quasi-conformal sector, allowing us to model this sector with just a few parameters.

To keep things simple, we assume there is only one SM scalar singlet $S$ that plays the role of the portal, with a large mass $m_S$ that lies well within the energy range where the dark sector is quasi-conformal. 
We consider both $s$-channel annihilation through an $S$ resonance and annihilation into a pair of $S$ fields, e.g.~through a $t$-channel process. 
In the case of an $s$-channel process through an off-shell $S$-mediator, DM annihilation produces a single SUEP of mass $m_\mathrm{SUEP} = m_{S^*} = 2 m_{DM}$ (ignoring the small DM velocity) which is at rest in the DM annihilation center of mass frame. 
In the case of on-shell production of a pair of $S$ fields, DM annihilation produces two SUEPs, each with mass $m_\mathrm{SUEP} = m_S < m_\mathrm{DM}$ and  boost factor $m_\mathrm{DM}/m_S$ in the DM annihilation center of mass frame.

A quasi-conformal $SU(N_D)$ gauge theory needs a moderately large number of dark quark flavors (dark quarks and anti-quarks), $N_f$, which we take to be close to the upper edge of the conformal window, $N_f \sim (11/2)\,N_D$. The dark gauge coupling will remain quasi-conformal so long as all of the dark quark flavors are lighter than the energy injected into the dark shower,
\begin{equation}
 m_{S^{(\star)}} \gtrsim 
 \mu \gtrsim m_q:\qquad \lambda(\mu) \gtrsim 1 \,.
\end{equation}
Splittings in the dark shower are unsuppressed throughout this energy interval.

For our purposes, we split the dark quarks into two sets: one set of heavier dark quark flavors and one light dark quark flavor. As we will discuss below, we assume that the meson composed of this light dark quark flavor can decay promptly into SM fields, and we discuss explicit theory constructions that can realize decay rates and decay modes that maximize the antimatter signal.

As the shower evolves to lower scales and crosses the heavy-quark threshold, the heavy flavors decouple and only the single light dark quark remains dynamical. The effective $N_f$ is then reduced, causing the theory to exit the conformal window, become strongly coupled, and confine at a scale $\Lambda_D$. The dark shower ultimately hadronizes into a large multiplicity of dark hadrons.
In this way, the decay of $S^{(\star)}$ ``cascades'' through a quasi-conformal dark shower to produce an approximately spherical, high-multiplicity SUEP-like final state composed of dark hadrons.

The dark-hadron momentum distributions are expected to be thermal to a reasonable approximation~\cite{Hagedorn:1965st,Bjorken:1969wi,VanApeldoorn:1981gx,Hatta:2008tn}. This means that the majority of produced dark hadrons are the lightest dark mesons consisting of the lightest quark, with a momentum distribution that we assume  follows a relativistic Boltzmann form,
\begin{equation}
    \frac{dN}{d^3\bm{p}} \sim \exp\Bigg[-\frac{\sqrt{\bm{p}^2+m_{\pi_D}^2}}{T_{\rm SUEP}} \Bigg],
\label{eq:Beq}
\end{equation}
where $T_{\rm SUEP}$ is the Hagedorn temperature of the confining dark sector, with $T_{\rm SUEP}\sim\Lambda_D$~\cite{Blanchard:2004du,Noronha-Hostler:2010nut}, and $m_{\pi_D}$ is the mass of the lightest dark meson.

The low-energy theory, consisting of a confined $SU(N_D)$ gauge theory with one flavor of dark quark, is a unique and well-known regime of a strongly coupled theory (for a few recent dark-sector examples, see \cite{Morrison:2020yeg,Fleming:2024flc}). In this theory, there is one light meson, the analog of the $\eta'$ of QCD, whose mass is expected to scale as~\cite{Veneziano:1976wm,Witten:1979vv}
\begin{eqnarray}
m_{\pi_D}^2 &\sim& \frac{\Lambda_D^2}{N_D} \, .
\end{eqnarray}
In the one-flavor theory the dark $\pi_D$ is not a pseudo-Goldstone boson; instead, it acquires its mass from the $U(1)$ axial anomaly, similar to the QCD $\eta'$. Nevertheless, $m_{\pi_D}$ can be parametrically smaller than the other mesonic states, while the lightest glueball is expected to be $m_{0^{++}} \sim c\,\Lambda_D$ (with $c \sim 6$ for $SU(3)$ \cite{Morningstar:1999rf}). For the purposes of simulations, we assume $N_D \sim \mathcal{O}({\rm few})$, so that $m_{\pi_D} \ll m_{0^{++}}$. We also allow for up to an order-of-magnitude hierarchy between $T_{\rm SUEP}\sim\Lambda_D$ and $m_{\pi_D}$, since the precise value of the Hagedorn temperature relative to $m_{\pi_D}$ is not precisely known. A larger hierarchy could be achieved by raising the light dark quark flavor mass to be of order $\Lambda_D$, thereby increasing the mass of $\pi_D$ to be dominated by the dark quark current mass.

\subsection{SUEP Production from DM Annihilation}
\label{sec:simulation}

\begin{figure}[t]
  \centering
  \includegraphics[width=\linewidth]{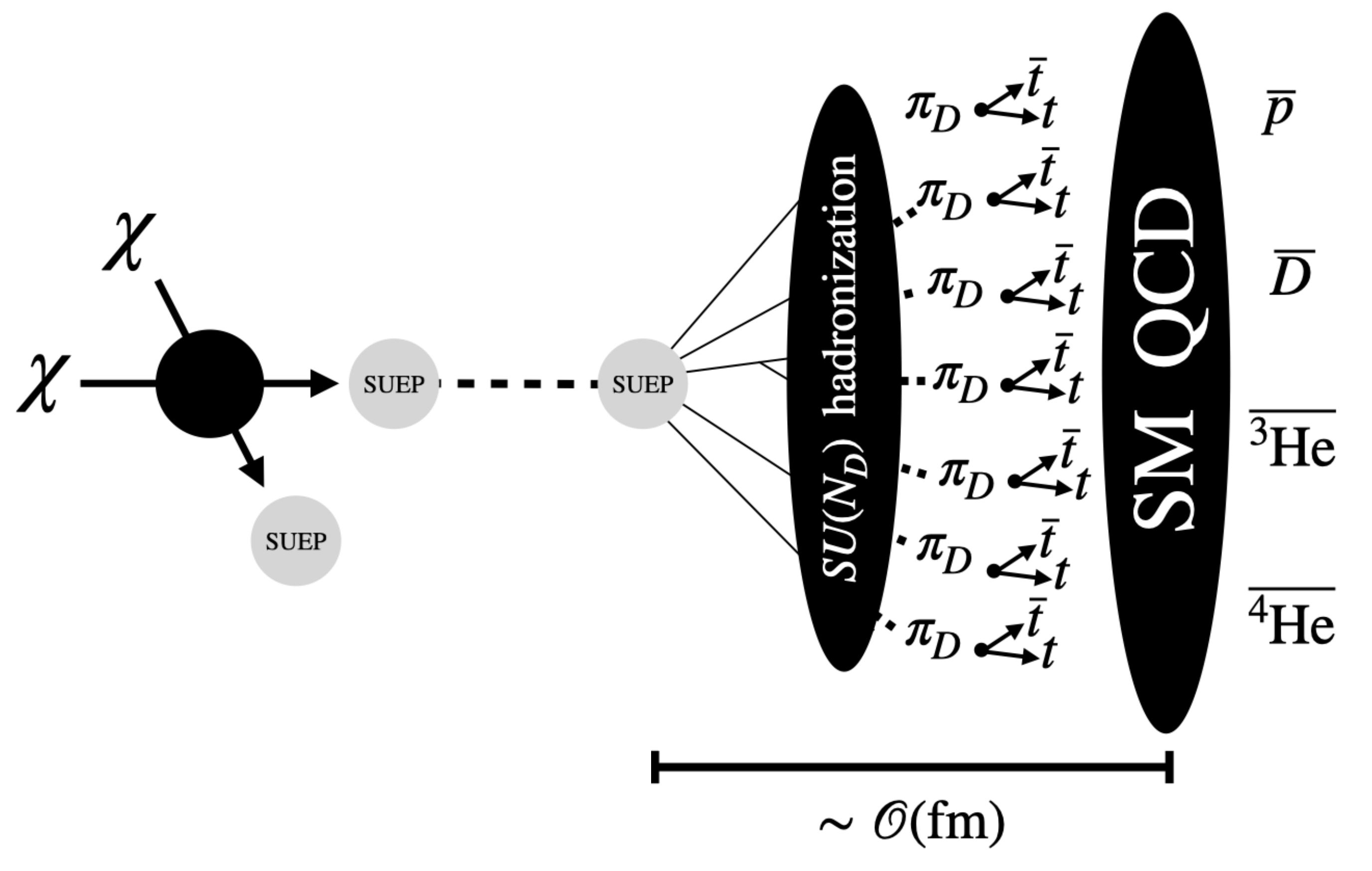}
  \caption{Scheme of the DM annihilation process into SM antinuclei, with relevant distance scales for the case of producing two boosted SUEPs. Only one SUEP decay is shown for simplicity, but the other SUEP state will decay in the same manner. DM annihilation directly to dark quarks gives rise to a single SUEP, which evolves analogously.}
  \label{fig:SUEP}
\end{figure}

To generate events we use the \texttt{suep\_generator} plugin~\cite{suepcode} for \textsc{Pythia 8.243}~\cite{Sjostrand:2014zea}\footnote{The \texttt{suep\_generator} plugin \cite{suepcode} is not maintained by the \textsc{Pythia} developers; therefore we use the \textsc{Pythia} version for which the plugin was originally written.} to simulate the decay of the heavy scalar $S$.
The total energy of the SUEP shower is $m_{\rm SUEP}$, which for the specific portal interaction involving $S$ that we consider is $m_{\rm SUEP}=m_S^{(\star)}$. The dark-sector shower is assumed to produce a large multiplicity of only the lightest dark meson $\pi_D$. In this plugin, the dark mesons are generated isotropically, and their momenta follow the relativistic Boltzmann distribution in Eq.~(\ref{eq:Beq}). Thus, for fixed $m_{\pi_D}$, the kinematics of the dark-meson ensemble are completely determined by $m_{\rm SUEP}$ and the Hagedorn temperature $T_{\rm SUEP}$ of the confining dark sector.

Once the dark mesons have been generated in the rest frame of $S$, their decays and the subsequent SM parton shower and hadronisation are implemented using \textsc{Pythia 8.309}~\cite{Bierlich:2022pfr}. From these events we extract the spectra and momenta of antiprotons and antineutrons. We then apply the coalescence criteria discussed in Sec.~\ref{sec:coalescence} to determine the antideuteron and antihelium yields in the rest frame of the SUEP system. A schematic of the full decay chain, with the relevant physical scales, is shown in Fig.~\ref{fig:SUEP}.

Depending on the mass hierarchy of the DM and the heavy scalar $S$, there are two kinematic possibilities for the DM annihilation event. In this work we consider both:

\noindent\textbf{(1) DM DM $\boldsymbol{\rightarrow S^*}$ (off-shell SUEP mediator).} In the regime where $m_{S} > m_{\rm DM}$, the scalar $S$ cannot be produced on shell and instead acts as an off-shell mediator for DM annihilation. In this case the DM annihilates to dark quarks in the quasi-conformal dark sector, which initiates the SUEP shower. In our setup, this can be straightforwardly modeled by assuming that $S$ is the $s$-channel mediator and computing the SUEP final state by setting $m_{S}=m_{\rm SUEP}=2\,m_{\rm DM}$. We refer to this case in the following as {\tt 1 SUEP}.

\medskip
\noindent\textbf{(2) DM DM $\boldsymbol{\rightarrow SS}$ (two boosted SUEPs).} If instead $m_{S}\leq m_{\rm DM}$, the DM will dominantly annihilate into two on-shell heavy scalars $S$. These scalars then decay, producing SUEP showers again with $m_{\rm SUEP}=m_S$, boosted with respect to the Galactic rest frame. Each $S$ obtains a boost factor
\[
\Gamma = \frac{m_{\rm DM}}{m_{S}},
\]
assuming non-relativistic DM in the Galactic halo and isotropic annihilation. To generate the resulting antinuclei spectrum, the spectra are first computed in the rest frame of $S$, and then Lorentz-boosted by the above factor, taking into account the multiplicity of two $S$'s in the final state. We refer to this case in the following as {\tt 2 SUEPs}.

\subsection{Dark Meson Decay Portals}
\label{sec:pion_decays}

For a signal to be generated, the lightest dark meson must decay to the SM through some portal coupling. Various benchmark portals are usually considered~\cite{Holdom:1985ag,Patt:2006fw,Falkowski:2009yz,Knapen:2021eip}; however, to enhance antinuclei production in SUEP decays, the dark mesons must decay promptly, i.e.\ within a SM hadronization length\footnote{In practice, the relevant length scale is the SM hadronization (confinement) length, $\ell_{\rm had} \sim \Lambda_{\rm QCD}^{-1} \simeq 1~\mathrm{fm}$,
so prompt decays correspond to $c\tau_{\pi_D}\lesssim \ell_{\rm had}$ (up to modest boosts).
Equivalently, this requires a width of order $\Gamma_{\pi_D} \gtrsim \frac{\hbar c}{\ell_{\rm had}}
\sim \Lambda_{\rm QCD}$,
i.e.\ $\Gamma_{\pi_D}\gtrsim \mathcal{O}(0.1\text{--}0.3)\ \mathrm{GeV}$.}.
Coalescence requires antinucleons to be close both in momentum and in configuration space: their relative separation at formation must be of order a few femtometres or less, and their relative momentum below the coalescence scale $p_c$. If the dark mesons travel a distance much larger than $\mathcal{O}(\mathrm{fm})$ before decaying, the event effectively fragments into many spatially separated ``clusters'' of antinucleons, each originating from a different dark meson decay. In that case, antinucleons from different clusters are too far apart to coalesce, so they can only form antinuclei with other daughters of the same decay, rather than with all antinucleons in the event. This removes the main source of enhancement from the high overall multiplicity of dark mesons and reduces the situation to many quasi-independent, low-multiplicity sources.

To avoid this, the decay length of the dark meson in the coalescence frame, $\ell_{\pi_D} \simeq \gamma_{\pi_D} c \tau_{\pi_D}$, must be comparable to or smaller than the nuclear formation scale, $\ell_{\pi_D} \lesssim \mathcal{O}(\mathrm{fm})$, for the typical boosts $\gamma_{\pi_D}$ in our setup. This translates into a requirement for a decay width of order the QCD confinement scale,
\begin{equation}
  \Gamma_{\pi_D} \gtrsim \mathcal{O}(0.1~\mathrm{GeV}) \sim \Lambda_{\rm QCD} \,,
\end{equation}
up to factors of a few from the precise choice of coalescence radius and typical boost. Achieving such a large width typically requires either sizable portal couplings or multi-hundred-GeV/TeV-scale dark meson masses, which naturally motivates heavy dark mesons in SUEP-like scenarios.

\begin{figure}
  \centering
  \includegraphics[width=\linewidth]{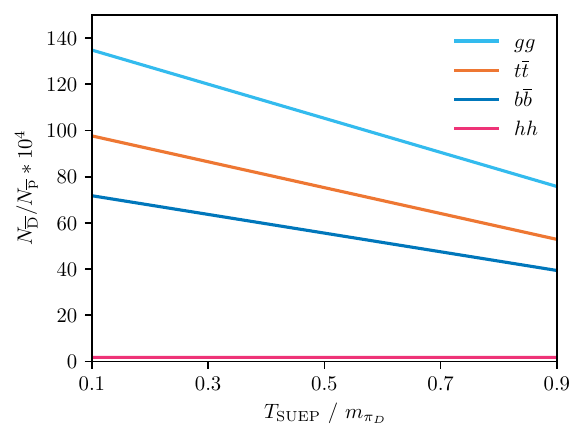}
  \caption{Value of the ratio $N_{\overline{\rm{D}}}/N_{\overline{p}}$ in the SUEP model as a function of $T_{\rm SUEP}/m_{\pi_D}$ for different decay products. This plot is for $m_{\rm SUEP}=80$~TeV and $m_{\pi_D}=380$~GeV. The numbers on the vertical axis approximately correspond to the enhancement for $N_{\overline{\rm{D}}}/N_{\overline{p}}$ with respect to the standard WIMP case of Eq. \ref{eq:ratios} for which $N_{\overline{\rm{D}}}/N_{\overline{p}}\sim 10^{-4}$.
  Comparable rates are found for hadronic channels with up to $\mathcal{O}(1)$ differences, while a Higgs portal leads to no enhancement due to the long lifetime of an on-shell Higgs.}
  \label{fig:decays}
\end{figure}

This restricts the possible decay portals: they must be primarily hadronic and must not proceed dominantly via long-lived intermediate states. We explore a set of benchmark decay portals in Fig.~\ref{fig:decays}, where we show the enhancement of the antideuteron-to-antiproton ratio, $N_{\overline{\rm D}}/N_{\overline p}$, in the SUEP scenario relative to a canonical WIMP with SM hadronization (for which $N_{\overline{\rm D}}/N_{\overline p}\simeq 10^{-4}$). We adopt this ratio as our reference observable to quantify the ability of SUEP dynamics to increase antinuclei production at fixed antiproton yield; throughout the paper we refer to this relative increase as the \emph{enhancement}. We consider the case with $m_{\rm SUEP}=80$~TeV and $m_{\pi_D}=380$~GeV. In particular, portals that preferentially produce leptons or photons are disfavoured for antinuclei production, and portals that generate a long-lived mediator violate the prompt-decay requirement above. This excludes a common benchmark, the Higgs portal, because for the mass ranges needed to have prompt meson decays the dark mesons would decay to on-shell Higgs bosons, whose subsequent decays occur on length scales much larger than $\sim\mathrm{fm}$. For the hadronic portals ($b,t,g$) we find comparable production rates among themselves, leading to enhancement factors $N_{\overline{\rm D}}/N_{\overline{p}}$ as high as $10^2$ with respect to the WIMP case, with the largest antideuteron enhancement arising from gluon decay products. However, the gluonic portal would introduce couplings to SM colored particles that would lead to observable collider signals for the dark meson mass range we consider. We thus conclude that dark mesons must decay dominantly to a top--antitop pair.

Dominant couplings to tops are possible in models of gaugephilic Stealth Dark Matter~\cite{Kribs:2018ilo}, where the dark mesons have $\sim 100\%$ branching fractions to top quarks for masses $\gtrsim 380~\mathrm{GeV}$. Ref.~\cite{Fleming:2024flc} contains a study of one UV completion of such a model where it is possible to achieve $\tau_{\pi_D} \sim 1~\mathrm{fm}/c$ in the large-color limit for the lightest meson. However, this is only one example of a possible UV completion, and other models may also provide dominant couplings to top quarks.

\section{Coalescence Model of Antinuclei Production}
\label{sec:coalescence}

In Ref.~\cite{DiMauro:2024kml} we considered several coalescence prescriptions that reproduce the $\overline{\mathrm D}$ multiplicity measured by \textsf{ALEPH} in hadronic $Z$ decays from $e^+e^-$ collisions~\cite{2006192,ALEPH:2006qoi}, which are assumed to mimick the production of quarks from DM particle annihilation. Two baseline options impose simple geometric cuts in phase space: (i) a momentum--space criterion based on the relative momentum only, and (ii) a mixed criterion that requires both a relative momentum of about $0.2$~GeV and an antinucleon spatial separation of order a few fm. In addition, two more refined implementations, based on quantum mechanics, employ Wigner functions for the deuteron bound state, using either a Gaussian wave function or an Argonne-inspired~\cite{Wiringa:1994wb} one to capture possible correlations in coordinate and momentum space. When tuned to the \textsf{ALEPH} data, all these schemes yield comparable $\overline{\mathrm D}$ spectra within statistical uncertainties~\cite{DiMauro:2024kml}.

In the current paper we adopt a simple coalescence model that imposes criteria on the momentum difference ($\Delta p < p_c$) and spatial separation ($\Delta r < r_c$) of the $\overline{p}$--$\bar{n}$ pair in their center-of-mass frame. We adopt a full event--by--event Monte Carlo approach implemented in \textsc{Pythia 8.309}~\cite{Bierlich:2022pfr}\footnote{Ref.~\cite{DiMauro:2024kml} shows that, once tuned to the \textsf{ALEPH} $Z$--resonance data~\cite{ALEPH:2006qoi}, simple $(\Delta p,\Delta r)$ cuts and Wigner--function--based models lead to very similar $\overline{\mathrm D}$ spectra. Hence, the main conclusions of this work are not especially sensitive to the specific coalescence prescription.}.
For each simulated event we identify all $(\bar p,\bar n)$ pairs and evaluate their relative three-momentum and spatial separation in the pair rest frame,
\begin{equation}
\Delta p \equiv \tfrac{1}{2}\,\big|\bm p_{\bar p}-\bm p_{\bar n}\big|\,,\qquad
\Delta r \equiv \big|\bm r_{\bar p}-\bm r_{\bar n}\big|\,.
\end{equation}
A candidate pair is promoted to an antideuteron if it lies inside the coalescence domain,
\begin{equation}
\Delta p < p_c\,,\qquad \Delta r < r_c\,,
\label{eq:pccond}
\end{equation}
with $r_c\simeq 3\,\mathrm{fm}$ and $p_c=0.212$~GeV the coalescence momentum (a nuisance parameter fixed by \textsf{ALEPH} data).
The values of $p_c$ and $r_c$ are taken from Ref.~\cite{DiMauro:2024kml}. Once the condition is satisfied, we construct the $\overline{\mathrm D}$ four-momentum from the parent antinucleons and record its kinematics in the DM particle-pair annihilation frame~\cite{DiMauro:2024kml}. Antinuclei formed from displaced decays (off-vertex sources) are treated on the same footing; the explicit $\Delta r$ requirement is crucial in that case.

After processing all events, the kinetic-energy spectrum per DM annihilation event is obtained by histogramming the generated antideuterons in bins $[K_i, K_i+\Delta K]$ as
\begin{equation}
\frac{dN_{\rm DM}}{dK_i}
= \frac{N_i(K\in[K_i,\,K_i+\Delta K])}{\Delta K}\,,
\end{equation}
where $N_i$ denotes the number of antideuterons in the $i$th bin.

For the production of ${}^3\overline{\mathrm{He}}$ we proceed analogously: we scan over all triplets containing two antiprotons and one antineutron, and require that the conditions in Eq.~(\ref{eq:pccond}) are satisfied simultaneously for each pair within the triplet. Triplets that satisfy this three-body coalescence condition are promoted to ${}^3\overline{\mathrm{He}}$, and their total four-momentum is then recorded.

In Ref.~\cite{DiMauro:2025vxp}, some of the authors of this work applied the same coalescence model described above to the production of ${}^3\overline{\mathrm{He}}$ measured by \textsf{ALICE} in $pp$ collisions at $\sqrt{s}=7$~TeV~\cite{PhysRevC.97.024615}. Fixing $\Delta r<3$~fm, the best-fit coalescence momentum found by fitting the \textsf{ALICE} data is $0.20 \pm 0.01$~GeV, which is compatible within $1\sigma$ with the value obtained using the \textsf{ALEPH} data~\cite{DiMauro:2024kml}, $0.21\pm0.02$~GeV. This motivates us to apply this coalescence model with $\Delta p < 0.212\,\mathrm{GeV}$ and $\Delta r < 3$~fm throughout this paper, for both $\overline{\mathrm D}$ and ${}^3\overline{\mathrm{He}}$ production.

\section{Propagation of cosmic rays in the Galaxy}
\label{sec:propagation}

The model we apply to evaluate the transport of cosmic-ray (CR) $\bar p$, $\overline{\mathrm D}$, and ${}^3\overline{\mathrm{He}}$ in the Galaxy is briefly summarized in this section, with more technical details deferred to Appendix~\ref{subsec:transport-eq}.
We work in the standard two-zone, axisymmetric diffusion model with a thin gaseous disk of half-height $h\simeq 100\,\mathrm{pc}$ embedded in a cylindrical magnetic halo of half-height $L$ and radius $R$~\cite{Strong:2007nh,Maurin:2001sj,Donato:2001ms,Genolini:2021doh,DiMauro:2021qcf}.
The model is characterized by several free parameters, which are fitted to the latest \textsf{AMS-02} data for several primary and secondary CR species, as reported in Ref.~\cite{DiMauro:2023jgg} and further discussed in Appendix~\ref{subsec:transport-eq}.

In what follows, we assume steady state, homogeneous diffusion within the magnetic halo, and neglect energy losses and gains.
This approximation is well justified for nuclei and is further discussed in the appendix. In this limit, the transport equation admits an analytic solution, as first shown in Ref.~\cite{Barrau:2001ev}. The interstellar differential flux for an antinucleus $j$ produced by DM annihilation can then be written as
\begin{equation}
\label{eq:fluxprim}
\frac{d\Phi_{j}}{dK}\left(K,\vec r_\odot\right)
= \frac{\beta}{4\pi}\,
\left(\frac{\rho_\odot}{m_{\rm DM}}\right)^{\!2}
\, \frac{1}{2}\,\langle\sigma v\rangle\,
\frac{dN_{j}}{dK} \cdot \mathcal{G}_j(K),
\end{equation}
where $\beta=v/c$ is the particle velocity, $\langle \sigma v \rangle$ is the DM annihilation cross section, and $dN_{j}/dK$ is the source spectrum.
We adopt an NFW DM density profile in the Galaxy, $\rho(\vec{r})$~\cite{1997ApJ...490..493N}, with parameters chosen such that the local DM density and total DM mass are $\rho_\odot=0.4$~GeV\,cm$^{-3}$ and $M^{\rm DM}_{200}=10^{12}\,M_\odot$ (consistent with recent estimates; see, e.g., Refs.~\cite{2019JCAP...10..037D,deSalas:2020hbh,2025MNRAS.542.2987S}).
The impact of changing the DM profile is discussed in the appendix.
The function $\mathcal{G}_j(K)$ encapsulates the CR transport physics in the Galaxy. In our case, $\mathcal{G}_j(K)$ includes diffusion, convection, and inelastic scattering processes.

\begin{figure}
  \centering
  \includegraphics[width=0.99\linewidth]{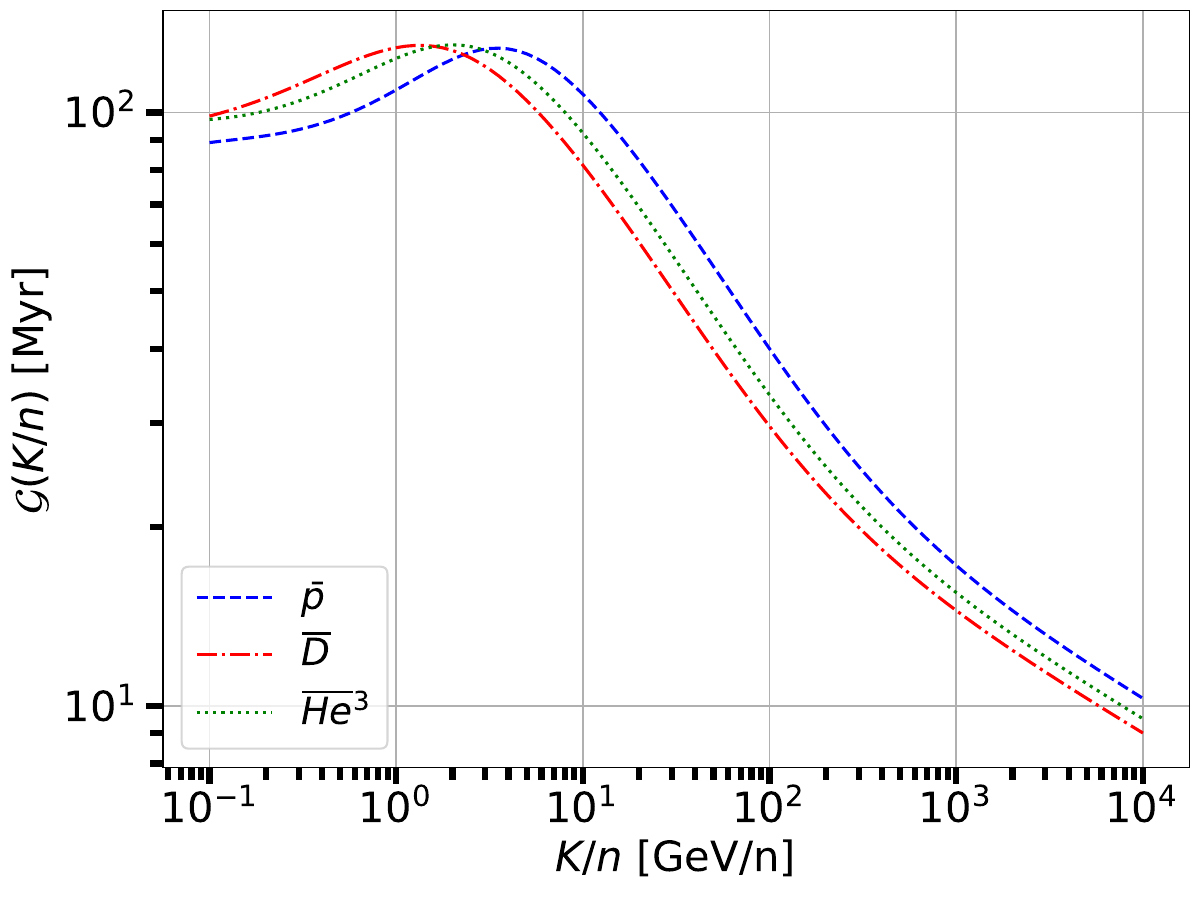}
  \caption{Comparison of the propagation function $\mathcal{G}(K/n)$ for $\overline{p}$, $\overline{\mathrm D}$ and ${}^3\overline{\mathrm{He}}$ (see Eq. \ref{eq:fluxprim}), for an NFW dark matter density profile.}
\label{fig:RKp}
\end{figure}

The propagation setup and parameter values implemented here are taken from Ref.~\cite{DiMauro:2023jgg}.
A relevant assumption is the choice of the spatial diffusion coefficient $D(R)$, for which we adopt a double broken power law in rigidity with two breaks: one at $R=4$~GV and the other at $R=185$~GV. The slopes of $D(R)$, from low to high rigidities, are $-0.80$, $0.72$, and $0.12$, respectively. The diffusion coefficient is normalized at 1~GV to $2.8 \times 10^{28}$~cm$^2$/s. Convection is modeled as a constant wind perpendicular to the disk, $\bm V_c=\mathrm{sign}(z)\,V_c\,\hat{\bm z}$~\cite{Maurin:2001sj}, with $V_c=13$~km/s.
Fig.~\ref{fig:RKp} shows $\mathcal{G}_j(K)$ as a function of kinetic energy per nucleon for $\overline{p}$, $\overline{\mathrm D}$, and ${}^3\overline{\mathrm{He}}$, for an NFW DM density profile. We find that $\mathcal{G}_j(K)$ peaks at a few GeV/n, where it reaches values of order $100$~Myr, and decreases by about an order of magnitude as the energy increases by roughly three decades.
The interstellar flux is then solar-modulated in the force-field approximation with a Fisk potential $\phi_\odot$ (sign- and $Z$-dependent effects beyond the force-field approximation may matter at $\mathcal R\lesssim$ a few GV)~\cite{GleesonAxford:1968}.

\section{Source spectra}
\label{sec:source}

\begin{figure*}
  \centering
  \includegraphics[width=0.49\linewidth]{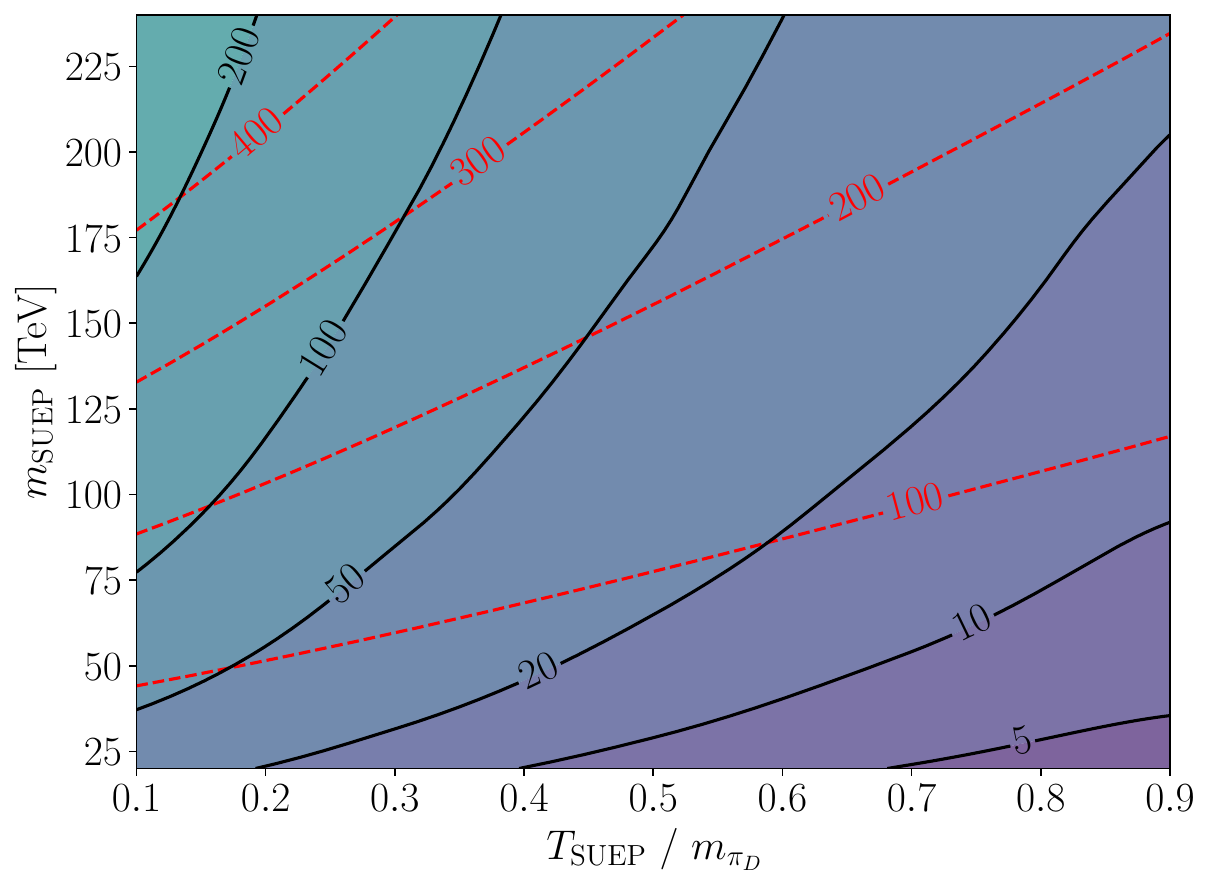}
  \hfill
  \includegraphics[width=0.49\linewidth]{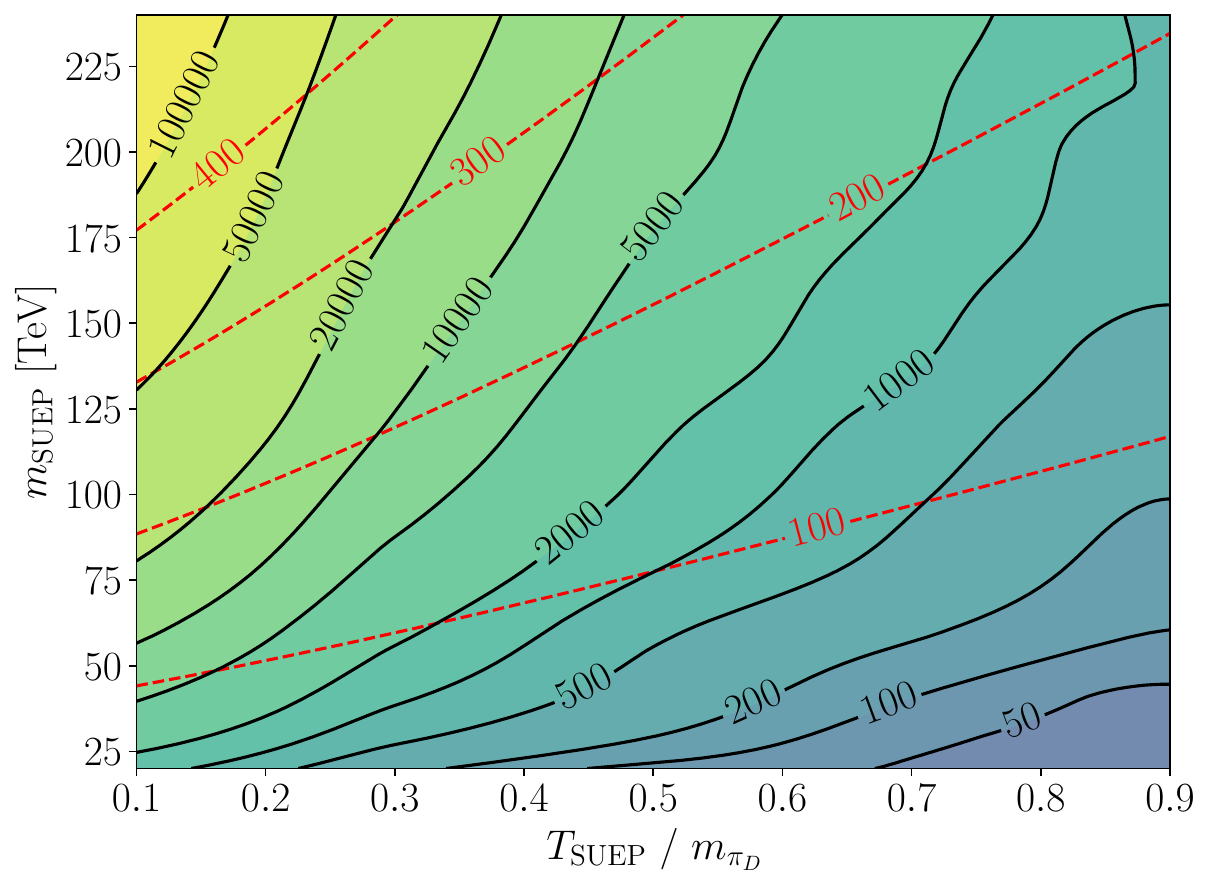}
  \caption{Black contours: enhancements of the ratio $N_{\overline{\rm{D}}}/N_{\overline{p}}$ (left panel) and $N_{^3\overline{\rm{He}}}/N_{\overline{p}}$ (right panel) in SUEP decays for a dark meson mass $m_{\pi_D}=380$~GeV and decay portal to $t \bar t$, with respect to the standard WIMP expectation for the same antinuclei ratio, i.e.~$10^{-4}$ and $10^{-8}$ for $\overline{\rm{D}}$ and $^3\overline{\rm{He}}$ respectively. Red dashed contours show the average multiplicity of dark {\meson}s in a SUEP shower. There are slight fluctuations in the contours in the $^3\overline{\rm{He}}$ enhancement plot which are not present in the antideuteron plot; this is just an artefact of the lower $^3\overline{\rm{He}}$ statistics and interpolating function.}
  \label{fig:dbar_He3bar_enhance}
\end{figure*}

In this section we discuss the relative production of antideuterons and antihelium--3 with respect to antiprotons, evaluated for the \texttt{1 SUEP} scenario in the rest frame of the decaying heavy state $S$, with mass $m_{\rm SUEP}$.
In this case, DM annihilation $\chi\chi\to S \to {\rm SUEP}$ produces a SUEP approximately at rest in the Galactic frame (neglecting the small DM velocity dispersion), so the SUEP rest frame coincides with the Galactic rest frame. These ratios are particularly informative because they highlight the ability of the SUEP dynamics to enhance the production of heavy antinuclei without simultaneously overproducing antiprotons. While the observable flux at Earth depends on the DM mass, annihilation cross section, and Galactic propagation, the \emph{source-level} enhancement depends only on the SUEP parameters $(m_{\rm SUEP},\,T_{\rm SUEP},\,m_{\pi_D})$ and on the coalescence prescription.
The same source-level antinuclei multiplicities and enhancement factors apply to the \texttt{2 SUEPs} scenario; in that case each $S$ is produced boosted in the Galactic frame, but this affects only the boost of the spectra, not the particle yields computed in the $S$ rest frame.

Figure~\ref{fig:dbar_He3bar_enhance} shows the enhancement factors for $\overline{\mathrm{D}}$ and ${}^3\overline{\mathrm{He}}$ relative to the production ratios expected in conventional WIMP annihilation and SM hadronization. The colored contours illustrate how the enhancement varies across the $(T_{\rm SUEP}/m_{\pi_D},\,m_{\rm SUEP})$ plane, while the red dashed contours denote the average number of dark {\meson}s produced in each SUEP event.
We scan the range $T_{\rm SUEP}/m_{\pi_D}\in[0.1,0.9]$; at higher values the enhancement rapidly diminishes, while for $T_{\rm SUEP}/m_{\pi_D}<0.1$ the underlying assumptions of the SUEP generator begin to break down~\cite{suepcode}.

The plots make clear that, although a larger multiplicity of dark {\meson}s generally correlates with an increased antinuclei yield, the dominant driver of the enhancement is the temperature parameter $T_{\rm SUEP}$. Lower temperatures produce softer dark {\meson} momenta, which in turn generate final-state antinucleons that are more closely packed in phase space, significantly boosting the probability for coalescence. In contrast, the region with high $T_{\rm SUEP}$ and low multiplicity correctly reproduces little to no enhancement, as expected in a quasi-QCD-like regime.
In particular, for $T_{\rm SUEP} \sim 0.1\,m_{\pi_D}$ and $m_{\rm SUEP} \geq 80$~TeV, the enhancement reaches $\mathcal{O}(10^2)$ for $\overline{\mathrm{D}}$ and $\mathcal{O}(10^4)$ for ${}^3\overline{\mathrm{He}}$, relative to the SM hadronization present in the WIMP case. For this SUEP model setup, the shower generates $\sim 200$ dark {\meson}s. This region of parameter space therefore maximizes antinuclei production and will be the main focus of the subsequent analysis.

\section{SUEP Signals of antinuclei at Earth}
\label{sec:results}
In this section we present the resulting spectra from DM annihilation to SUEPs once propagated to Earth's position. Firstly, in subsection \ref{sec:sigmav} we determine the maximum allowed DM cross-section from unitarity, antiproton constraints from \textsf{AMS-02} and $\gamma$-ray constraints from Fermi-LAT data, which we then use to show the maximal antinuclei fluxes compared to benchmark WIMP spectra in subsection \ref{sec:fluxes}. Next we discuss the sensitivity estimates of \textsf{GAPS} and \textsf{AMS-02} in \ref{sec:sensitivity} before calculating, in \ref{sec:results}, the expected number of events that \textsf{GAPS} and \textsf{AMS-02} could possibly detect in the future.

\subsection{Choice of the annihilation cross section}
\label{sec:sigmav}

At very large masses, the annihilation cross section of thermal relic DM is constrained by partial-wave unitarity, which limits the maximal rate at which heavy particles can annihilate. This bound provides a physically motivated upper limit on $\langle\sigma v\rangle$ as a function of $m_{\rm DM}$, and must be taken into account when exploring the SUEP parameter space. For this reason, for a given DM mass we first fix the annihilation cross section $\langle\sigma v\rangle$ to the value set by the unitarity bound reported in Ref.~\cite{Smirnov:2019ngs}, which can be approximated as
\begin{equation}
    \langle \sigma v \rangle_{\rm UB} \simeq 10^{-21}\; \mathrm{cm}^3\,\mathrm{s}^{-1}
    \left( \frac{m_{\rm DM}}{10^4~\mathrm{GeV}} \right)^{-2}.
\end{equation}
However, if this value produces an antiproton flux that is too large compared with \textsf{AMS-02} measurements, we rescale the cross section as follows.
We compute, energy bin by energy bin, the ratio between the DM prediction and the \textsf{AMS-02} antiproton data of Ref.~\cite{AGUILAR2020}. If for at least one data point
\begin{equation}
\frac{\phi_{\rm theory}}{\phi_{\rm AMS\text{-}02}} > 0.15,
\end{equation}
we renormalize the value of $\langle\sigma v\rangle$ (initially set by the unitarity bound) such that the largest ratio becomes exactly $0.15$.

The choice of a maximum DM contribution of $15\%$ to the observed antiproton flux is motivated by the fact that the bulk of the antiproton signal is expected to be of secondary astrophysical origin, as shown in
Refs.~\cite{Boudaud:2019efq,Cuoco:2019kuu,Heisig:2020nse,DiMauro:2021qcf,DiMauro:2023jgg}.
Theoretical uncertainties on the secondary antiproton flux arise from several sources.
First, the antiproton production cross sections carry an uncertainty of about $15$--$20\%$~\cite{diMauro:2014zea,Kappl:2014hha,Korsmeier:2018gcy}.
Second, the choice of CR propagation model introduces an additional $20$--$30\%$ variation~\cite{Boudaud:2019efq,DiMauro:2023jgg}.
Therefore, adopting $\phi_{\rm theory}/\phi_{\rm AMS\text{-}02}=0.15$ as the maximum allowed DM contribution is a conservative requirement, because the combined uncertainty from propagation and antiproton production cross sections is at least $\sim 30\%$.

As an illustration, for the off-shell SUEP mediator case ({\tt 1 SUEP}) with $m_{\rm DM}=80$~TeV and $T_{\rm SUEP}/m_{\pi_D}=0.1$ (the setup that, as we will show below, gives the largest enhancement for the antimatter flux at Earth), we must rescale the unitarity cross section by a factor 0.953, corresponding to a physical cross section of $1.5\times 10^{-23}$~cm$^3$/s.
Similarly, for the two-SUEP case that we will show maximizes the antimatter flux, with $m_{\rm DM}=90$~TeV, $m_{\rm SUEP}=80$~TeV, and $T_{\rm SUEP}/m_{\pi_D}=0.1$, we must rescale the unitarity cross section by a factor 0.956, corresponding to a physical cross section of $1.2\times 10^{-23}$~cm$^3$/s. For $m_{\rm DM}\gtrsim 90~\mathrm{TeV}$, the antiproton flux stays below $15\%$ of the measured \textsf{AMS-02} flux, and no rescaling is required.

The take-home message is that, for each choice of SUEP parameters, we adopt an annihilation cross section that is either equal to the unitarity bound or rescaled such that the DM-induced antiproton flux does not exceed $15\%$ of the \textsf{AMS-02} antiproton data.

DM particles annihilating into SUEP states can also produce $\gamma$ rays, so indirect constraints from $\gamma$-ray observations should be considered. In this respect, Milky Way dwarf spheroidal galaxies (dSphs) provide one of the cleanest targets for DM searches: they contain little interstellar gas and have low star-formation activity, so their astrophysical $\gamma$-ray emission is expected to be minimal. A number of analyses have searched for a DM-induced $\gamma$-ray signal from a combined set of dSphs without finding evidence for an excess (see, e.g.,~\cite{Fermi-LAT:2016uux,Hoof:2018hyn,DiMauro:2021qcf,DiMauro:2022hue,McDaniel:2023bju}). Here we test the corresponding constraints on the annihilation cross section using the recent results of Ref.~\cite{McDaniel:2023bju}, for which we have access to the likelihood profiles in flux space derived from 14 years of \emph{Fermi}-LAT data.\footnote{We cannot directly adopt published limits on $\langle\sigma v\rangle$ obtained under the standard WIMP assumption of $100\%$ annihilation into a single SM two-body final state, because in the SUEP scenario the $\gamma$-ray yield arises from a cascade and is therefore model-dependent (see Sec.~\ref{sec:model}).}
We compute upper limits on $\langle\sigma v\rangle$ for $m_{\rm DM}$ between 10 and 150~TeV and find that, for the benchmark scenarios considered, the resulting bounds are of the same order as those implied by our antiproton requirement. For instance, in the {\tt 2 SUEPs} case for $m_{\rm DM}=10~(90)$~TeV, $m_{\rm SUEP}=9~(80)$~TeV, and $T_{\rm SUEP}/m_{\pi_D}=0.1$, we obtain $\langle\sigma v\rangle \lesssim 9.4\times10^{-25}~(8.5\times10^{-24})~\mathrm{cm^3\,s^{-1}}$. Since these limits are of the same order as the bounds obtained with our antiproton-based prescription, in the remainder of the paper we adopt the annihilation cross sections determined by the unitarity bound and by the \textsf{AMS-02} antiproton condition described above. We have also tested that this approach is appropriate for the {\tt 1 SUEP} case.

Tab.~\ref{tab:sigmav_limits} illustrates how our benchmark annihilation cross sections are set by the most constraining requirement among partial-wave unitarity, the \textsf{AMS-02} antiproton bound, and (where evaluated) \emph{Fermi}-LAT dwarf-spheroidal limits. In practice, the antiproton constraint forces sizeable suppressions with respect to unitarity at lower masses, while for $m_{\rm DM}\gtrsim \mathcal{O}(100~\mathrm{TeV})$ the adopted values approach the unitarity bound.

\begin{table*}[t]
\centering
\begin{tabular}{c c | c c c | c c c}
\hline\hline
$m_{\rm DM}$ &
$\langle\sigma v\rangle_{\rm UB}$ &
$M_{\rm SUEP}$ &
$\langle\sigma v\rangle^{2\,{\rm SUEPs}}_{\bar p}$ &
$\langle\sigma v\rangle^{2\,{\rm SUEPs}}_{\rm dSph}$ &
$m_{\rm SUEP}$ &
$\langle\sigma v\rangle^{1\,{\rm SUEP}}_{\bar p}$ &
$\langle\sigma v\rangle^{1\,{\rm SUEP}}_{\rm dSph}$ \\
$[\mathrm{TeV}]$ &
$[\mathrm{cm^{3}\,s^{-1}}]$ &
$[\mathrm{TeV}]$ &
$[\mathrm{cm^{3}\,s^{-1}}]$ &
$[\mathrm{cm^{3}\,s^{-1}}]$ &
$[\mathrm{TeV}]$ &
$[\mathrm{cm^{3}\,s^{-1}}]$ &
$[\mathrm{cm^{3}\,s^{-1}}]$ \\
\hline
10  & $1.0\times 10^{-21}$ & 8   & $1.3\times 10^{-24}$ & $9.4\times 10^{-25}$ & 20  & $1.8\times 10^{-24}$ & $2.0\times 10^{-24}$ \\
30  & $1.1\times 10^{-22}$ & 20  & $3.4\times 10^{-24}$ & $2.8\times 10^{-24}$             & 60  & $5.5\times 10^{-24}$ & $6.0\times 10^{-24}$ \\
50  & $4.0\times 10^{-23}$ & 40  & $6.1\times 10^{-24}$ & $6.9\times 10^{-24}$             & 100 & $9.3\times 10^{-24}$ & $9.9\times 10^{-24}$ \\
90  & $1.2\times 10^{-23}$ & 80  & $1.2\times 10^{-23}$ & $8.5\times 10^{-24}$  & 180 & $1.7\times 10^{-23}$ & $1.8\times 10^{-23}$ \\
150 & $4.4\times 10^{-24}$ & 120 & $2.1\times 10^{-23}$ & $1.4\times 10^{-23}$  & 300 & $2.8\times 10^{-23}$ & $3.0\times 10^{-23}$ \\
\hline\hline
\end{tabular}
\caption{Benchmark annihilation cross sections adopted throughout the analysis for
$T_{\rm SUEP}/m_{\pi_D}=0.1$, shown separately for the {\tt 2 SUEPs} and {\tt 1 SUEP} kinematic scenarios.
For each DM mass (first column), the second column reports the partial-wave unitarity upper bound
$\langle\sigma v\rangle_{\rm UB}=10^{-21}\,\mathrm{cm^3\,s^{-1}}\,(m_{\rm DM}/10^4\,\mathrm{GeV})^{-2}$,
which reproduces the scaling in Ref.~\cite{Smirnov:2019ngs}.
The two blocks of columns correspond to the benchmark SUEP mass used in each scenario
(third and sixth columns) and to the resulting maximal allowed annihilation cross section after applying
(i) the antiproton constraint (fourth and seventh columns), implemented by requiring that the DM-induced
$\bar p$ flux does not exceed $15\%$ of the \textsf{AMS-02} data (Sec.~\ref{sec:sigmav}), and
(ii) the dwarf-spheroidal $\gamma$-ray limits from the \emph{Fermi}-LAT likelihood analysis of
Ref.~\cite{McDaniel:2023bju} (fifth and eighth columns), where evaluated.
Whenever the antiproton requirement is weaker than unitarity we set
$\langle\sigma v\rangle_{\bar p}=\langle\sigma v\rangle_{\rm UB}$.}
\label{tab:sigmav_limits} 
\end{table*}

\subsection{Fluxes of antinuclei from SUEP}
\label{sec:fluxes}

\begin{figure}
  \centering
  \includegraphics[width=0.99\linewidth]{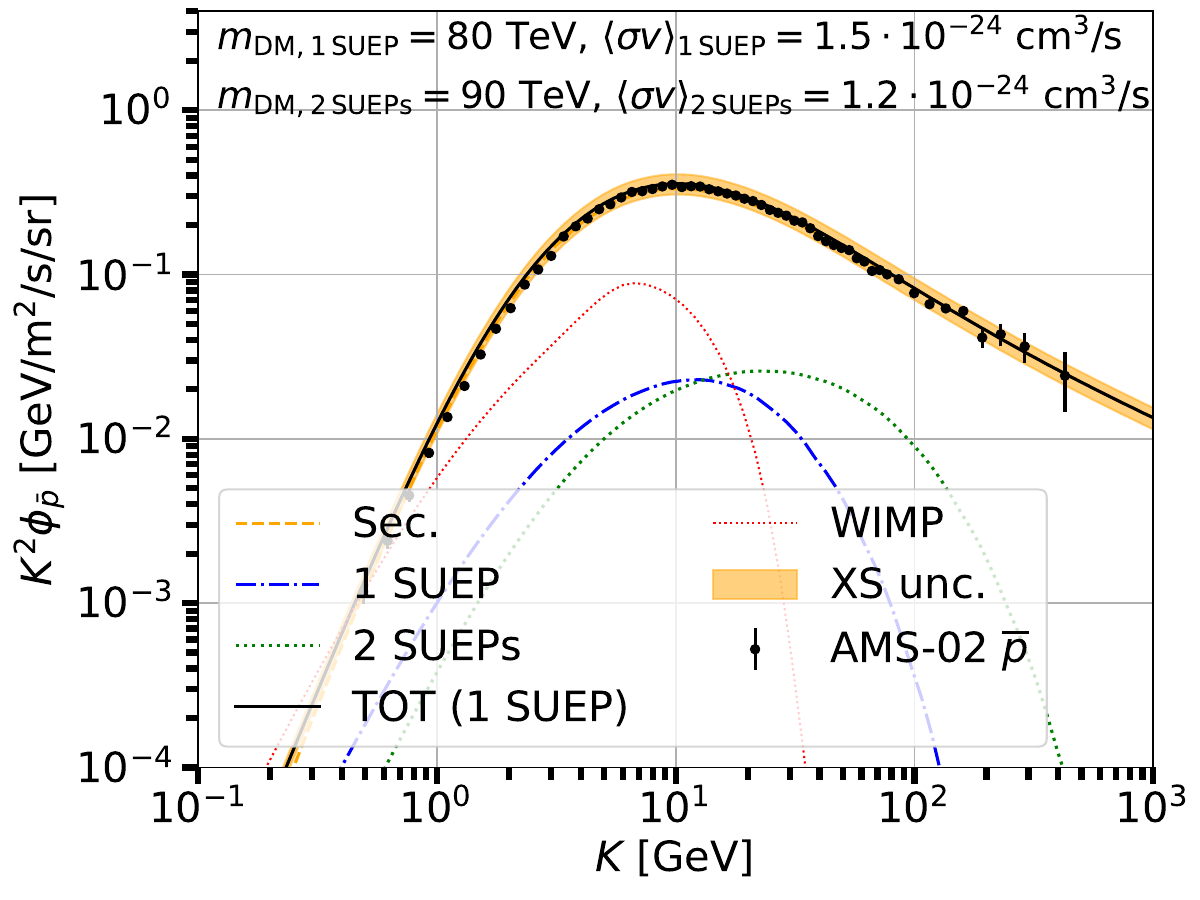}
  \includegraphics[width=0.99\linewidth]{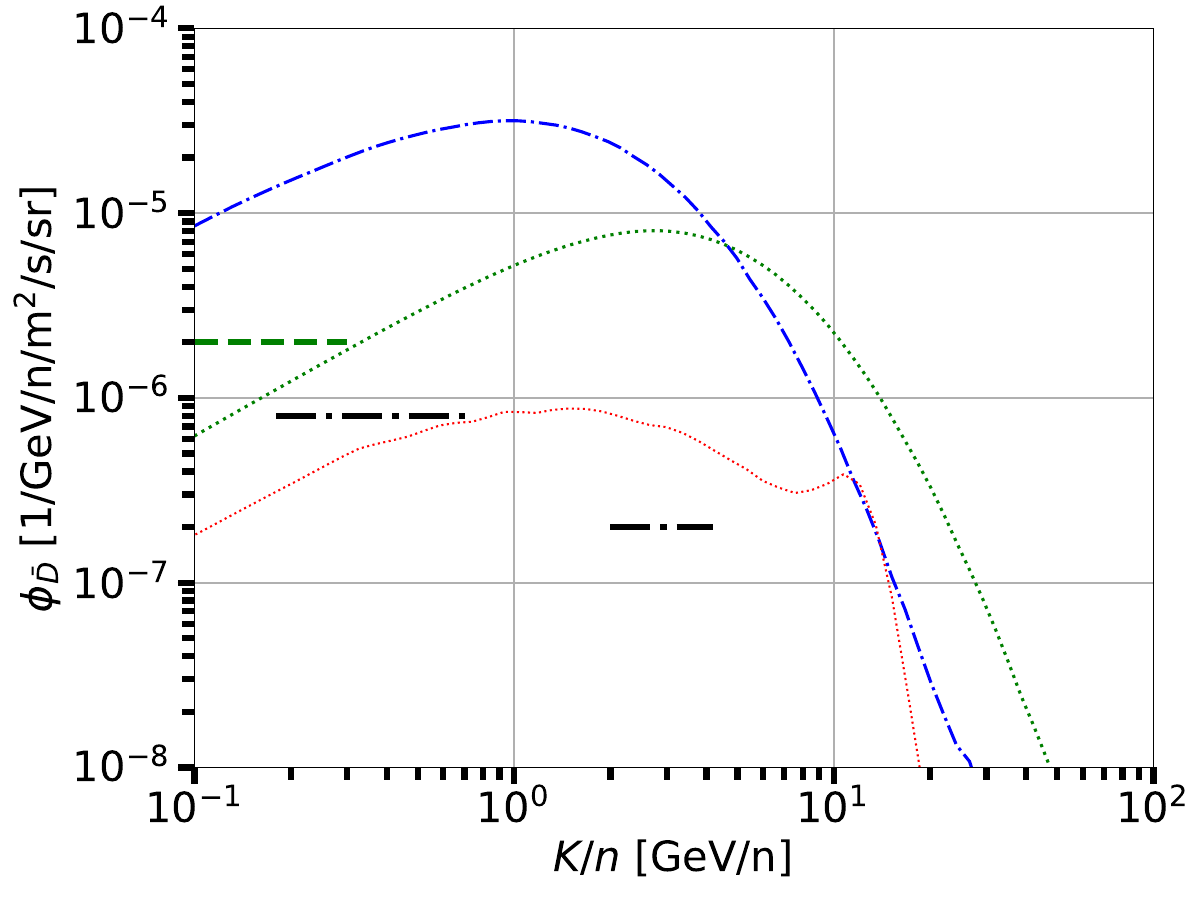}
  \includegraphics[width=0.99\linewidth]{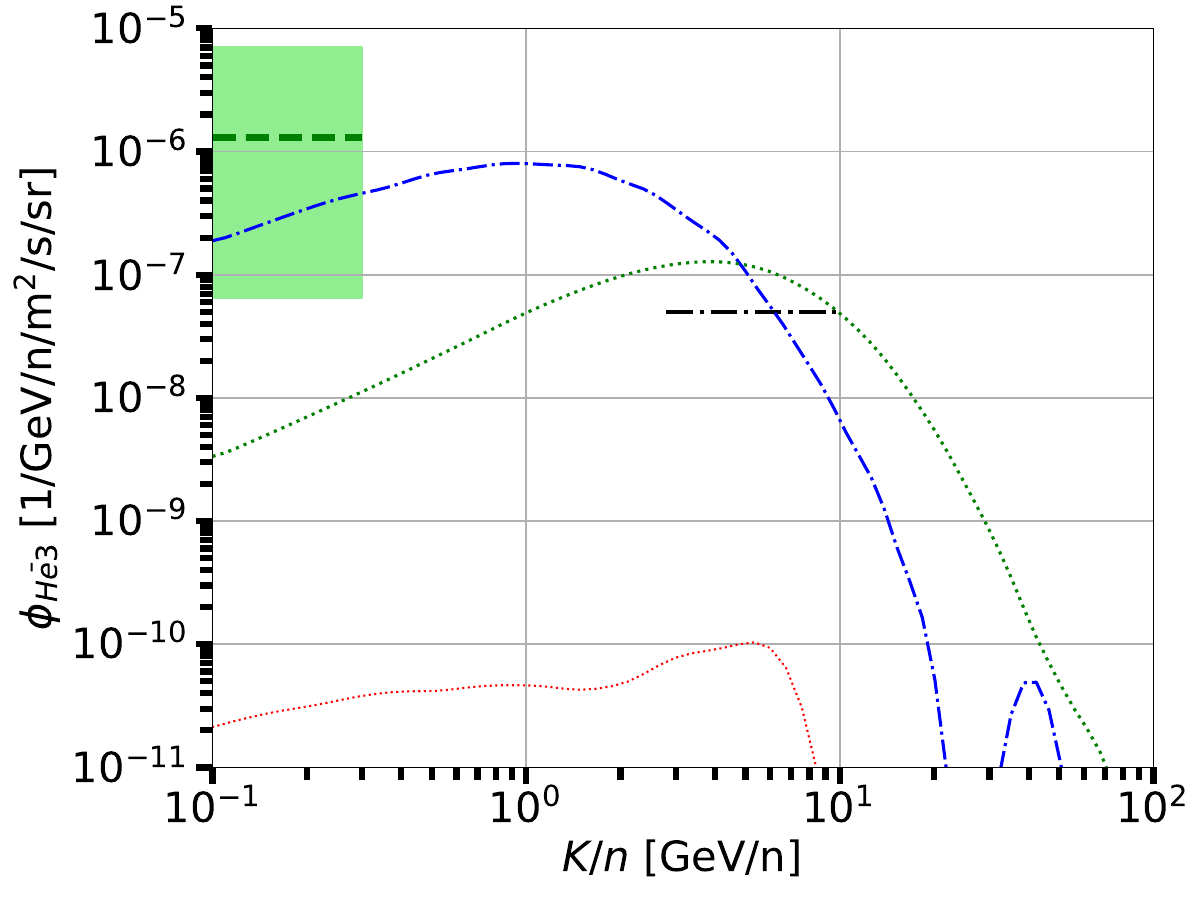}
  \caption{Comparison of the flux of antimatter from DM annihilation in two scenarios: (1) $m_{\rm DM}=80$~TeV and $T/m_{\pi_D}=0.1$ in the {\tt 1 SUEP} scenario (blue dot-dashed curves),
  and (2) $m_{\rm DM}=90$~TeV and $T/m_{\pi_D}=0.1$ in the {\tt 2 SUEPs} scenario (green dotted curves), together with the \textsf{GAPS} (green dashed) and \textsf{AMS-02} sensitivities (black dot-dashed) lines. We show the cases of $\overline{p}$ (top panel), $\overline{\mathrm{D}}$ (central panel) and ${}^3\overline{\mathrm{He}}$ (bottom panel). For reference, we also show the standard WIMP expectation for DM annihilating into $b\bar b$ with $m_{\rm DM}=50$~GeV and a thermal cross section (red dotted curves), and the secondary $\overline{p}$ production (orange dashed curve).}
\label{fig:SUEP_fluxes}
\end{figure}

Here we present example spectra for parameter that maximize the antinuclei fluxes for each SUEP kinematic case in Fig.~\ref{fig:SUEP_fluxes}. For the off-shell mediator case ({\tt 1 SUEP}), the maximum fluxes occur for $m_{\rm DM}=80$~TeV and $T_{\rm SUEP}/m_{\pi_D}=0.1$, while for the boosted SUEP case ({\tt 2 SUEPs}), the maximum is at $m_{\rm DM}=90$~TeV, $m_{\rm SUEP}=80$~TeV, and $T_{\rm SUEP}/m_{\pi_D}=0.1$. We also compare to benchmark antinuclei fluxes for WIMP dark matter, specifically for $m_{\rm DM}=50$~GeV annihilating to a $b\bar{b}$ pair with a cross section that gives the correct thermal relic abundance.

The top panel of Fig.~\ref{fig:SUEP_fluxes} shows the DM and secondary $\overline{p}$ fluxes compared with the \textsf{AMS-02} data. The SUEP spectra for both cases are relatively similar, peaking around $K \sim 20$~GeV, and remain below $\sim 15\%$ of the measured flux, as designed. The secondary component, as expected, fits the data across the full energy range. We also report the theoretical uncertainties associated with the antiproton production cross sections, at the level of $\sim 20\%$. The antiproton prediction for a WIMP particle with a thermal cross section gives a flux that is higher than the SUEP cases up to around 20~GeV. In particular, such a WIMP particle is typically severely constrained by \textsf{AMS-02} $\bar{p}$ data (see, e.g.,~\cite{DiMauro:2021qcf}).

The central and bottom panels display the DM $\overline{\rm D}$ and ${}^3\overline{\mathrm{He}}$ fluxes for the {\tt 1 SUEP} and {\tt 2 SUEPs} cases compared to the WIMP case with standard QCD hadronization. We first note that the {\tt 1 SUEP} case produces antinuclei fluxes that are about one order of magnitude larger than the {\tt 2 SUEPs} case at low energies. At energies above $\sim 4$~GeV/n, the latter becomes larger. This is expected because the {\tt 2 SUEPs} scenario produces boosted $\overline{\rm D}$ and ${}^3\overline{\mathrm{He}}$ events with larger kinetic energies. Moreover, the WIMP case predicts $\overline{\rm D}$ and ${}^3\overline{\mathrm{He}}$ fluxes that are smaller by a factor of $\sim 30$ and $\sim 10^4$, respectively, compared to the {\tt 1 SUEP} predictions. This is consistent with the results for the source spectra shown in Fig.~\ref{fig:dbar_He3bar_enhance}.

To summarize, for the off-shell mediator case ({\tt 1 SUEP}) we obtain a ratio between the different species of
\begin{equation}
\label{eq:ratiosnew1}
\overline{p} : \overline{\mathrm{D}} : {}^3\overline{\mathrm{He}} \;\sim\; 1 : 2.6\times 10^{-2} : 6.2\times 10^{-4}~,
\end{equation}
corresponding to an enhancement of about two/four orders of magnitude for $\overline{\mathrm{D}}$/${}^3\overline{\mathrm{He}}$ relative to the WIMP case in Eq.~\ref{eq:ratios}. 
For the boosted SUEP ({\tt 2 SUEPs}) we obtain a similar hierarchy:
\begin{equation}
\label{eq:ratiosnew2}
\overline{p} : \overline{\mathrm{D}} : {}^3\overline{\mathrm{He}} \;\sim\; 1 : 1.4\times10^{-2} : 1.8\times10^{-4}~.
\end{equation}
In contrast, for $T_{\rm SUEP}/m_{\pi_D}$ closer to 1 and small $m_{\rm SUEP}$, the ratios between the different antimatter species approach the WIMP expectation. Thus, the relative enhancements at the source level shown in Sec.~\ref{sec:source} are preserved after CR propagation.

\subsection{AMS-02 and \textsf{GAPS} flux sensitivity to $\overline{\rm D}$ and $^3\overline{\mathrm{He}}$}
\label{sec:sensitivity}

The \textsf{GAPS} sensitivity to antideuterons and antihelium-3 has been reported in
Refs.~\cite{Aramaki:2015laa,GAPS:2020axg}.
These estimates are based on full \emph{in situ} instrument simulations,
event reconstruction, and realistic atmospheric modeling.
For the long-duration balloon (LDB) program (35~days $\times$ 3 flights),
the $\overline{\rm D}$ flux sensitivity is
$\sim 2\times 10^{-6}\;\mathrm{(m^{2}\,s\,sr\,GeV/n)}^{-1}$
in the kinetic-energy-per-nucleon window $K/n\in[0.1,0.3]~\mathrm{GeV/n}$.
This corresponds to the 99\%~CL for observing $\ge 2$ antideuteron events,
given an expected background of $\approx 0.014$ events.
In the same energy window and for the same LDB exposure, the 95\%~CL sensitivity
(corresponding to $\simeq 1$ expected event) for the detection of antihelium-3 is
$1.3\times 10^{-6}\;\mathrm{(m^{2}\,s\,sr\,GeV/n)}^{-1}$,
with a possible range from $6\times 10^{-8}$ to $7\times 10^{-6}$ due to
systematic uncertainties in the sensitivity evaluation.

The \textsf{AMS-02} sensitivity for $\overline{\rm D}$ and ${}^3\overline{\mathrm{He}}$ was
originally derived from idealized simulations that assumed an \textsf{AMS-02}
configuration different from the one ultimately realized, namely a superconducting magnet with a $0.86$~T field and a somewhat different layout of the silicon tracker planes~\cite{2008ICRC....4..765C}.
The flight configuration is therefore different.
Recently, the \textsf{AMS-02} Collaboration has presented at several conferences (e.g.~\cite{Oliva:JENAA:2024}) projected flux sensitivities for 2030 (after the acceptance upgrade): for $\overline{\rm D}$, about
$8\times 10^{-7}\;\mathrm{(m^{2}\,s\,sr\,GeV/n)}^{-1}$ in
$K/n\in[0.2,0.7]~\mathrm{GeV/n}$ and
$2\times 10^{-7}\;\mathrm{(m^{2}\,s\,sr\,GeV/n)}^{-1}$ in
$K/n\in[2,4]~\mathrm{GeV/n}$.

The \textsf{AMS-02} sensitivity to antihelium was estimated pre-launch relative to the
helium CR flux~\cite{Spada:2008xk}.
In particular, the 95\%~CL sensitivity to the ratio
${}^3\overline{\mathrm{He}}/\mathrm{He}$ was quoted as
$\sim(1\text{--}2)\times 10^{-9}$.
Many subsequent theory works (e.g.~\cite{Carlson:2014ssa,Cirelli:2014qia,Korsmeier:2017xzj,DeLaTorreLuque:2024htu}) rescaled this ratio to an absolute
${}^3\overline{\mathrm{He}}$ flux, obtaining sensitivities of order
$10^{-7}$, $10^{-9}$, and a few $\times 10^{-11}\;
\mathrm{(m^{2}\,s\,sr\,GeV/n)}^{-1}$ at 1, 10, and 100~GeV/n, respectively.
The steeply decreasing shape of this ``flux sensitivity'' is, however, an
artifact of expressing the estimate as a ratio ${}^3\overline{\mathrm{He}}/\mathrm{He}$: given the energy dependence of
the helium acceptance and other detector effects, a multi-order-of-magnitude
drop between 1 and 100~GeV/n is unrealistic.

We therefore adopt a more realistic estimate based on the expected background
and the number of signal events associated with a given confidence level,
\begin{equation}
  \phi_{\mathrm{sens}}^{\mathrm{AMS\text{-}02}}
  = \frac{N_{\mathrm{events}}}{\mathrm{Acc}\times \mathrm{Time}\times \mathrm{Surv}}\,,
  \label{eq:ams_sens}
\end{equation}
where $\mathrm{Acc}$ is the acceptance (taken to be similar to helium),
$\mathrm{Time}$ is the detector livetime, and $\mathrm{Surv}$ is the survival
probability of ${}^3\overline{\mathrm{He}}$ against inelastic scattering in the instrument material.
With this approach, a 95\%~CL value for the flux sensitivity (associated with at least two events) is estimated to be
$\phi_{\mathrm{sens}}^{\mathrm{AMS\text{-}02}}
 \sim 5\times 10^{-8}\;\mathrm{(m^{2}\,s\,sr\,GeV/n)}^{-1}$
for $K/n\in[1,10]~\mathrm{GeV/n}$,\footnote{The energy-dependent
$\phi_{\mathrm{sens}}^{\mathrm{AMS\text{-}02}}$ used here was provided via
private communication by V.~Choutko. In any case, using an acceptance for He of $0.012~\mathrm{m^{2}\,sr}$ as reported in Ref.~\cite{HabibyAlaoui:2016:PhD}, 12~years of data taking, and neglecting inelastic scattering in the \textsf{AMS-02} detector yields
$\phi_{\mathrm{sens}}^{\mathrm{AMS\text{-}02}}\!\sim\!7\times 10^{-8}\; \mathrm{(m^{2}\,s\,sr\,GeV/n)}^{-1}$, consistent with the private communication once the detailed energy dependence of acceptance, livetime, and survival probability is included.}
We focus on this energy range because at both lower and higher energies the
\textsf{AMS-02} sensitivity worsens significantly.
This value of the \textsf{AMS-02} ${}^3\overline{\mathrm{He}}$ flux sensitivity is a factor of 4 (50) higher at $K/n=3$ (10)~GeV/n than the one originally estimated in Ref.~\cite{Spada:2008xk} and used, for example, in Refs.~\cite{Carlson:2014ssa,Cirelli:2014qia,Korsmeier:2017xzj,DeLaTorreLuque:2024htu}.

\subsection{Predicted number of events at \textsf{AMS-02} and GAPS}
\label{sec:results}

\begin{figure*}
  \centering
  \includegraphics[width=0.32\linewidth]{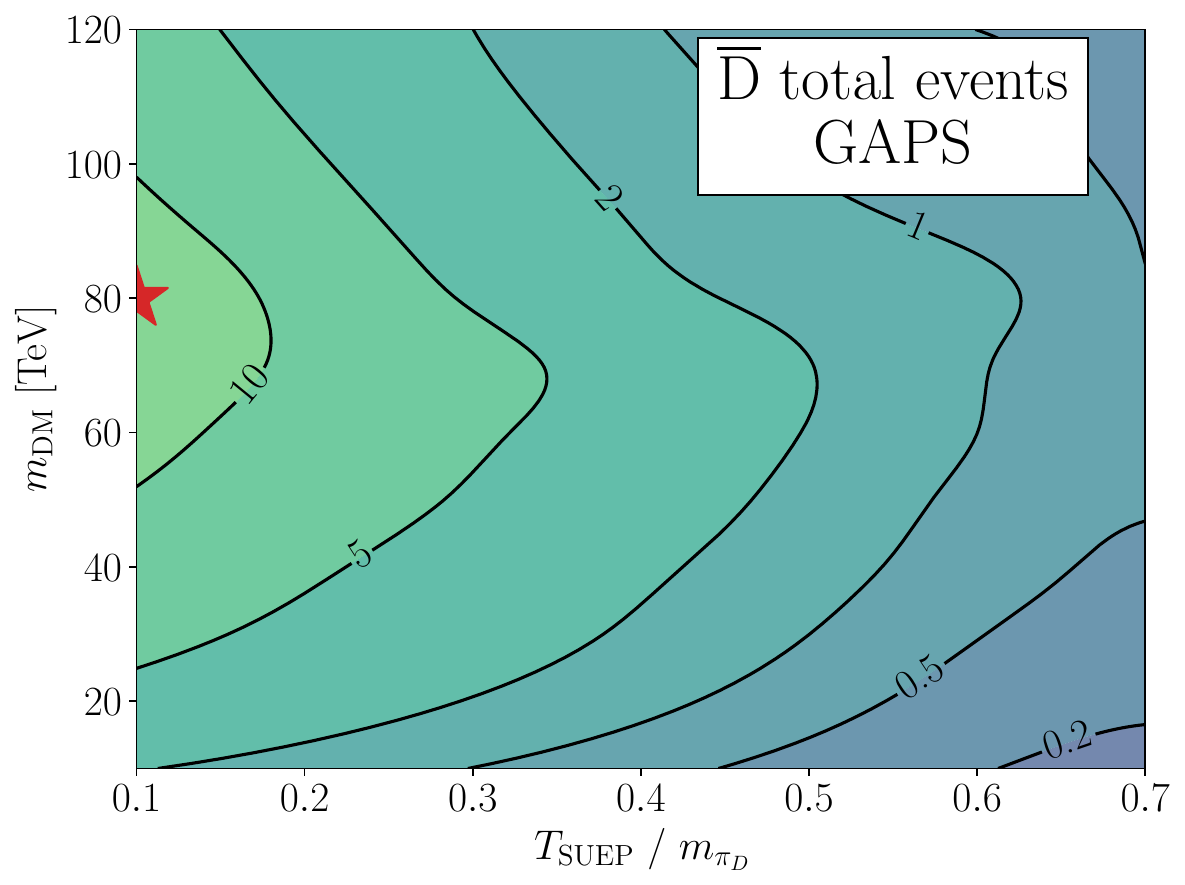}
  \includegraphics[width=0.32\linewidth]{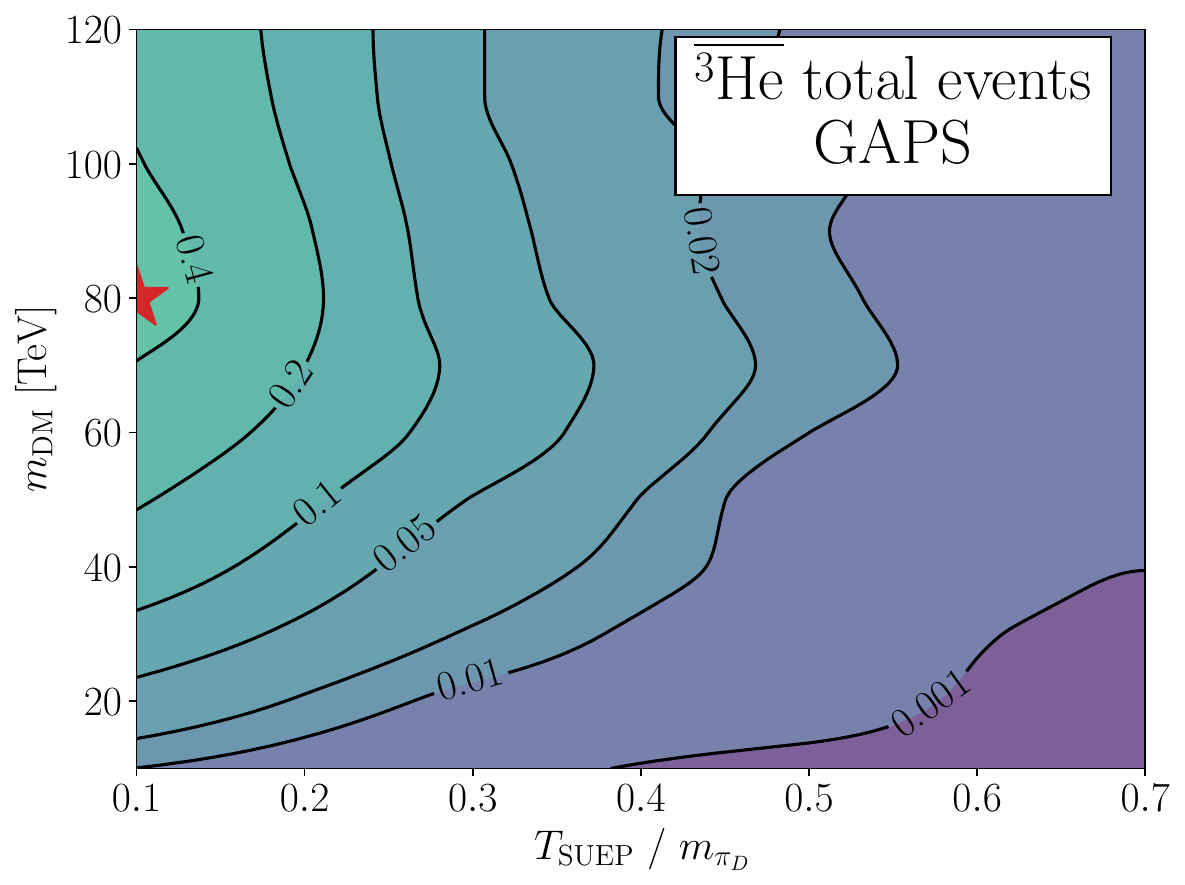}
  \includegraphics[width=0.32\linewidth]{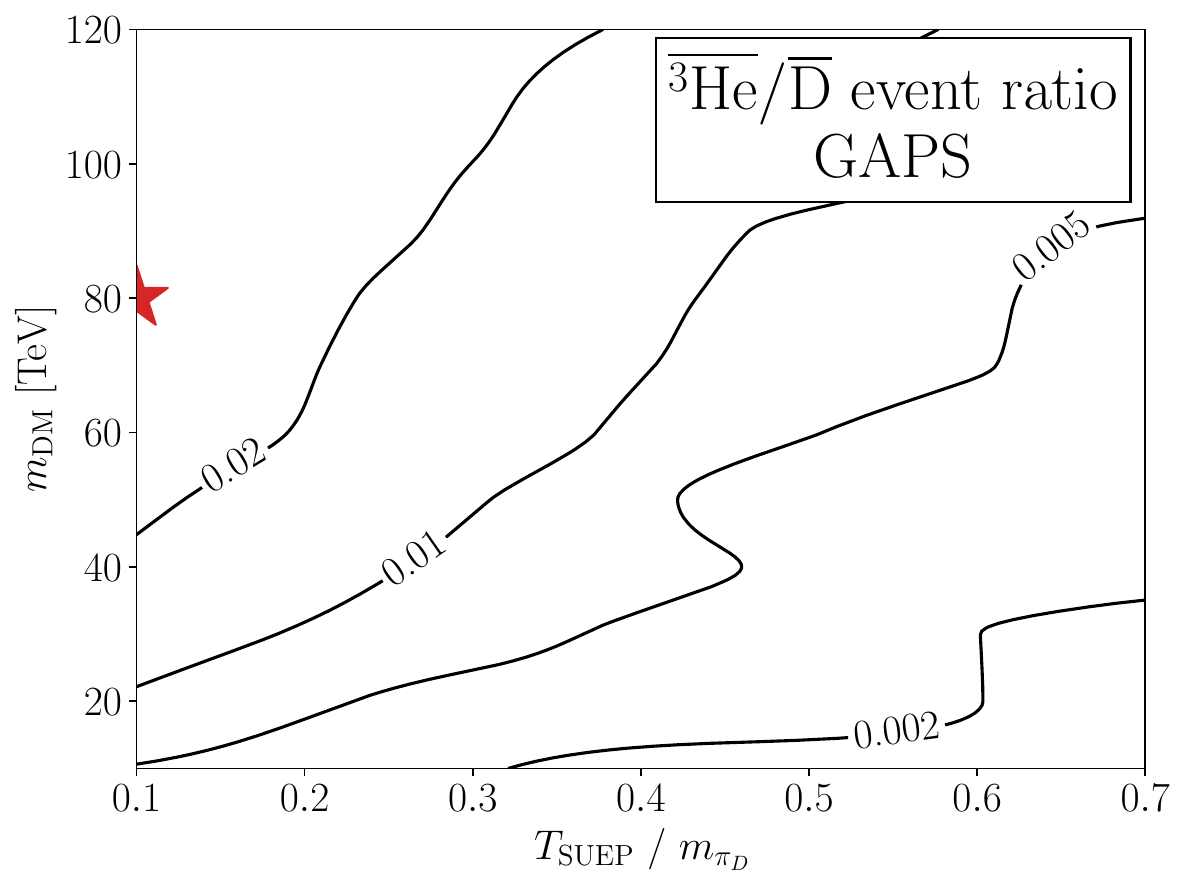}\\
  \includegraphics[width=0.32\linewidth]{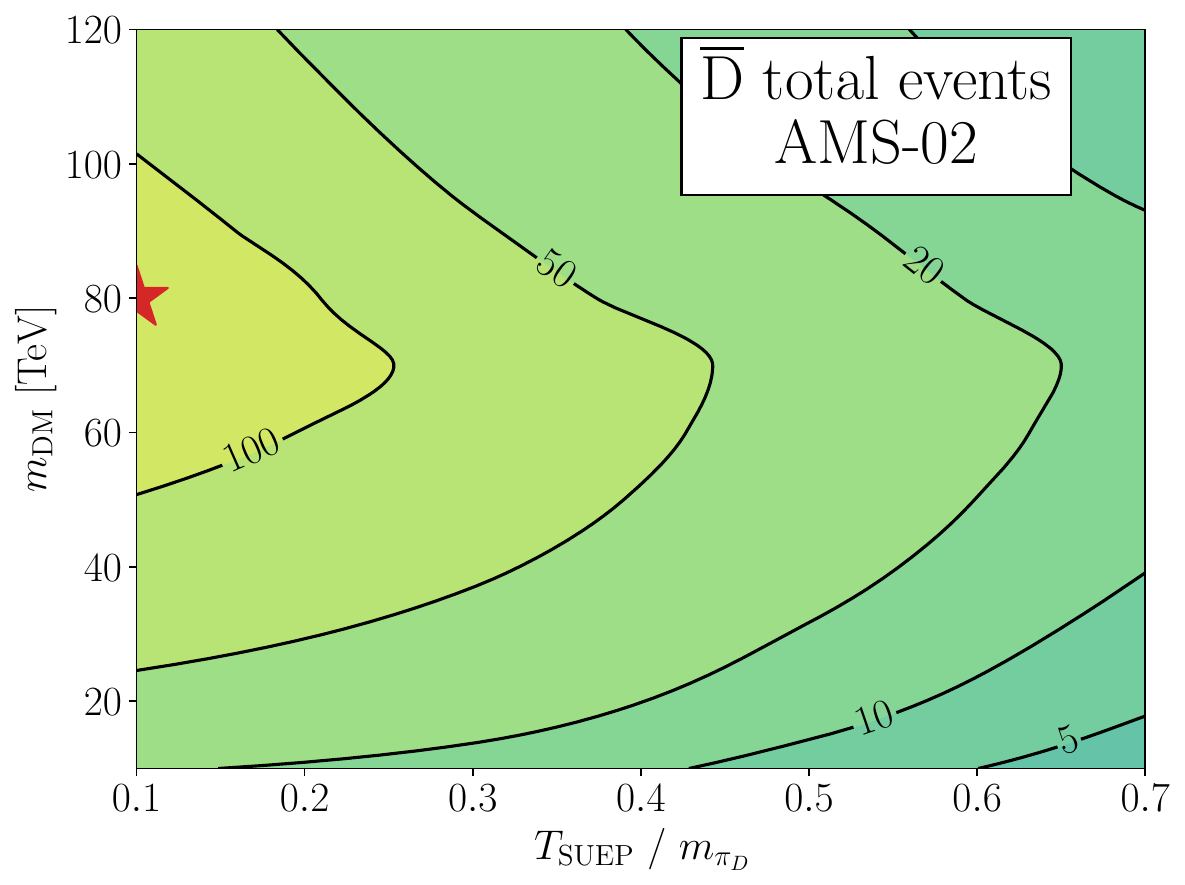}
  \includegraphics[width=0.32\linewidth]{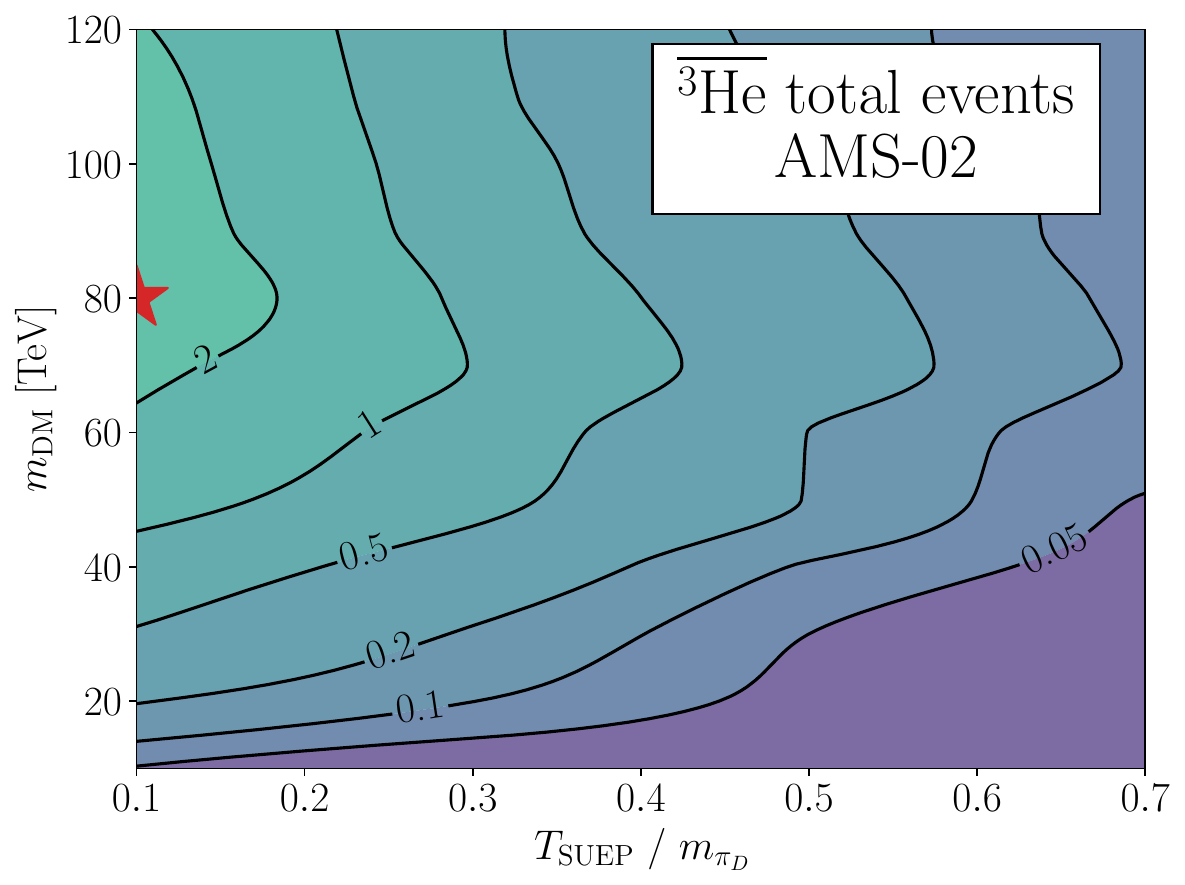}
  \includegraphics[width=0.32\linewidth]{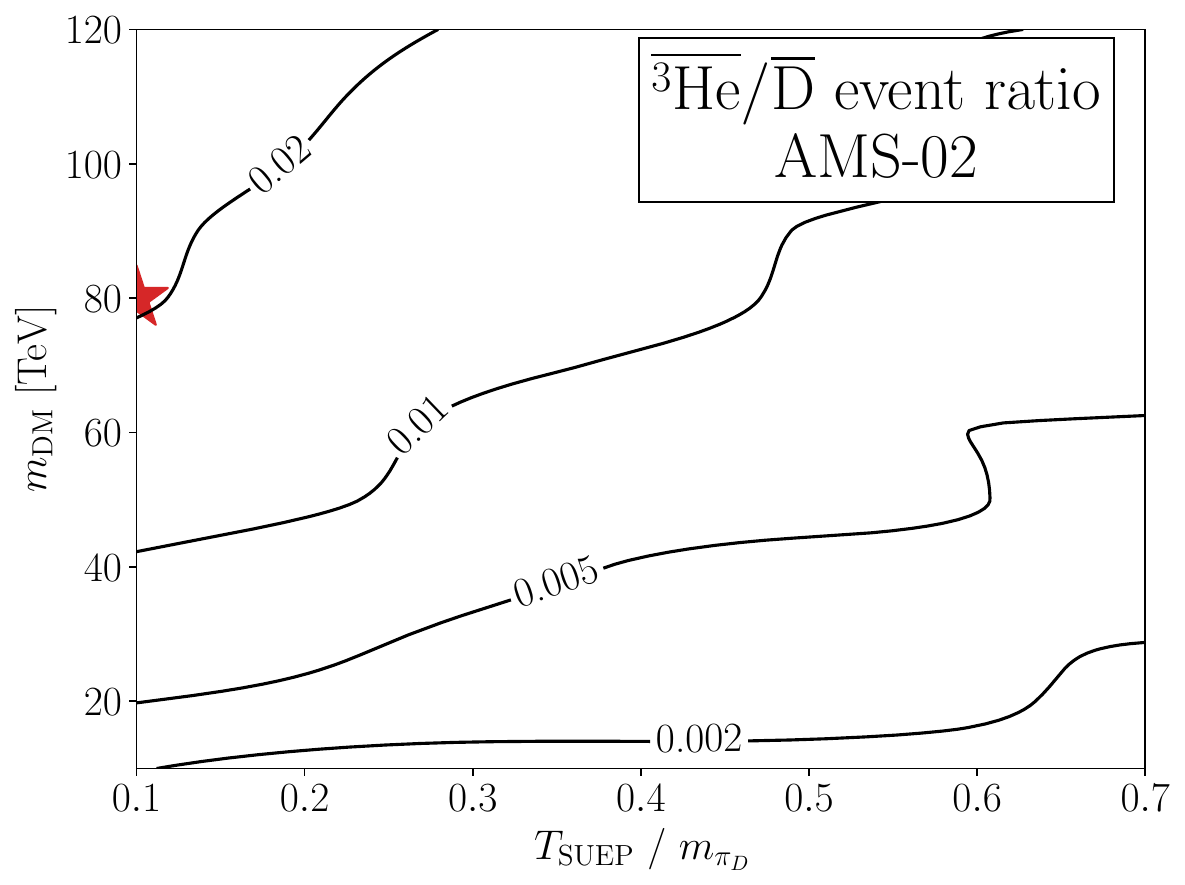}
  \caption{Expected event yields for the \texttt{1 SUEP} scenario, shown as a function of the DM mass $m_{\rm DM}$ and the SUEP temperature ratio $T_{\rm SUEP}/m_{\pi_D}$. The top (bottom) row refer to the \textsf{GAPS} (AMS-02) flux sensitivity. Left and middle panels show the total number of $\overline{\mathrm D}$ and $^3\overline{\mathrm{He}}$ events, respectively, while the right panels show the corresponding event ratio $N_{^3\overline{\mathrm{He}}}/N_{\overline{\mathrm D}}$. The red star indicates the benchmark point used in Fig.~\ref{fig:SUEP_fluxes}, where the propagated spectra are shown.}
  \label{fig:n_events_contours}
\end{figure*}

In Fig.~\ref{fig:SUEP_fluxes} we compare representative propagated fluxes in the SUEP framework with the \textsf{AMS-02} and \textsf{GAPS} sensitivity estimates. The benchmark points are chosen in the region of parameter space that maximises antinuclei production (low $T_{\rm SUEP}/m_{\pi_D}$ and $m_{\rm SUEP}$ close to the kinematic scale set by the annihilation), while the annihilation cross sections are fixed according to Sec.~\ref{sec:sigmav} so as to satisfy the unitarity bound and not exceed the adopted \textsf{AMS-02} antiproton requirement.

\begin{figure*}
  \centering
  \includegraphics[width=0.32\linewidth]{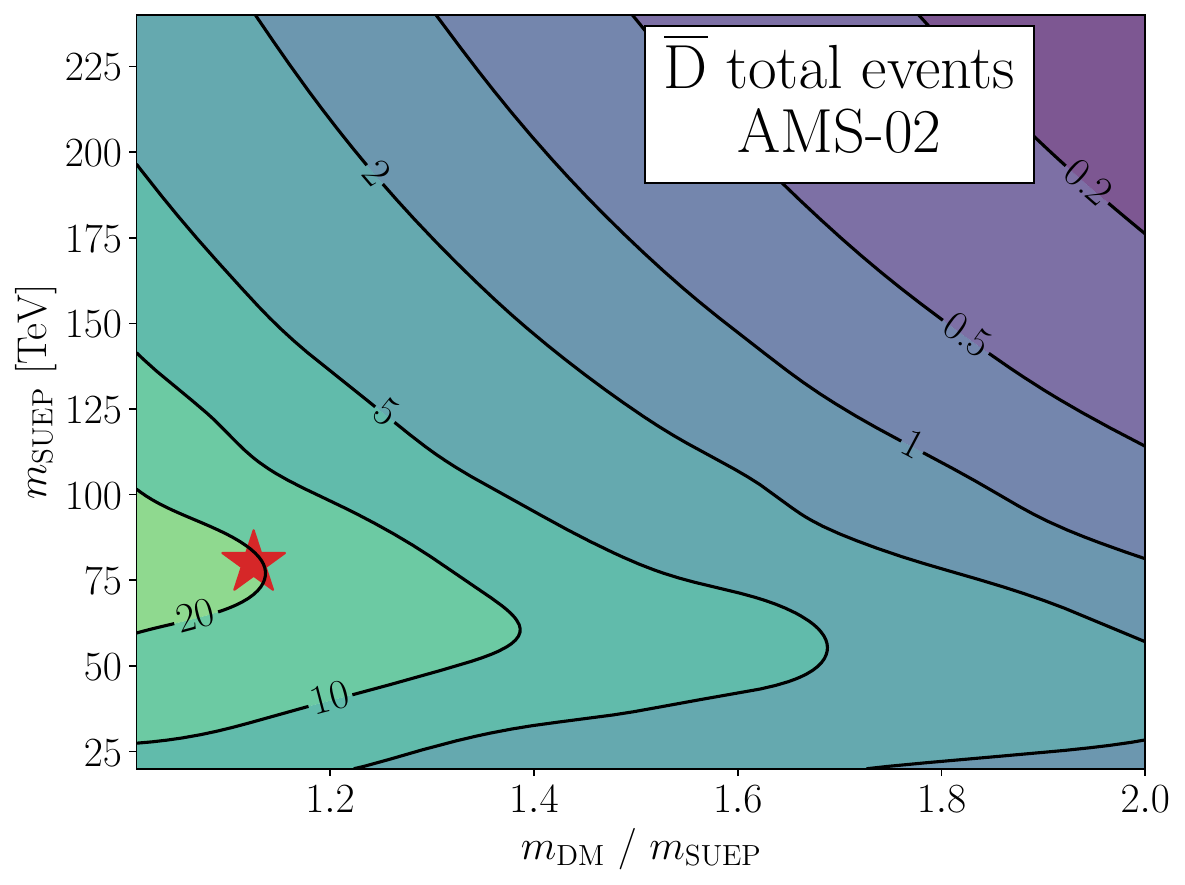}
  \includegraphics[width=0.32\linewidth]{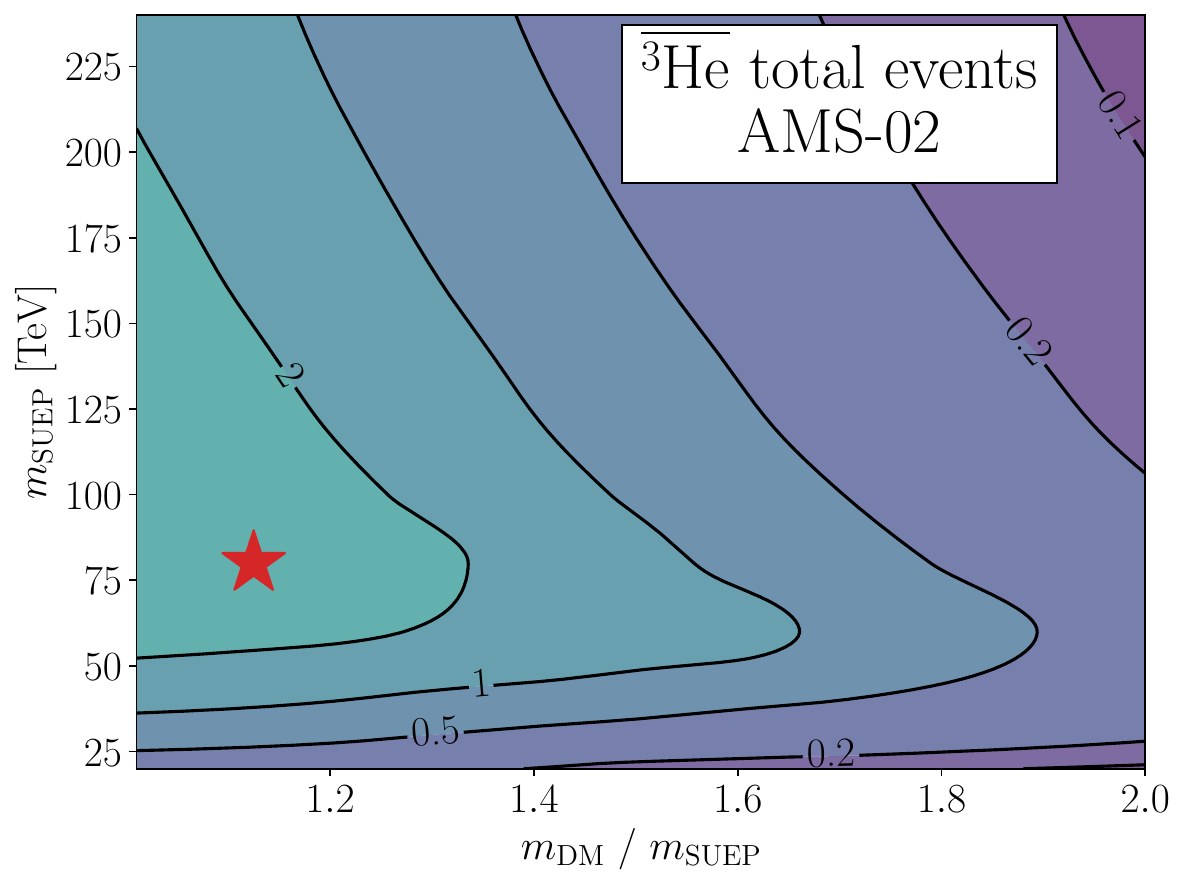}
  \includegraphics[width=0.32\linewidth]{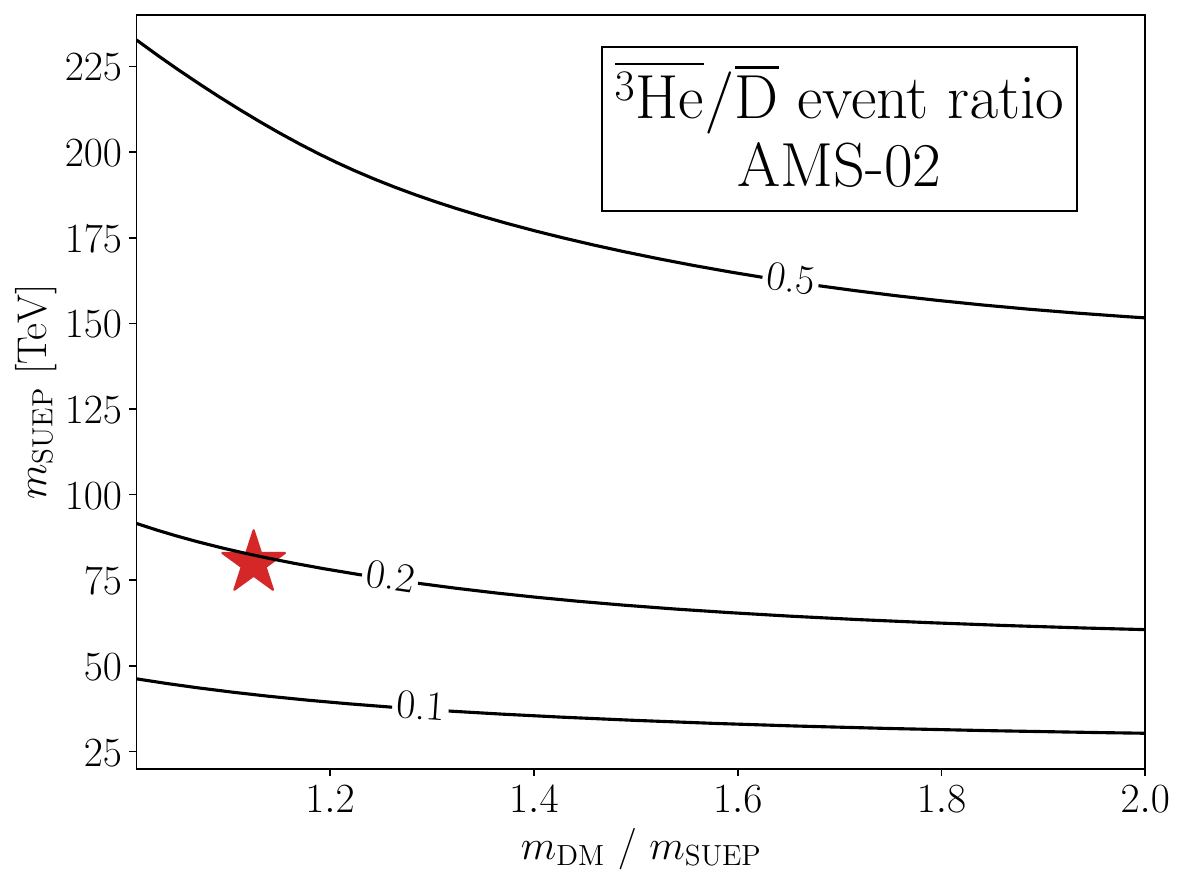}
  \caption{Expected event yields for the \texttt{2 SUEPs} scenario, shown as a function of the SUEP mass $m_{\rm SUEP}$ and the boost parameter $m_{\rm DM}/m_{\rm SUEP}$ (with $T_{\rm SUEP}/m_{\pi_D}=0.1$ fixed). Only \textsf{AMS-02} is shown, since the boosted spectra lie largely outside the low-energy window relevant for \textsf{GAPS}. Left and middle panels show the total number of $\overline{\mathrm D}$ and $^3\overline{\mathrm{He}}$ events, and the right panel shows $N_{^3\overline{\mathrm{He}}}/N_{\overline{\mathrm D}}$. The red star indicates the benchmark point used in Fig.~\ref{fig:SUEP_fluxes}.}
  \label{fig:n_events_contours_boosted}
\end{figure*}

For $\overline{\mathrm D}$ (middle panel of Fig.~\ref{fig:SUEP_fluxes}), the \texttt{1 SUEP} case yields a comparatively softer spectrum, with sizeable flux already below $1~\mathrm{GeV/n}$. In this low-energy region, the predicted $\overline{\mathrm D}$ flux can lie $\mathcal{O}(10)$ above the \textsf{AMS-02} sensitivity estimate and is the only scenario among those shown that can plausibly enter the \textsf{GAPS} sensitivity window. At higher kinetic energies, where \textsf{AMS-02} has its best sensitivity to antideuterons (in particular around $K/n\simeq 2$--$4~\mathrm{GeV/n}$), both the \texttt{1 SUEP} and \texttt{2 SUEPs} benchmarks can exceed the \textsf{AMS-02} sensitivity by large factors.
Instead, the WIMP benchmark remains below the experiment sensitivites for energies below 1 GeV/n while it could be detectable by \textsf{AMS-02} between 2 and 4 GeV/n. We stress, however, that the interpretation of sub-GeV/n fluxes is subject to larger propagation systematics (solar modulation, energy losses, and possible low-rigidity breaks), whereas the multi-GeV/n region is comparatively more robust.

The bottom panel of Fig.~\ref{fig:SUEP_fluxes} shows the $^3\overline{\mathrm{He}}$ flux. Here the two SUEP scenarios populate different energy ranges: the \texttt{1 SUEP} spectrum is softer and can marginally approach the optimistic \textsf{GAPS} sensitivity estimate at sub-GeV/n energies, while the \texttt{2 SUEPs} case is shifted to higher kinetic energies and can therefore be more relevant for \textsf{AMS-02} in the $K/n\simeq 3$--$10~\mathrm{GeV/n}$ range. The WIMP benchmark flux remains orders of magnitude below both \textsf{AMS-02} and \textsf{GAPS} sensitivity estimates across the full energy range shown.

We now translate these fluxes into expected event yields considering that the \textsf{AMS-02} and \textsf{GAPS} sensitivities refer to the detection of 2 events. Fig.~\ref{fig:n_events_contours} shows the predicted number of events for the \texttt{1 SUEP} scenario. For $\overline{\mathrm D}$ we adopt the \textsf{AMS-02} and \textsf{GAPS} sensitivity curves shown in Fig.~\ref{fig:SUEP_fluxes}. For $^3\overline{\mathrm{He}}$ we use the \textsf{AMS-02} sensitivity estimate presented in Sec.~\ref{sec:sensitivity} and, for \textsf{GAPS}, the expected sensitivity value of $1.3\times 10^{-6}\;\mathrm{(m^{2}\,s\,sr\,GeV/n)}^{-1}$. In addition, at each parameter point we use the maximal annihilation cross section consistent with Sec.~\ref{sec:sigmav} (unitarity and the antiproton requirement, and compatibility with the $\gamma$-ray bounds discussed there). The resulting event numbers therefore represent \emph{maximal} yields for our setup; for smaller values of $\langle\sigma v\rangle$ they scale linearly.

Overall, Fig.~\ref{fig:n_events_contours} shows that large regions of parameter space can yield observable antinuclei counts at \textsf{AMS-02} without producing an antiproton excess. In general we observe that decreasing $T_{\rm SUEP}/m_{\pi_D}$ increases the event rate, as a softer dark {\meson} spectra enhances antinuclei coalescence. Additionally an intermediate mass range (60 TeV $\lesssim m_{\rm{DM}}\lesssim$ 100 TeV) is identified that maximises the yield; for $m_{\rm DM}\lesssim 60~\mathrm{TeV}$ the enhancement is reduced due to fewer dark {\meson}s for smaller SUEP masses and, in our prescription, stronger rescaling of $\langle\sigma v\rangle$ due to antiproton constraints, while for $m_{\rm DM}\gtrsim 100~\mathrm{TeV}$ the annihilation flux becomes increasingly suppressed by the DM mass scaling. Similar trends are found for the \textsf{GAPS} expected event rate for $\overline{\rm{D}}$, but with only marginal sensitivity for $^3\overline{\rm{He}}$. Thus, the final number of expected $^3\overline{\rm{He}}$ events in \textsf{GAPS} is highly dependent on the finalized sensitivity which currently is affected by more than one-order-of-magnitude uncertainty (see Fig.~\ref{fig:SUEP_fluxes}).

Focusing on the benchmark point indicated by the red star in Fig.~\ref{fig:n_events_contours} (and shown in Fig.~\ref{fig:SUEP_fluxes}), we obtain $\sim 155$ $\overline{\mathrm D}$ events and $\sim 3$ $^3\overline{\mathrm{He}}$ events at \textsf{AMS-02}, corresponding to
$N_{^3\overline{\mathrm{He}}}/N_{\overline{\mathrm D}}\simeq 0.02$.
This ratio can increase mildly at larger $m_{\rm DM}$ for fixed low $T_{\rm SUEP}/m_{\pi_D}$, since the $\overline{\mathrm D}$ yield decreases faster than the $^3\overline{\mathrm{He}}$ yield.
For the same theory parameters, \textsf{GAPS} would instead observe $\sim 15$ $\overline{\mathrm D}$ events and $\sim 0.5$ $^3\overline{\mathrm{He}}$ events, i.e.\ $N_{^3\overline{\mathrm{He}}}/N_{\overline{\mathrm D}}\simeq 0.03$.
A key, model-independent point is that these \emph{relative} event rates are approximately insensitive to $\langle\sigma v\rangle$, since both $\overline{\mathrm D}$ and $^3\overline{\mathrm{He}}$ fluxes scale linearly with the annihilation cross section.

For the \texttt{2 SUEPs} scenario, Fig.~\ref{fig:n_events_contours_boosted} shows the expected antinuclei yields at \textsf{AMS-02}. As in the \texttt{1 SUEP} case, we apply the same sensitivity assumptions and cross-section prescription (Sec.~\ref{sec:sigmav}). Here $m_{\rm SUEP}$ is a free parameter and the ratio $m_{\rm DM}/m_{\rm SUEP}$ controls the boost, and therefore the location of the spectral peak in kinetic energy per nucleon. We find again that sizeable regions of parameter space can lead to observable event rates, with an intermediate $m_{\rm SUEP}$ range maximising the total yield and with mild boosts ($m_{\rm DM}$ slightly larger than $m_{\rm SUEP}$) generally preferred.

At the benchmark point highlighted in Fig.~\ref{fig:n_events_contours_boosted} ($m_{\rm SUEP}=80~\mathrm{TeV}$, $m_{\rm DM}=90~\mathrm{TeV}$; see Fig.~\ref{fig:SUEP_fluxes}), we predict $\sim 19$ $\overline{\mathrm D}$ events and $\sim 4$ $^3\overline{\mathrm{He}}$ events, corresponding to $N_{^3\overline{\mathrm{He}}}/N_{\overline{\mathrm D}}\simeq 0.19$.
The ratio can be increased to $N_{^3\overline{\mathrm{He}}}/N_{\overline{\mathrm D}}\gtrsim 0.5$ by moving to larger $m_{\rm SUEP}$ and larger $m_{\rm DM}/m_{\rm SUEP}$, at the expense of a smaller overall number of expected events.

\section{Conclusions}
\label{sec:conclusion}

The strong phase-space suppression associated with forming heavy antinuclei in ordinary QCD showers makes it difficult for standard WIMP annihilation to yield simultaneously observable fluxes of $\overline{\mathrm D}$ and ${}^3\overline{\mathrm{He}}$ without overshooting the antiproton measurements from \textsf{AMS-02}. This tension motivates mechanisms in which hadronisation proceeds in a qualitatively different way, enhancing the probability that antinucleons are produced nearby in phase space and can coalesce into heavier antinuclei.

SUEPs provide a concrete realization of such non-standard dynamics. In confining dark sectors with a large and approximately quasi-conformal ’t~Hooft coupling over an extended range of scales, the shower is dominated by the emission of a very large multiplicity of soft dark hadrons (here, dark {\meson}s). The resulting high-multiplicity, low-momentum final state is qualitatively distinct from QCD and increases the likelihood that antinucleons emerge within the coalescence domain, thereby enhancing antideuteron and antihelium production. In this sense, antinuclei offer a potentially unique indirect probe of hidden confining dynamics.

In this work we performed a dedicated study of antinuclei production in SUEP scenarios by combining:
(i) a full Monte Carlo treatment of dark-{\meson} production and decay into SM quarks,
(ii) an event-by-event coalescence prescription validated against \textsf{ALEPH} and \textsf{ALICE} data,
and (iii) Galactic propagation of the resulting antiproton, antideuteron, and antihelium fluxes to Earth.

We considered two kinematic regimes depending on the DM--SUEP mass hierarchy:
\texttt{1 SUEP} ($\mathrm{DM}\,\mathrm{DM}\to S^\ast$), in which DM annihilates through an off-shell mediator and the antinuclei spectra are comparatively soft, peaking at low kinetic energies; and
\texttt{2 SUEPs} ($\mathrm{DM}\,\mathrm{DM}\to S\,S$), in which two on-shell mediators are produced and the antinuclei spectra are boosted to higher energies.
Across broad regions of parameter space in both cases, we find a substantial enhancement of the \emph{relative} yields $N_{\overline{\mathrm D}}/N_{\bar p}$ and $N_{{}^3\overline{\mathrm{He}}}/N_{\bar p}$ compared to ordinary QCD showers in standard WIMP annihilation.

The annihilation cross sections used in our predictions are chosen to ensure compatibility with existing indirect constraints: we require that the DM-induced antiproton contribution remains well below the \textsf{AMS-02} measurements, and we verify that the associated $\gamma$-ray emission from Milky Way dwarf spheroidal galaxies is consistent with \emph{Fermi}-LAT limits. In addition, for DM masses above a few tens of TeV we enforce the partial-wave unitarity upper bound on $\langle\sigma v\rangle$.

At the source level, the enhancement is maximized for low SUEP temperatures, $T_{\rm SUEP}/m_{\pi_D}\sim 0.1$, and for masses in the tens-to-hundreds of TeV range. For representative parameters around $m_{\rm DM}\sim 80$--$90~\mathrm{TeV}$ and $T_{\rm SUEP}/m_{\pi_D}\sim 0.1$, we obtain a two--orders-of-magnitude enhancement of the antideuteron yield relative to the standard WIMP expectation, and up to $10^5$ enhancements for ${}^3\overline{\mathrm{He}}$ (reflecting the stronger baseline coalescence suppression in QCD showers). The enhancement decreases as $T_{\rm SUEP}/m_{\pi_D}$ increases, as $m_{\rm DM}$ decreases, and (in the \texttt{2 SUEPs} case) for strong boosts corresponding to $m_{\rm SUEP}\ll m_{\rm DM}$.

Moreover, determining the precise enhancement in antinuclei production is subject to uncertainties in modeling the SUEP\@.  We have assumed the SUEP is dominated by just high multiplicity $\pi_D$ emission.  However, for lower values of $T_{\rm SUEP}/m_{\pi_D}$, we expect the lightest glueball mass to be closer to $m_{\pi_D}$, allowing the SUEP to emit some dark sector glueball mesons as well,  diluting the fraction of the SUEP that ends in just $\pi_D \rightarrow t\bar{t}$.  Also, the momentum distribution of the $\pi_D$s was assumed to be thermal, and this is subject to uncertainties in both the  coefficient (that we took to be $1$) relating $T_{\rm SUEP}$ and $\Lambda_D$, as well as deviations from a purely thermal spectrum.

Beyond source-level yields, we provided predictions for the \emph{detectable} number of events at \textsf{AMS-02} and \textsf{GAPS}. A crucial ingredient is our updated and more realistic estimate of the \textsf{AMS-02} ${}^3\overline{\mathrm{He}}$ sensitivity, based on acceptance, livetime, and survival probabilities rather than pre-launch flux-ratio extrapolations. Using this sensitivity, we find that in the \texttt{1 SUEP} scenario an observable flux of both $\overline{\mathrm D}$ and ${}^3\overline{\mathrm{He}}$ can be achieved for low temperatures $T_{\rm SUEP}/m_{\pi_D}\sim 0.1$ and sufficiently heavy DM, $m_{\rm DM}\gtrsim 20~\mathrm{TeV}$. In this regime the predicted event ratio is typically $N_{{}^3\overline{\mathrm{He}}}/N_{\overline{\mathrm D}}\sim 10^{-2}$,
substantially larger than in standard WIMP scenarios (where it is $\sim 10^{-4}$), while still below the very tentative ratio suggested by past \textsf{AMS-02} conference presentations~\cite{Miapp2022Dbar,Oliva:JENAA:2024}.
In the \texttt{2 SUEPs} scenario, we find that $N_{{}^3\overline{\mathrm{He}}}/N_{\overline{\mathrm D}}$ can reach $\gtrsim 0.1$ for $m_{\rm SUEP}\gtrsim 50~\mathrm{TeV}$, with benchmark regions yielding event counts of order $\sim 29$ and $\sim 6$ for $\overline{\mathrm D}$ and ${}^3\overline{\mathrm{He}}$, respectively. This corresponds to an event ratio, $N_{{}^3\overline{\mathrm{He}}}/N_{\overline{\mathrm D}}\sim 0.2$, closer to the tentative \textsf{AMS-02} ratio~\cite{Miapp2022Dbar,Oliva:JENAA:2024}.

We also find that \textsf{GAPS} has (at best) marginal sensitivity to the \texttt{1 SUEP} scenario, potentially enabling a handful of $\overline{\mathrm D}$ events for low $T_{\rm SUEP}/m_{\pi_D}$ and large $m_{\rm DM}$, and (optimistically) potential sensitivity to ${}^3\overline{\mathrm{He}}$ depending on the assumed performance. Since \textsf{GAPS} is most sensitive below $\sim\!1~\mathrm{GeV/n}$, a careful interpretation will require controlling low-energy propagation uncertainties, including solar modulation and energy losses. Nevertheless, \textsf{GAPS} provides valuable complementarity to \textsf{AMS-02} and important cross-checks for any future signal.

Finally, collider searches for SUEP-like signatures---while challenging due to their soft and approximately isotropic final states---have begun to be explored at the LHC~\cite{Barron:2021SUEP,CMS:2024nca}. In the specific realizations relevant to this work, the dark {\meson} masses and portal structure (often involving heavy-flavor couplings, e.g.\ to tops) imply that production rates and trigger strategies are highly model-dependent, and that sensitivity may be limited at the LHC for the mass scales suggested by the most promising antinuclei regions. Future high-luminosity data, dedicated soft-object triggers, and/or higher-energy colliders could therefore play an important role in testing the broader class of confining dark sectors that give rise to SUEP-like phenomenology. With \textsf{AMS-02} continuing to take data until at least 2030, the just completed and possible future \textsf{GAPS} balloon flights, and growing experimental interest in exotic hadronisation phenomena, the coming years may provide decisive information on the origin of antinuclei candidates. If confirmed, an interpretation in terms of SUEP-like dynamics would provide a striking signature of hidden confining physics and open a new frontier in dark-sector phenomenology.

\begin{acknowledgments} 
MDM, FD and NF acknowledge support from the research grant {\sl TAsP (Theoretical Astroparticle Physics)} funded by Istituto Nazionale di Fisica Nucleare (INFN). MDM acknowledges support from the Italian Ministry of University and Research (MUR), PRIN 2022 ``EXSKALIBUR – Euclid-Cross-SKA: Likelihood Inference Building for Universe’s Research'', Grant No. 20222BBYB9, CUP I53D23000610 0006, and from the European Union -- Next Generation EU.
The work of DC was supported in part by Discovery Grants from the Natural Sciences and Engineering Research Council of Canada (NSERC), the Canada Research Chair program, the Alfred P. Sloan Foundation, the Ontario Early Researcher Award, and the University of Toronto McLean Award.
The work of GDK was supported in part by the US Department of Energy under grant number DE-SC0011640.
The work of AB is supported by the US Department of Energy, Office of Science Graduate Research (SCGSR) program, which is administered by the Oak Ridge Institute for Science and Education under contract number DE-SC0014664. 
FD thanks the Department of Theoretical Physics of CERN, where part of this work was carried on.
\end{acknowledgments}

\newpage

\onecolumngrid
\appendix
\section{Antinuclei transport in the galactic environment}
\label{subsec:transport-eq}

We model the Galactic transport of CR antiprotons ($\bar p$), antideuterons ($\overline{\mathrm D}$), and antihelium-3 (${}^3\overline{\mathrm{He}}$) in a two-zone, axisymmetric diffusion setup. A thin gaseous disk of half-thickness $h\simeq 100\,\mathrm{pc}$ is embedded in a cylindrical magnetic halo of radius $R$ and half-height $L$. We impose free-escape boundary conditions at $z=\pm L$ and $r=R$~\cite{Strong:2007nh,Maurin:2001sj,Donato:2001ms,Genolini:2021doh,DiMauro:2021qcf}. Unless stated otherwise, we use the kinetic energy per nucleon $K/n$ (or rigidity $\mathcal R\equiv pc/|Z|e$) as spectral variable, with $\beta\equiv v/c$.

For a stable species $X\in\{\bar p,\,\overline{\mathrm D},\,{}^3\overline{\mathrm{He}}\}$, and neglecting diffusive reacceleration and continuous energy losses (more on this below), the stationary transport equation for the differential density $\psi_X(\bm x,p)$ (number per unit momentum) reads:
\begin{eqnarray}
0 &=& \bm\nabla\!\cdot\!\big( D(\mathcal R)\,\bm\nabla\psi_X \big)
 - \bm\nabla\!\cdot\!\big( \bm V_c\,\psi_X \big)
 - \Gamma^{\rm inel}_X\,\psi_X
 + Q_X^{\rm prim}+Q_X^{\rm sec}+Q_X^{\rm ter}\, .
\label{eq:master}
\end{eqnarray}
Here $D(\mathcal R)$ is the spatial diffusion coefficient and $\bm V_c$ is the convective wind velocity, which we take as a constant flow perpendicular to the disk with $V_c=13~\mathrm{km/s}$ as in Ref.~\cite{DiMauro:2023jgg}. The diffusion coefficient follows the double-broken power-law form of the {\tt diff.brk} model~\cite{DiMauro:2023jgg}, motivated by recent precision measurements of primary and secondary CR spectra. Inelastic losses are confined to the disk and are encoded by
$\Gamma_X^{\rm inel}(p)= 2h\,\delta(z)\, n_{\rm ISM}\,\beta c\,\sigma_{X}^{\rm inel}(p)$, with $n_{\rm ISM}=n_\mathrm{H}+4^{2/3}n_\mathrm{He}$ accounting for H/He targets. For $\bar p$--H we take $\sigma^{\rm inel}_{\bar p}$ from Ref.~\cite{Tan:1983de}; for $\overline{\mathrm D}$ and ${}^3\overline{\mathrm{He}}$ we rescale as $\sigma^{\rm inel}_{X} = A_{\rm proj}^{0.8}\,\sigma^{\rm inel}_{\overline{p}}$, where $A_{\rm proj}$ is the projectile mass number (cf.~\cite{Korsmeier:2018gcy,Orusa:2022pvp}). The source term includes a primary component $Q_X^{\rm prim}$ (e.g.\ DM annihilation or astrophysical sources) and a secondary component $Q_X^{\rm sec}$ from hadronic production of antinuclei in CR--gas collisions. A tertiary contribution $Q_X^{\rm ter}$ from non-annihilating inelastic scattering redistributes particles toward lower momenta~\cite{Donato:2001ms}; however, we neglect this term in the following since it is subdominant in the energy range relevant for our DM signals (see, e.g.,~\cite{Korsmeier:2017xzj}).

\begin{figure}
  \centering
  \includegraphics[width=0.49\linewidth]{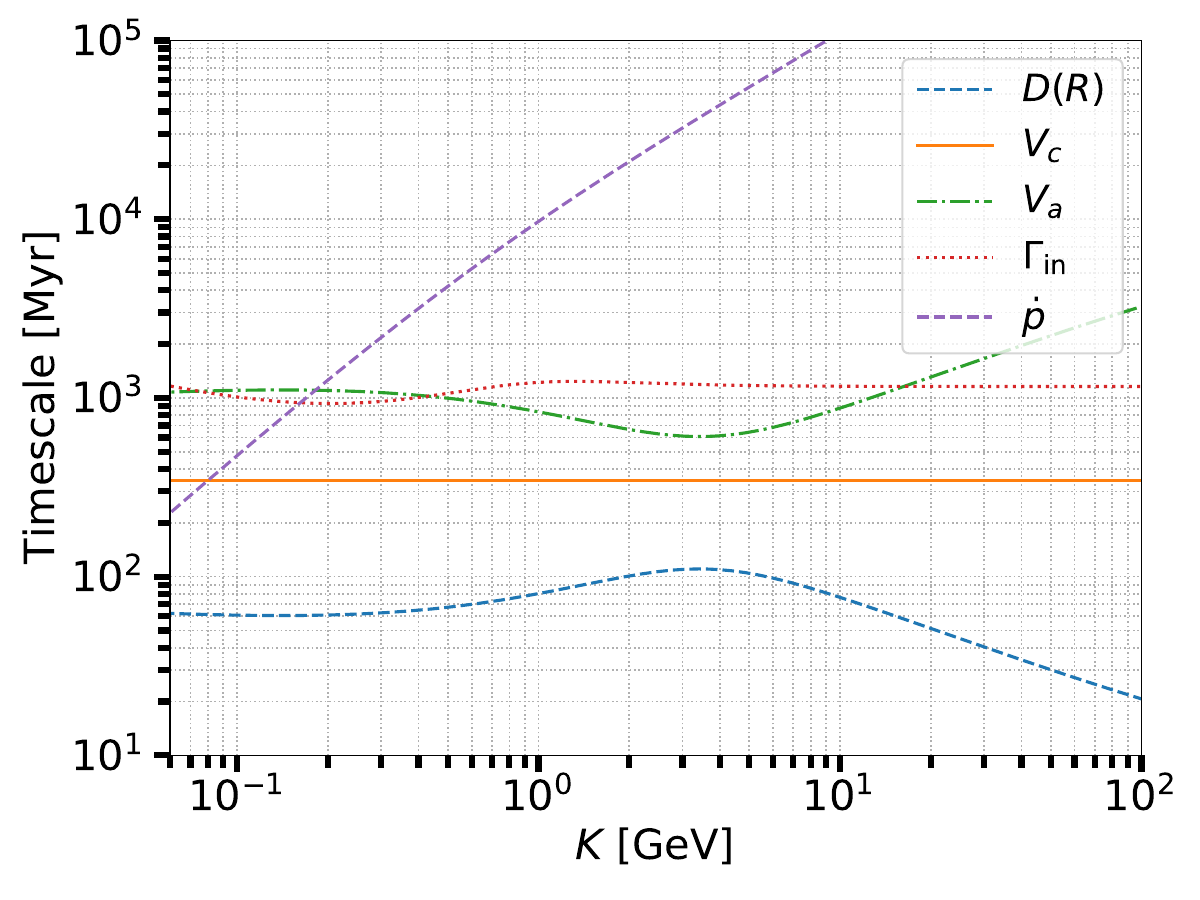}
  \caption{Typical time scales of spatial diffusion governed by $D(\mathcal{R})$ (blue dashed), convection with velocity $V_c$ (orange solid), inelastic scatterings encoded by $\Gamma_{\rm inel}$ (red dotted), continuous energy losses ($\dot{p}$, purple dashed) and reacceleration quantified by the Alfv\'en velocity $V_A$ (green dot-dashed), for the propagation setup described in the text.}
\label{fig:tauprop}
\end{figure}

As stated above, in this work we also neglect diffusive reacceleration and continuous energy losses, consistently with the {\tt diff.brk} setup in Ref.~\cite{DiMauro:2023jgg}, which fits the \textsf{AMS-02} nuclei data. As a further justification,
Fig.~\ref{fig:tauprop} compares the characteristic time scales associated with diffusion, convection, and inelastic interactions for the {\tt diff.brk} propagation setup~\cite{DiMauro:2023jgg}. We see that diffusion dominates over the full energy range of interest, while convection provides an approximately energy-independent competing sink; inelastic losses become most relevant at low energies, where $\beta$ is small and the grammage accumulated in the disk is the largest.

In order to determine the interstellar fluxes, we solve Eq.~\ref{eq:master} semi-analytically via a Bessel expansion. For the propagation parameters we adopt the {\tt diff.brk} setup of Ref.~\cite{DiMauro:2023jgg}, which provides a good description of \textsf{AMS-02} data for several primary and secondary CR species.

\begin{figure}
  \centering
  \includegraphics[width=0.49\linewidth]{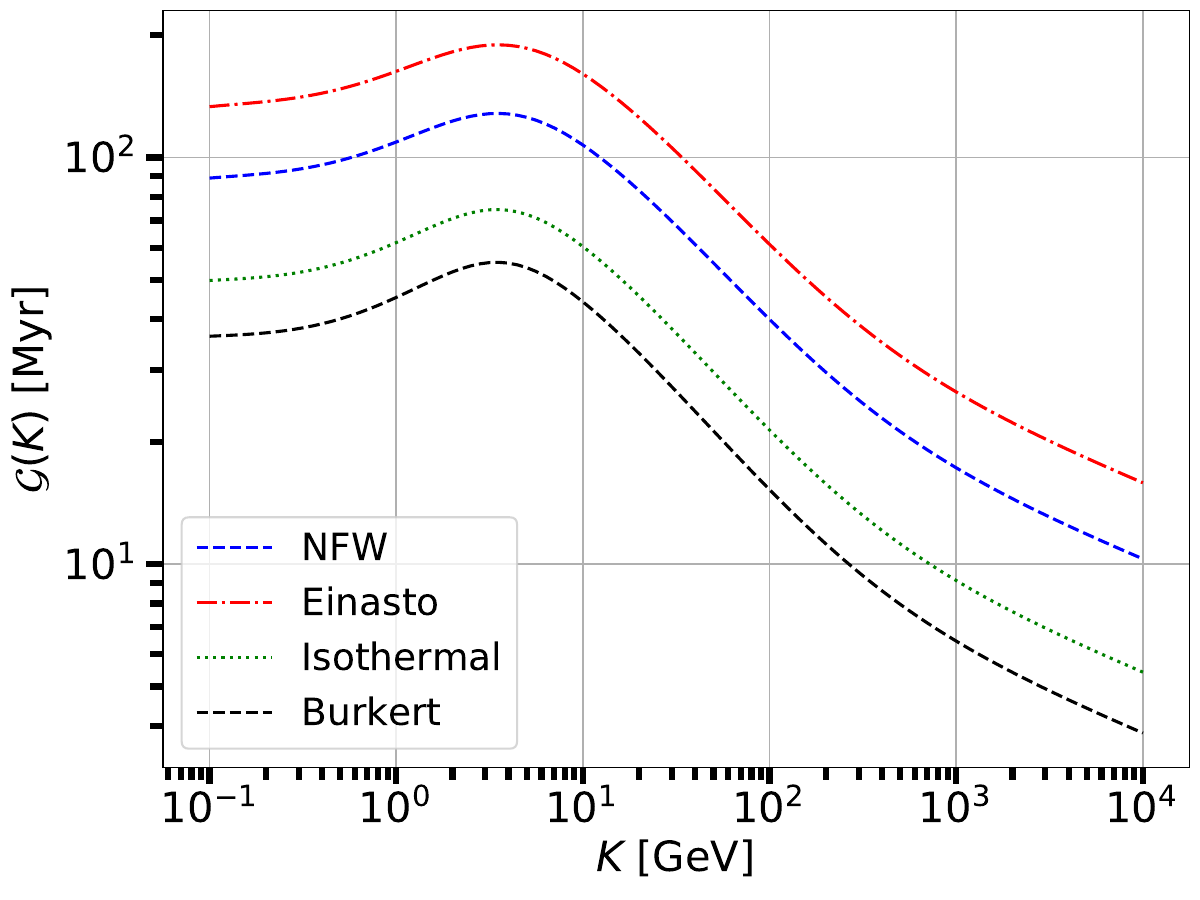}
  \includegraphics[width=0.49\linewidth]{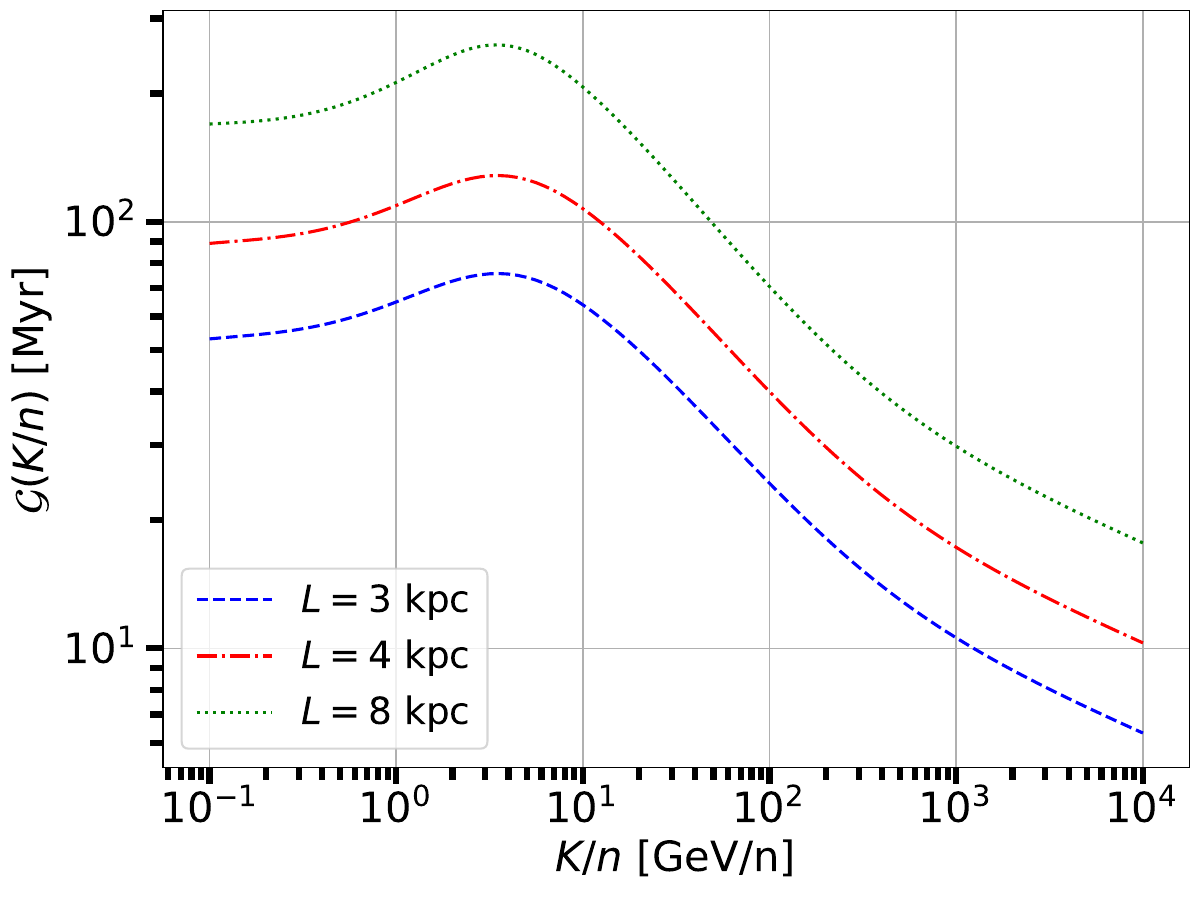}
  \caption{Left: Propagation function $\mathcal{G}(K)$ for antiprotons for different DM density profiles, for fixed halo height $L=4~\mathrm{kpc}$.
  Right: $\mathcal{G}(K/n)$ for three representative halo heights, $L=3,4,8~\mathrm{kpc}$. Each value of $L$ is associated with the corresponding best-fit diffusion normalization $D_0$ in the {\tt diff.brk} setup~\cite{DiMauro:2023jgg}.}
\label{fig:cases}
\end{figure}

Under the assumptions stated above, the transport equation in Eq.~\ref{eq:master} admits an analytic solution for primary sources located in the halo, as first demonstrated in Ref.~\cite{Barrau:2001ev} (see also Ref.~\cite{Maurin:2001sj}). Therefore, the factorized propagation function $\mathcal{G}(K)$, which encodes the transport effects, can be written as:
\begin{equation}
\mathcal{G}(K)
= \sum_{n=1}^{\infty}
J_0\!\left(\zeta_n\,\frac{r_\odot}{R}\right)\,
\exp\!\left[-\,\frac{V_{\rm c}\,L}{2\,D(\mathcal R)}\right]\,
\frac{ y_n(L) }{ A_n\,\sinh\!\big(S_n L/2\big) }\,,
\tag{28}
\label{eq:propfun}
\end{equation}
with
\begin{equation}
y_n(z)
= \frac{4}{J_1^2(\zeta_n)\,R^2}
\int_0^{R}\!dr\,r\,J_0\!\left(\zeta_n \frac{r}{R}\right)
\int_0^{L}\!dz'\,
\exp\!\left[\frac{V_{\rm c}\,(L-z')}{2D(\mathcal R)}\right]\,
\sinh\!\left(\frac{S_n\,(L-z')}{2}\right)
\left(\frac{\rho(r,z')}{\rho_\odot}\right)^{\!2},
\tag{29}
\label{eq:yn}
\end{equation}
and
\begin{eqnarray}
S_n &\equiv& \sqrt{\frac{V_{\rm c}^2}{D(\mathcal R)^2} + \frac{4\,\zeta_n^{\,2}}{R^2}}\;, \\
A_n &\equiv& 2h\,\Gamma^{\rm inel}_X(K) + D(\mathcal R)\,S_n\,\coth\!\big(S_n L/2\big),
\end{eqnarray}
where $\zeta_n$ are the zeros of $J_0$.
Fig.~\ref{fig:cases} shows $\mathcal{G}(K)$ as a function of kinetic energy per nucleon, illustrating its dependence on the DM density profile and on the propagation setup.

Solar modulation is modeled with the force-field approximation, using a Fisk potential $\phi_\odot$ (charge-sign and other effects beyond the force-field approximation may become relevant for $\mathcal R\lesssim \mathrm{few}\ \mathrm{GV}$)~\cite{GleesonAxford:1968}. For non-relativistic $\bar p$ (and analogously for $\overline{\mathrm D}$ and ${}^3\overline{\mathrm{He}}$), the flux at Earth is approximately related to the interstellar one by
\begin{equation}
\frac{d\Phi_{\bar p,\oplus}}{dK_\oplus}
= \frac{p_\oplus^2}{p^2}\,
\frac{d\Phi_{\bar p}}{dK}\,, \qquad
K = K_\oplus + |Z|\,e\,\phi_F\,,
\tag{31}
\end{equation}
where $\phi_F$ is the (possibly charge-sign dependent) Fisk potential, for which we use the values found in Ref.~\cite{DiMauro:2023jgg}.

Uncertainties related to the DM distribution enter through the assumed Galactic density profile and the local DM density value $\rho_\odot$. Varying the profile at fixed $\rho_\odot$ primarily rescales the propagation function $\mathcal{G}(K)$. Relative to an NFW profile, adopting Einasto, isothermal, or Burkert profiles with parameters from Ref.~\cite{Cirelli:2024ssz} changes the fluxes by at most $\sim\!1.5$ upward or $\sim\!0.4$ downward. Independently, varying the local density within $\rho_\odot=(0.4\pm0.1)\,\mathrm{GeV\,cm^{-3}}$~\cite{Cirelli:2024ssz} implies an overall factor $\sim\!1.8$ (upper) and $\sim\!0.6$ (lower). Combining these effects yields an approximate uncertainty band of $\sim\!2.7$ (upper) and $\sim\!0.25$ (lower) relative to our benchmark, making the DM distribution the dominant modeling uncertainty.

Within our diffusion--convection--inelastic framework, the leading propagation uncertainty is the halo half-height $L$. Isotopic beryllium measurements favor $L\simeq 3$--$8~\mathrm{kpc}$ (see, e.g., Refs.~\cite{Evoli:2019iih,Weinrich:2020ftb,Genolini:2021doh,DiMauro:2023oqx}), with $L=4~\mathrm{kpc}$ often adopted as a reference. Changing $L$ mainly induces a compensating change in the normalization $D_0$ of the diffusion coefficient while leaving the spectral shape of $\mathcal{G}(K)$ nearly unchanged~\cite{Evoli:2019iih,Weinrich:2020ftb,Genolini:2021doh,DiMauro:2023oqx}. In practice, $D_0$ scales approximately proportionally with $L$, so that $D_0/L$ remains roughly constant. Evaluating $\mathcal{G}(K)$ for $L=3,4,8~\mathrm{kpc}$ with the correspondingly rescaled $D_0$, we find a relative variation of order $-60\%$ to $+100\%$ around the reference setup.

Finally, uncertainties from the coalescence prescription are subdominant for our purposes. As discussed in Sec.~\ref{sec:coalescence}, Ref.~\cite{DiMauro:2024kml} finds that differences in antinuclei source spectra induced by alternative coalescence implementations are at the level of a few percent, comparable to Monte Carlo statistical uncertainties. The dominant systematics instead originate from the QCD modeling of the event generator (e.g.\ \textsc{Pythia}) and from the experimental data used for tuning (e.g.\ \textsf{ALEPH}), each contributing at the $\sim\!30\%$ level~\cite{ALEPH:2006qoi}.

\section{Results}
\label{app:results}

\subsection{Fluxes of antinuclei from SUEP for different DM masses}
\label{sec:fluxes_mdm}

\begin{figure*}
  \centering
  \includegraphics[width=0.32\linewidth]{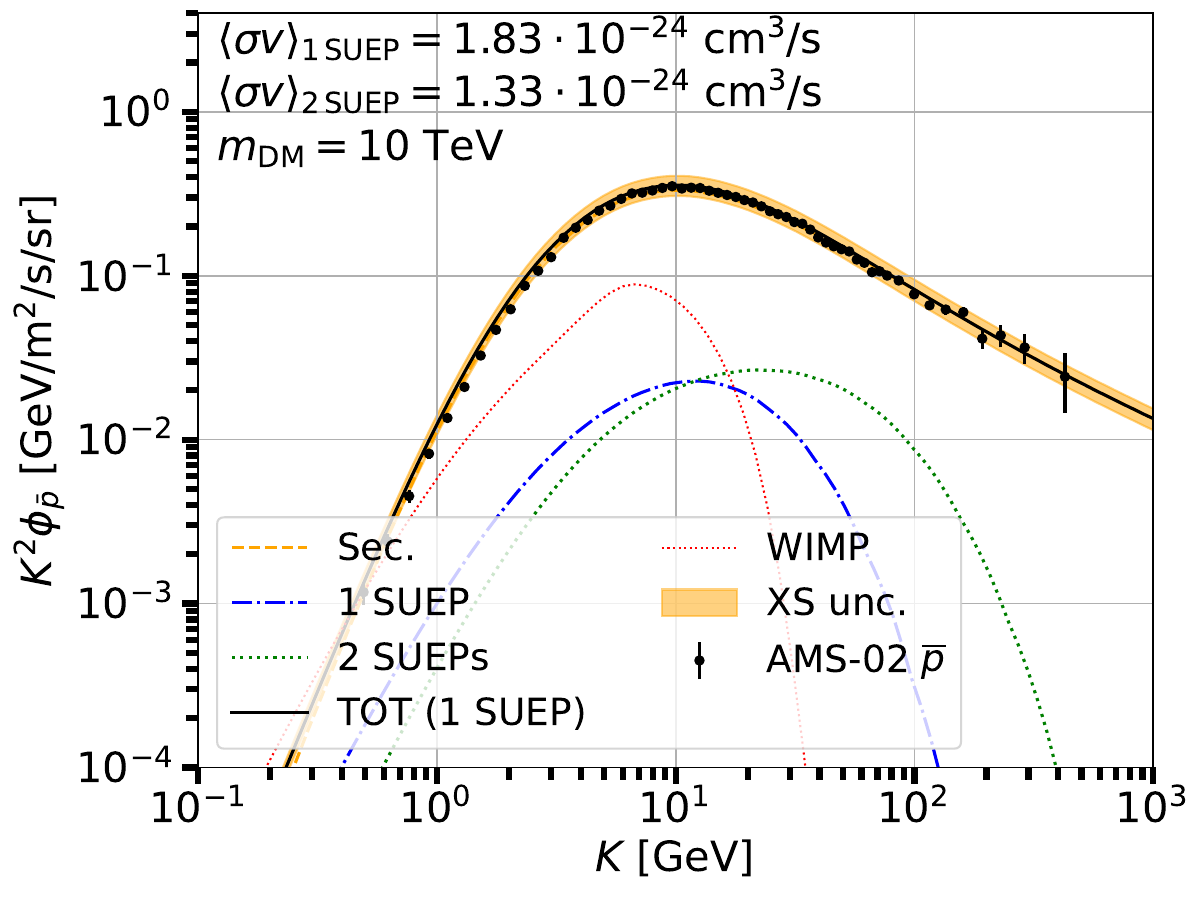}
  \includegraphics[width=0.32\linewidth]{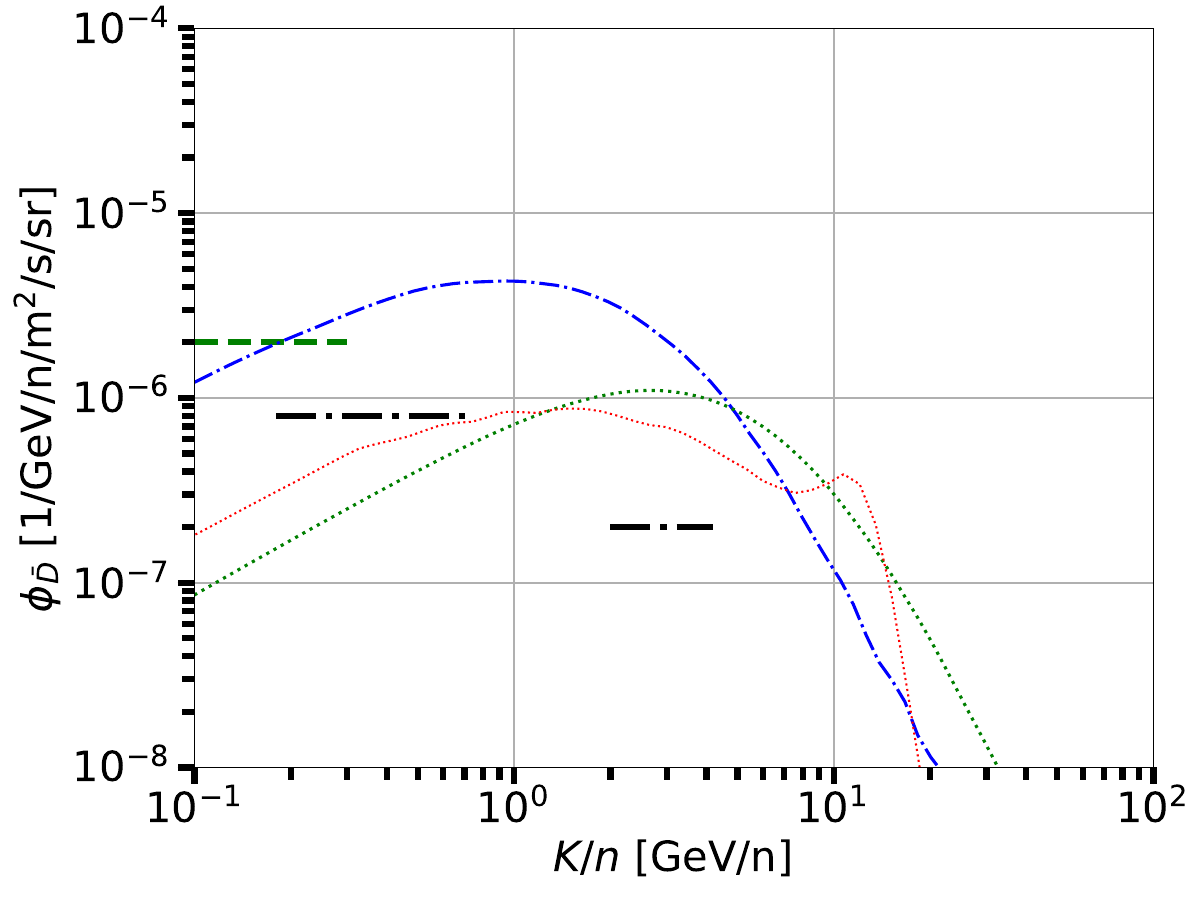}
  \includegraphics[width=0.32\linewidth]{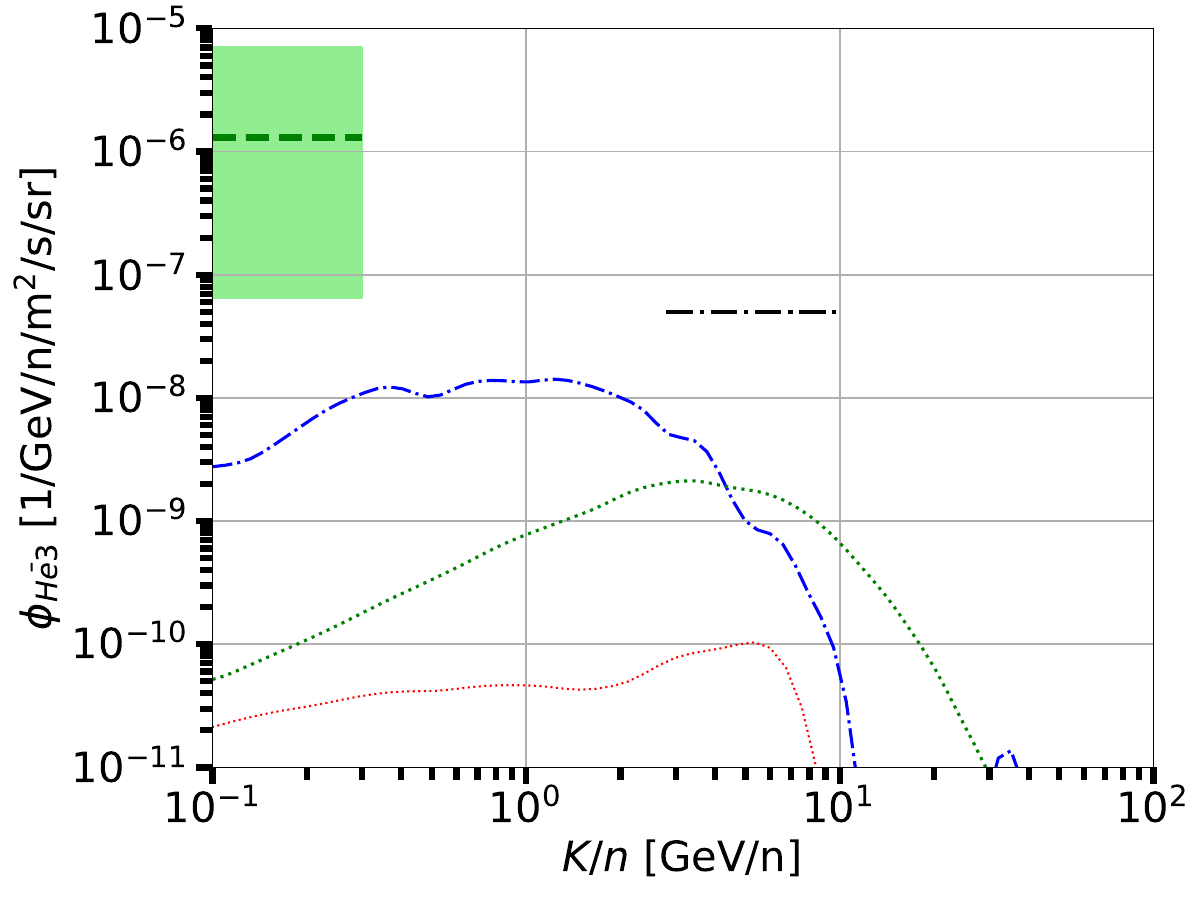}
  \includegraphics[width=0.32\linewidth]{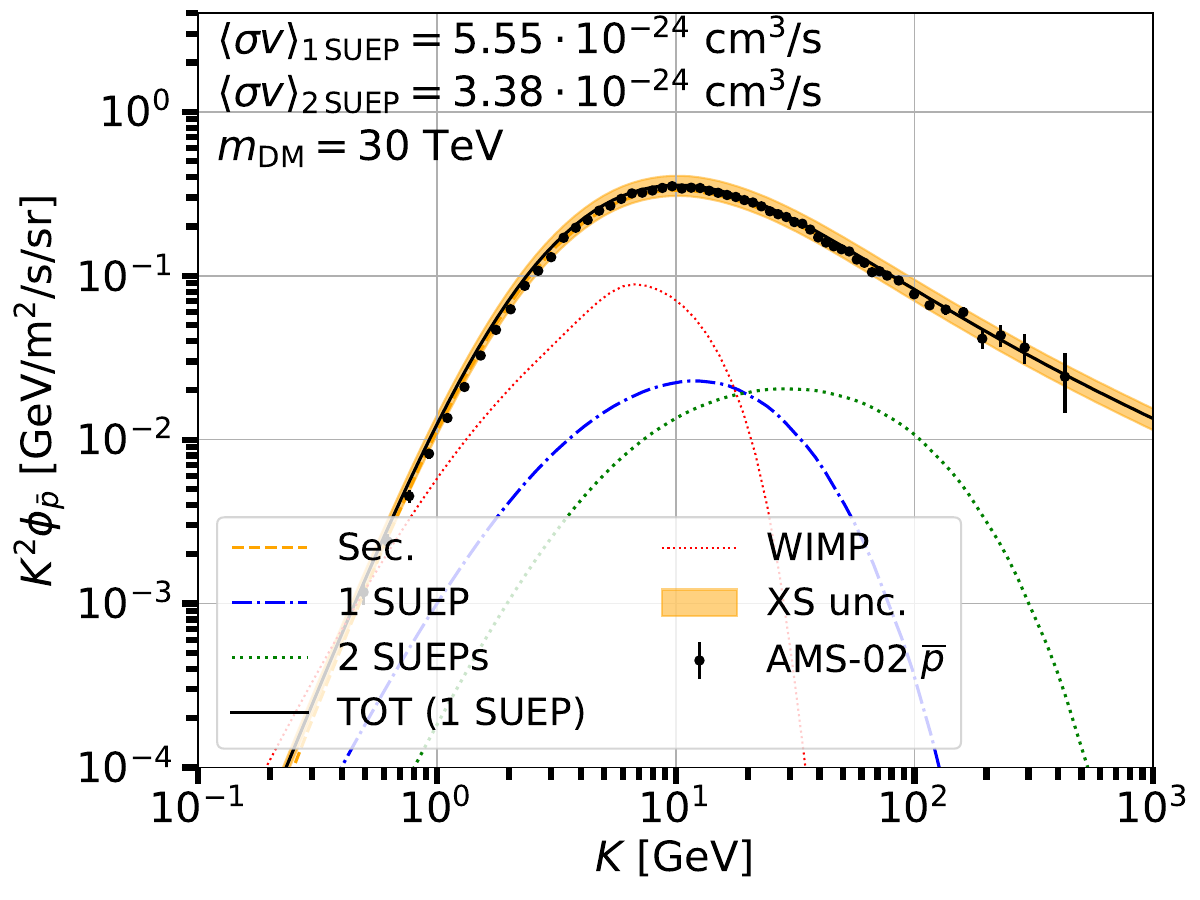}
  \includegraphics[width=0.32\linewidth]{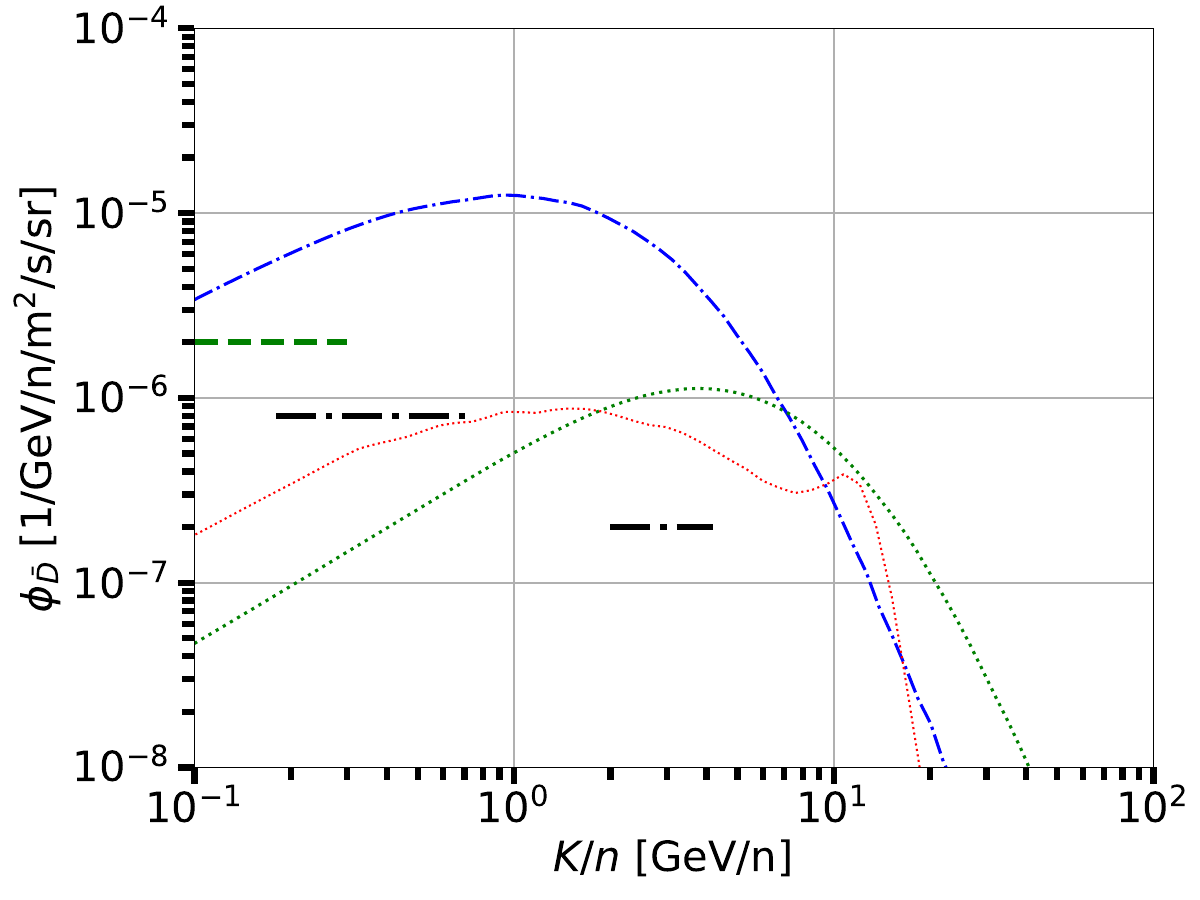}
  \includegraphics[width=0.32\linewidth]{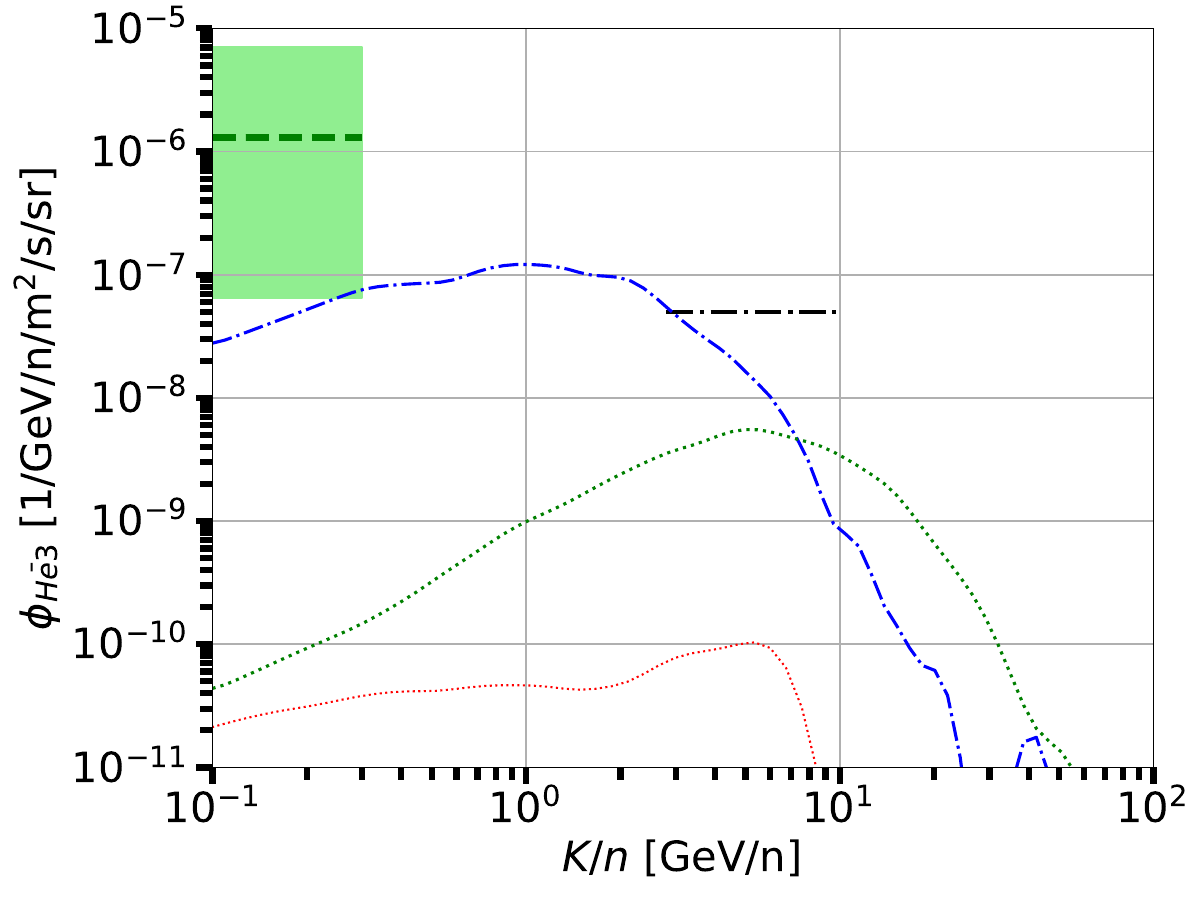}
  \includegraphics[width=0.32\linewidth]{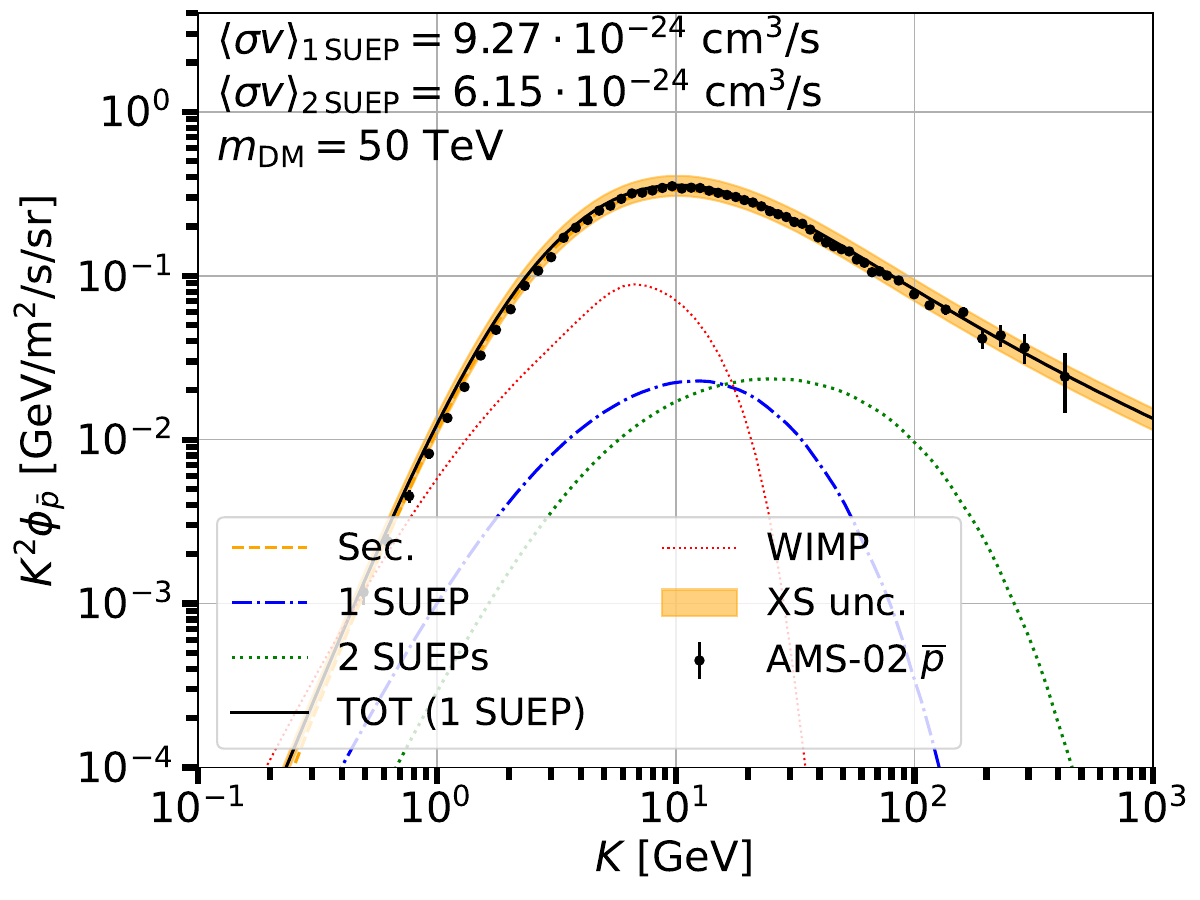}
  \includegraphics[width=0.32\linewidth]{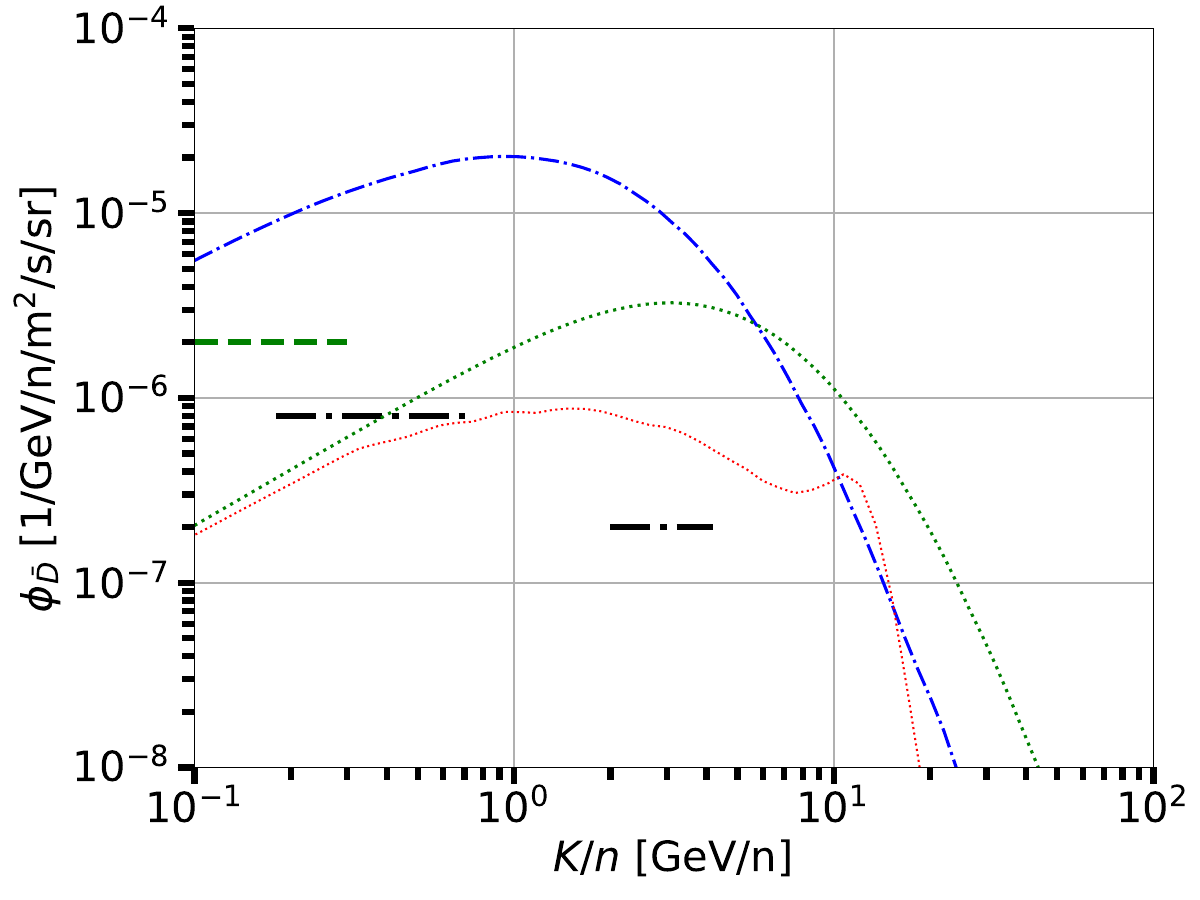}
  \includegraphics[width=0.32\linewidth]{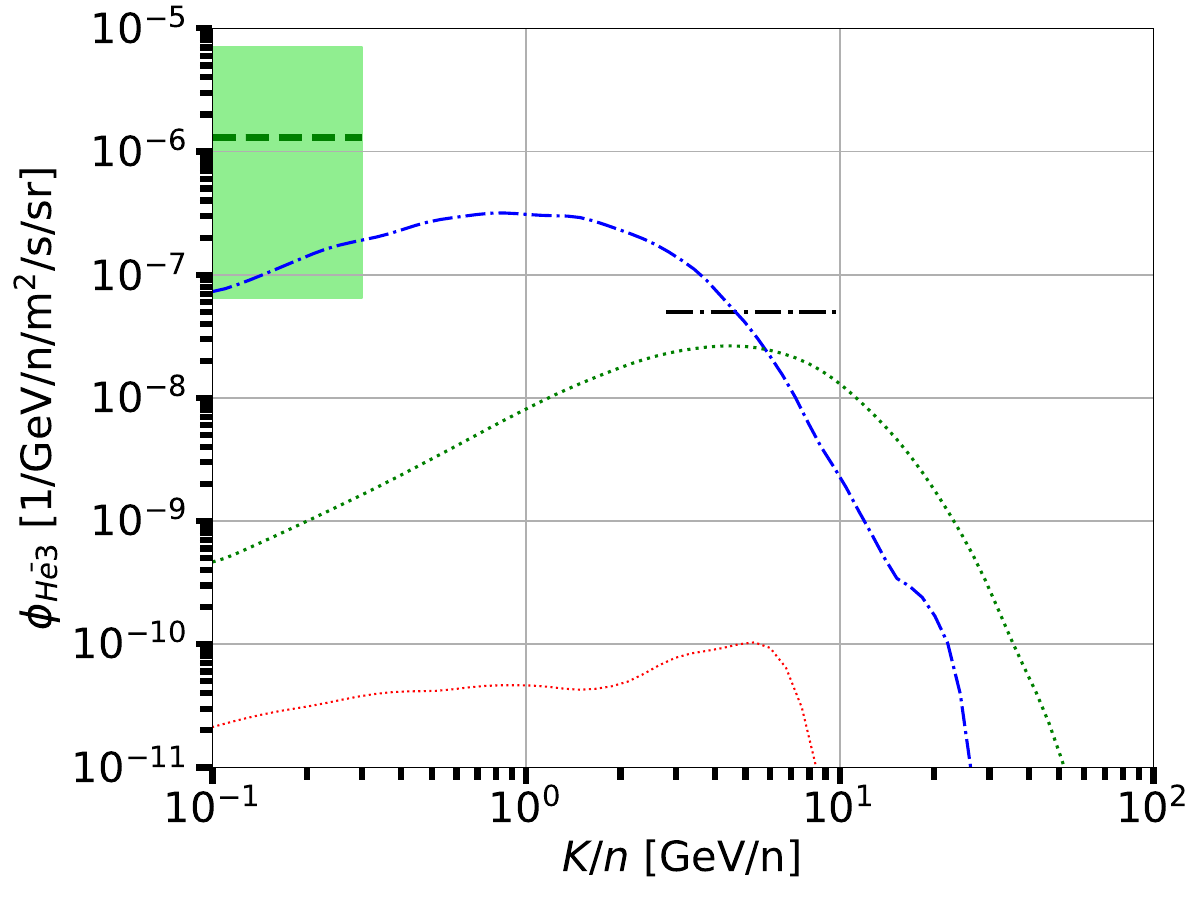}
  \includegraphics[width=0.32\linewidth]{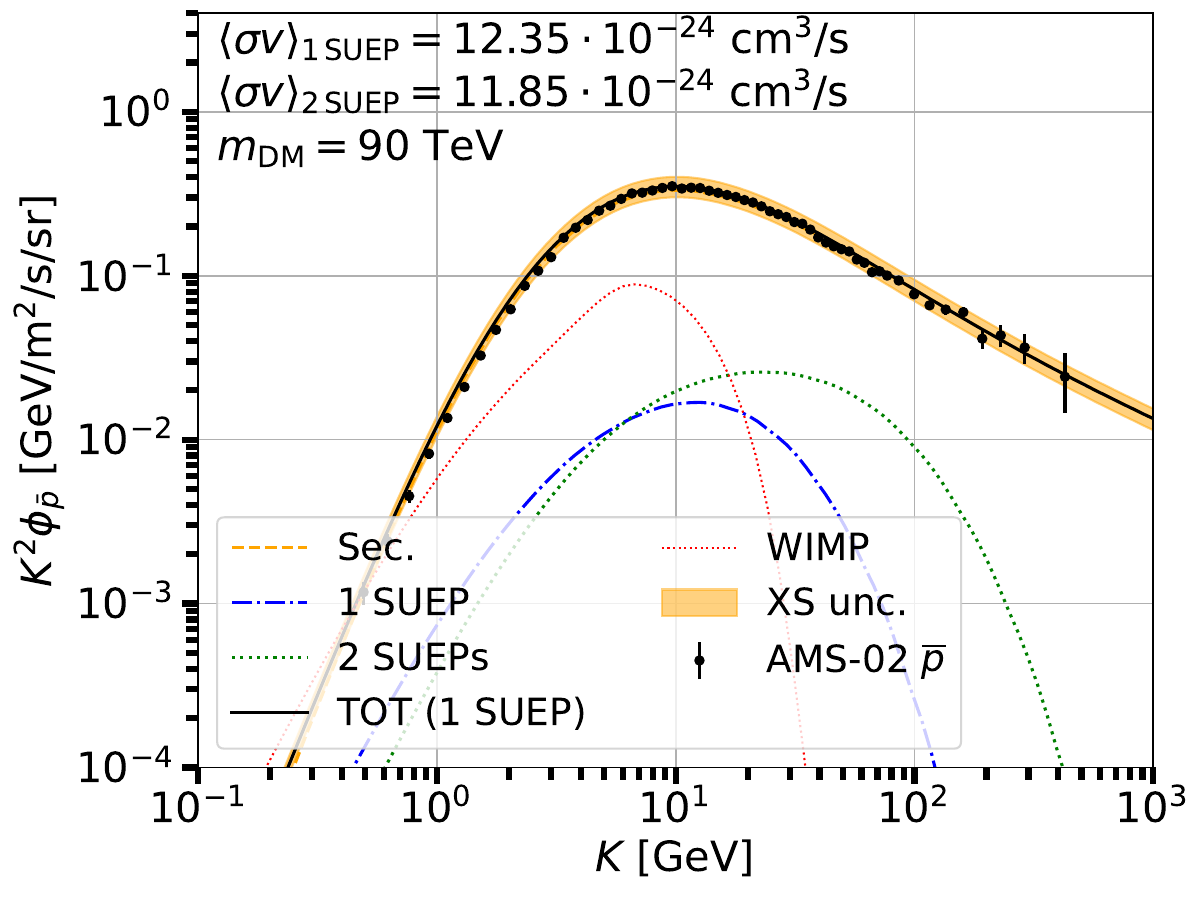}
  \includegraphics[width=0.32\linewidth]{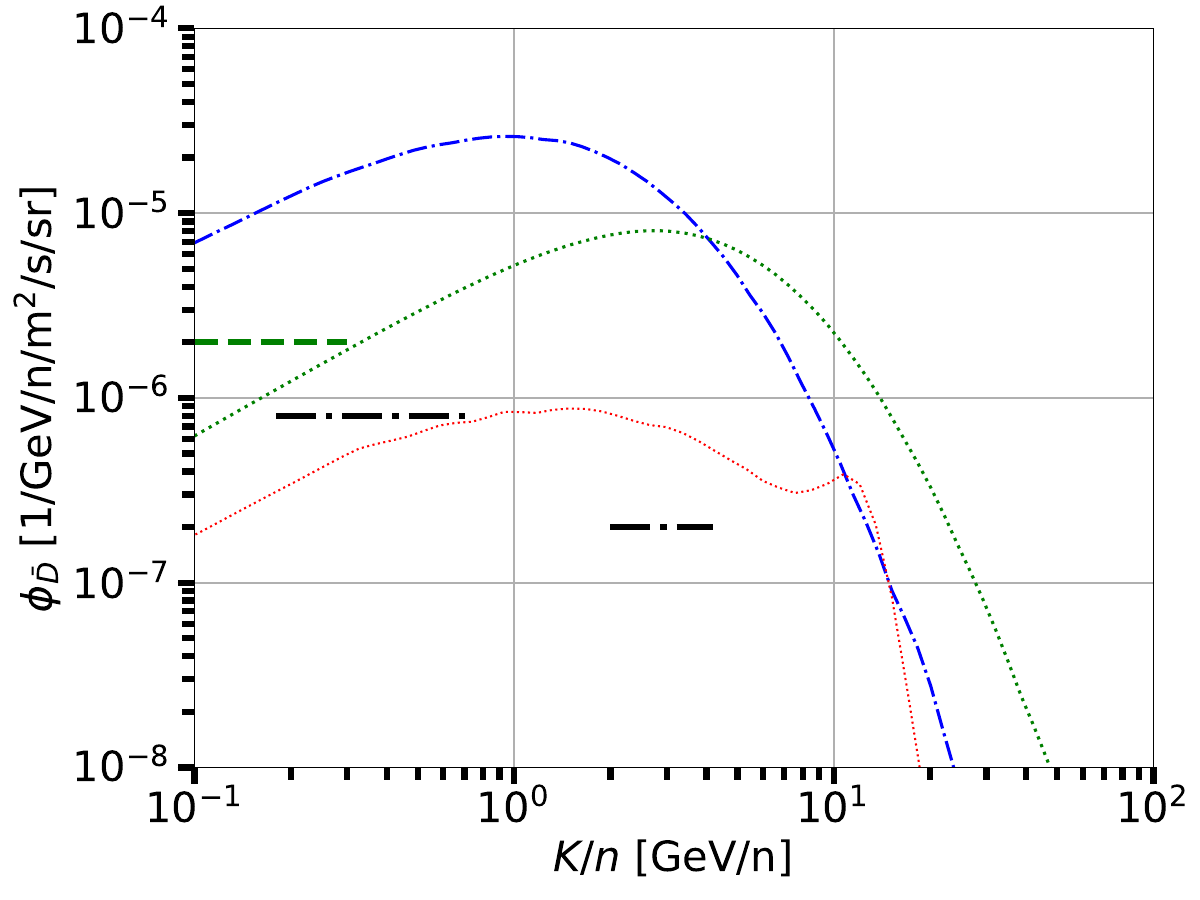}
  \includegraphics[width=0.32\linewidth]{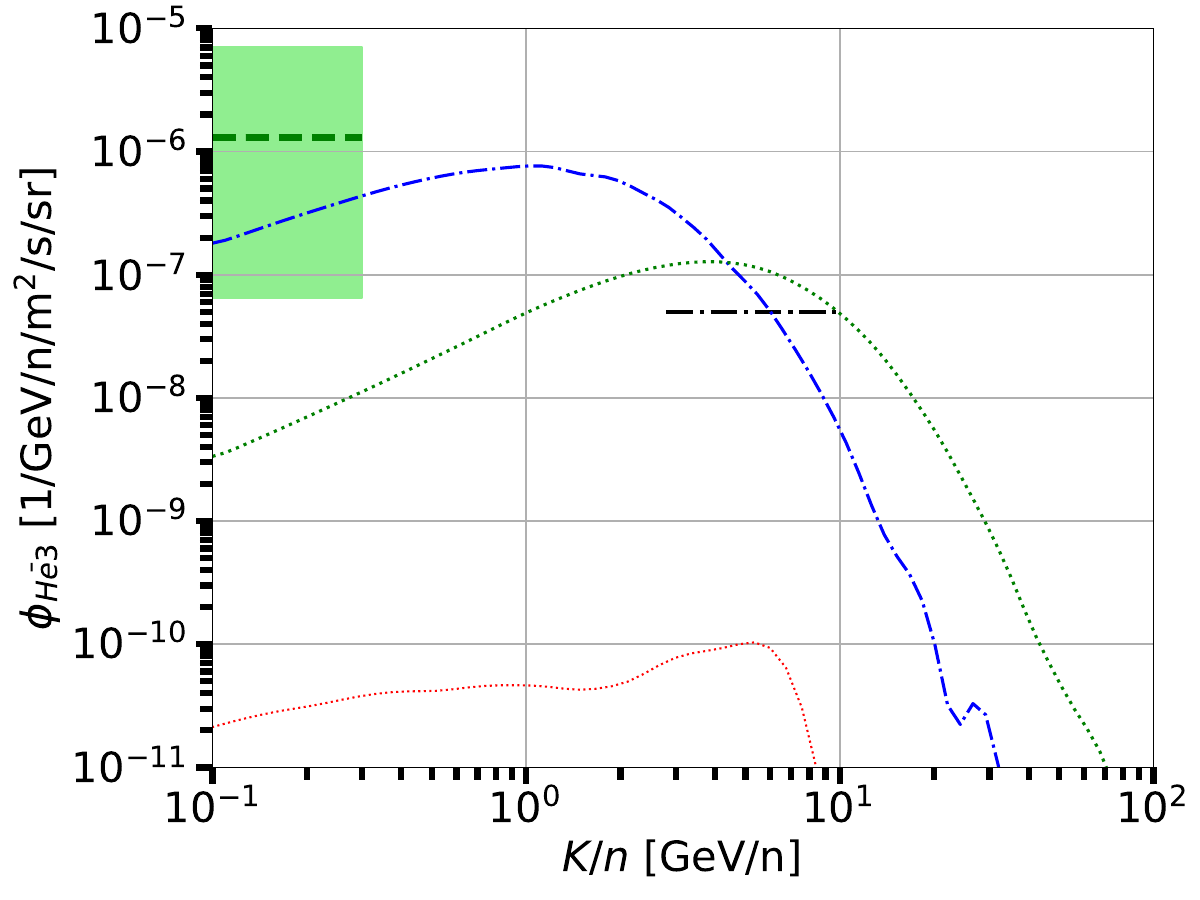}
  \includegraphics[width=0.32\linewidth]{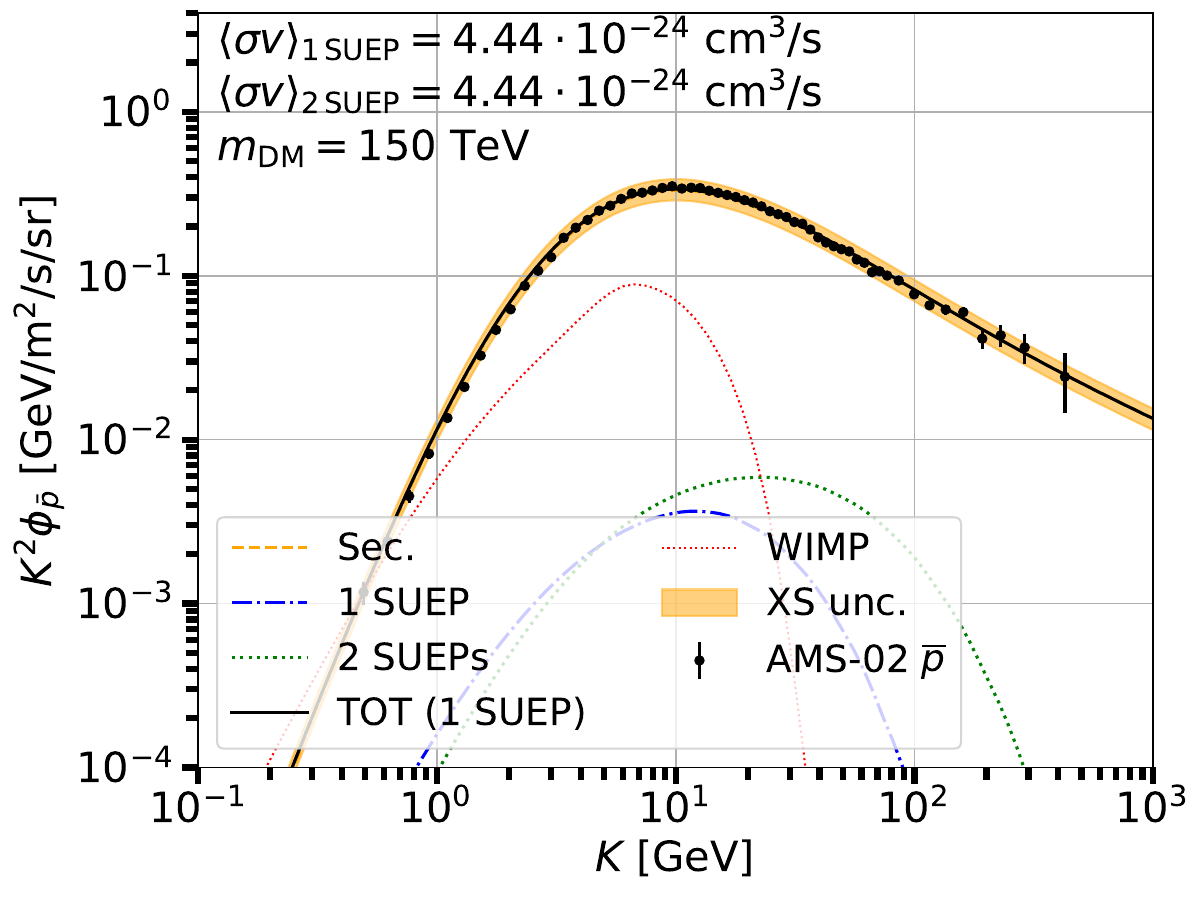}
  \includegraphics[width=0.32\linewidth]{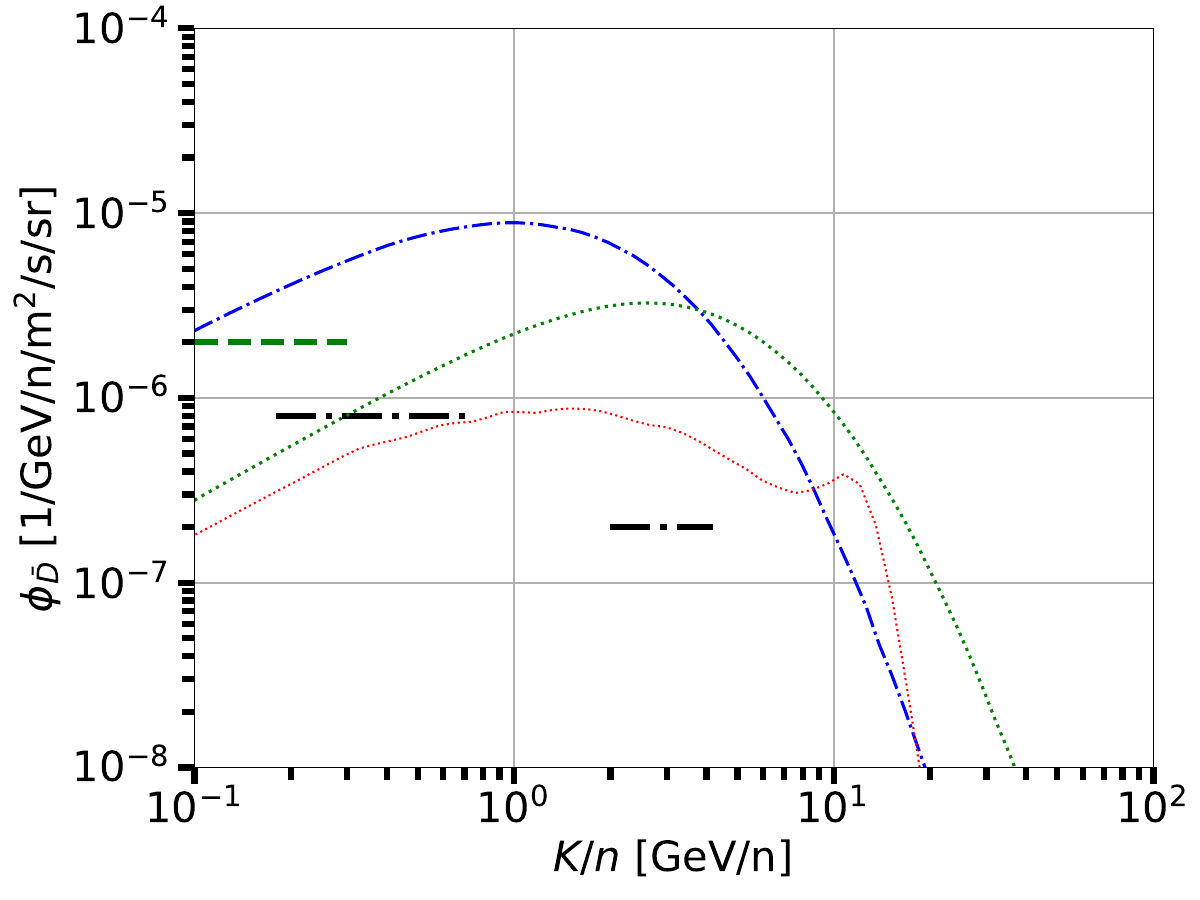}
  \includegraphics[width=0.32\linewidth]{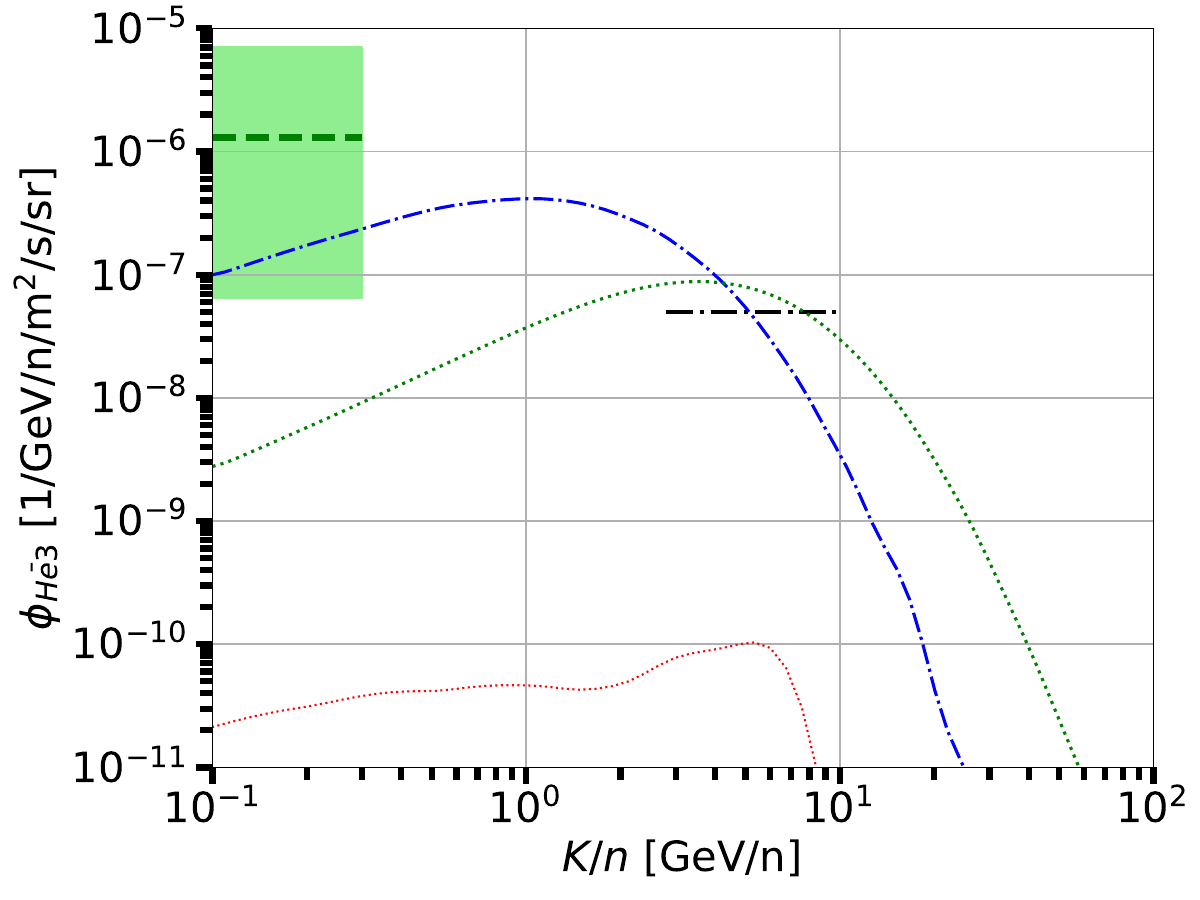}
  \caption{Comparison of the fluxes of antiprotons (left), antideuterons (middle), and antihelium-3 (right) predicted in our SUEP benchmark scenarios (blue dot–dashed curves) for DM masses between 10 and 150 TeV. For reference, we also show the standard WIMP expectation for DM annihilating into $b\bar b$ with $m_{\rm DM}=50$ GeV and a thermal cross section (red dotted curves). The \textsf{AMS-02} sensitivity is shown as black dot–dashed lines, while the \textsf{GAPS} sensitivity is shown as green dashed lines (for the antinuclei panels). The \textsf{AMS-02} antiproton data are overlaid in the left panels.For antiprotons, we also display the secondary contribution taken from Ref.~\cite{DiMauro:2023jgg} (grey line) and the total flux obtained by summing secondary and DM components (orange curve). The yellow band denotes the $\sim20\%$ theoretical uncertainty associated with antiproton production cross sections.} 
\label{fig:fluxesonesuep}
\end{figure*}

Fig.~\ref{fig:fluxesonesuep} extends the benchmark comparison of Fig.~\ref{fig:SUEP_fluxes} by showing how the propagated spectra change as the DM mass is varied, while keeping the SUEP shower in the region that maximizes antinuclei production (low $T_{\rm SUEP}/m_{\pi_D}$ and near-threshold kinematics in each scenario). The overall normalization for each row is fixed by the cross-section choice described in Sec.~\ref{sec:sigmav}, and the values of $\langle\sigma v\rangle$ adopted for {\tt 1 SUEP} and {\tt 2 SUEPs} are explicitly indicated in the $\bar p$ panels.

\paragraph{Antiprotons.}
For all masses, the DM-induced antiproton contribution (blue dot-dashed for {\tt 1 SUEP}, green dotted for {\tt 2 SUEPs}) remains subdominant with respect to the measured \textsf{AMS-02} flux by construction. The secondary component reproduces the data across the full energy range, and the total flux (secondary plus DM) stays within the allowed envelope once the antiproton production cross-section uncertainty is taken into account. As $m_{\rm DM}$ increases, the allowed annihilation cross section generally becomes less restricted by antiprotons (approaching the unitarity value in the highest-mass benchmarks), which increases the overall normalization of the DM component without spoiling agreement with \textsf{AMS-02}.

\paragraph{Antideuterons.}
In the $\overline{\mathrm D}$ panels (middle column), the {\tt 1 SUEP} scenario yields the largest flux over most of the displayed kinetic-energy range. This is expected because, for the same $m_{\rm DM}$ and comparable shower parameters, the effective hadronic activity per annihilation is larger in the single-mediator case, leading to larger antinucleon multiplicities and therefore more efficient coalescence.
Across the mass range shown, the predicted $\overline{\mathrm D}$ flux typically exceeds the projected \textsf{AMS-02} sensitivity in the few-GeV/n window where \textsf{AMS-02} is most sensitive (around $K/n\simeq 2$--$4~\mathrm{GeV/n}$), with the excess growing from the lowest-mass benchmark to the $m_{\rm DM}\sim\mathcal{O}(50$--$90)\,\mathrm{TeV}$ cases. At sub-GeV/n energies, only the {\tt 1 SUEP} spectra can approach (and, in the higher-mass rows, enter) the \textsf{GAPS} sensitivity window; this remains the most plausible configuration for a GAPS-relevant signal. As already noted, quantitative conclusions below $\sim$GeV/n are more sensitive to solar modulation and other low-rigidity systematics, whereas the multi-GeV/n region is comparatively more robust.

\paragraph{Antihelium-3.}
The ${}^3\overline{\mathrm{He}}$ fluxes (right column) remain the most challenging observable. For $m_{\rm DM}\lesssim 50~\mathrm{TeV}$, the predicted ${}^3\overline{\mathrm{He}}$ flux in both SUEP scenarios lies below the adopted \textsf{AMS-02} sensitivity estimate across most of the range shown, although the {\tt 1 SUEP} case moves progressively closer as the DM mass increases. In the highest-mass benchmarks displayed ($m_{\rm DM}=90/150~\mathrm{TeV}$), the {\tt 1 SUEP} spectrum becomes comparable to (and can marginally exceed) the \textsf{AMS-02} sensitivity in the few-GeV/n range, implying that an observable antihelium-3 signal is plausible only in the heavier-mass, low-temperature corner of parameter space. In contrast, the \textsf{GAPS} sensitivity to ${}^3\overline{\mathrm{He}}$ (even under optimistic assumptions) is not reached in these benchmarks.

Overall, Fig.~\ref{fig:fluxesonesuep} illustrates a key qualitative outcome of the SUEP framework: over a broad range of heavy DM masses, one can obtain an antideuteron signal that is readily testable at \textsf{AMS-02} (and, in favorable cases, potentially relevant for GAPS) while keeping antiprotons under control, whereas achieving a detectable ${}^3\overline{\mathrm{He}}$ flux requires pushing toward the heavier-mass, maximally enhanced region where coalescence benefits most strongly from the high-multiplicity, soft SUEP shower.

More quantitatively, focusing on the energy range where the predicted $\overline{\mathrm D}$ spectrum is maximal
($K/n\simeq 3$--$5~\mathrm{GeV/n}$, close to the optimal \textsf{AMS-02} reach), we find that for the {\tt 1 SUEP} case the peak flux exceeds
the projected \textsf{AMS-02} sensitivity by factors of
\[
\sim 10,\; 30,\; 50,\; 45,\;30
\quad\text{for}\quad
m_{\rm DM}=10,\;30,\;50,\;90,\;150~\mathrm{TeV},
\]
respectively (for the benchmark choice $T_{\rm SUEP}/m_{\pi_D}=0.1$ and $m_{\rm SUEP}$ just below the kinematic scale
set by the annihilation, with $\langle\sigma v\rangle$ fixed as in Sec.~\ref{sec:sigmav}).
Using the same exposure assumptions underlying the \textsf{AMS-02} sensitivity estimate (12~years) and the event-based
conversion described in Sec.~\ref{sec:sensitivity}, these ratios correspond to approximately
\[
N_{\overline{\mathrm D}} \simeq 20,\; 60,\; 100,\; 90, \; 60
\]
potentially detectable antideuteron events.
For the {\tt 2 SUEPs} model instead we find
\[
N_{\overline{\mathrm D}} \simeq 10,\; 20,\; 40,\; 80 \; 40
\]
We stress that these event numbers scale linearly with $\langle\sigma v\rangle$; they therefore represent the
maximal yields in our setup, given the simultaneous requirements from antiprotons and unitarity discussed in
Sec.~\ref{sec:sigmav}.
Because the DM-induced $\overline{\mathrm{D}}$ spectrum peaks at a few GeV/n, \textsf{AMS-02} is generally more sensitive than \textsf{GAPS}, although both experiments could observe a few events below $1$~GeV/n for $m_{\rm DM}\gtrsim 50$~TeV.

In summary, SUEP-induced antimatter fluxes are potentially observable at \textsf{AMS-02} over a broad region of parameter space, particularly at low SUEP temperatures and high DM masses.
Antideuterons provide the most promising signature, with tens to hundreds of detectable events in optimal scenarios.
Antihelium is more elusive but may still produce a few \textsf{AMS-02} events for $m_{\rm DM}\gtrsim 90$~TeV, offering a unique and highly informative probe of confining dark-sector dynamics.

\bibliographystyle{apsrev4-1}
\bibliography{main_deduped.bib}

@article{Curtin:2025ngf,
    author = "Curtin, David and Dreyer, Sascha and Fust{\'e} Costa, Max and Heim, Sarah and Kasieczka, Gregor and Moureaux, Louis and Rousso, David and Shih, David and Sommerhalder, Manuel",
    title = "{Soft-unclustered-energy patterns from quirks}",
    eprint = "2506.11192",
    archivePrefix = "arXiv",
    primaryClass = "hep-ph",
    reportNumber = "DESY-25-081-0, DESY-25-081, CERN-TH-2025-119",
    doi = "10.1103/g7c7-6qh2",
    journal = "Phys. Rev. D",
    volume = "113",
    number = "1",
    pages = "015010",
    year = "2026"
}

@article{DiMauro:2025jsb,
    author = "Di Mauro, Mattia and Wang, Yanhan",
    title = "{WIMP Shadows: Phenomenology of Secluded Dark Matter in Three Minimal BSM Scenarios}",
    eprint = "2510.23771",
    archivePrefix = "arXiv",
    primaryClass = "hep-ph",
    month = "10",
    year = "2025"
}

@article{DiMauro:2025uxt,
    author = "Di Mauro, Mattia",
    title = "{Two Puzzles, One Solution: Neutrino Mass and Secluded Dark Matter}",
    eprint = "2511.19622",
    archivePrefix = "arXiv",
    primaryClass = "hep-ph",
    month = "11",
    year = "2025"
}

@article{Cuoco:2019kuu,
    author = {Cuoco, Alessandro and Heisig, Jan and Klamt, Lukas and Korsmeier, Michael and Kr{\"a}mer, Michael},
    title = "{Scrutinizing the evidence for dark matter in cosmic-ray antiprotons}",
    eprint = "1903.01472",
    archivePrefix = "arXiv",
    primaryClass = "astro-ph.HE",
    reportNumber = "LAPTH-052/18, TTK-19-09, CP3-19-08",
    doi = "10.1103/PhysRevD.99.103014",
    journal = "Phys. Rev. D",
    volume = "99",
    number = "10",
    pages = "103014",
    year = "2019"
}

@article{Fermi-LAT:2016afa,
    author = "Charles, E. and others",
    collaboration = "Fermi-LAT",
    title = "{Sensitivity Projections for Dark Matter Searches with the Fermi Large Area Telescope}",
    eprint = "1605.02016",
    archivePrefix = "arXiv",
    primaryClass = "astro-ph.HE",
    reportNumber = "FERMILAB-PUB-16-179-AE",
    doi = "10.1016/j.physrep.2016.05.001",
    journal = "Phys. Rept.",
    volume = "636",
    pages = "1--46",
    year = "2016"
}

@article{DiMauro:2023tho,
    author = "Di Mauro, Mattia and Arina, Chiara and Fornengo, Nicolao and Heisig, Jan and Massaro, Daniele",
    title = "{Dark matter in the Higgs resonance region}",
    eprint = "2305.11937",
    archivePrefix = "arXiv",
    primaryClass = "hep-ph",
    reportNumber = "IRMP-CP3-23-21",
    doi = "10.1103/PhysRevD.108.095008",
    journal = "Phys. Rev. D",
    volume = "108",
    number = "9",
    pages = "095008",
    year = "2023"
}

@article{LZ:2024zvo,
    author = "Aalbers, J. and others",
    collaboration = "LZ",
    title = "{Dark Matter Search Results from 4.2{\,}{\,}Tonne-Years of Exposure of the LUX-ZEPLIN (LZ) Experiment}",
    eprint = "2410.17036",
    archivePrefix = "arXiv",
    primaryClass = "hep-ex",
    reportNumber = "FERMILAB-PUB-24-0796-V",
    doi = "10.1103/4dyc-z8zf",
    journal = "Phys. Rev. Lett.",
    volume = "135",
    number = "1",
    pages = "011802",
    year = "2025"
}

@article{Pospelov:2007mp,
    author = "Pospelov, Maxim and Ritz, Adam and Voloshin, Mikhail B.",
    title = "{Secluded WIMP Dark Matter}",
    eprint = "0711.4866",
    archivePrefix = "arXiv",
    primaryClass = "hep-ph",
    doi = "10.1016/j.physletb.2008.02.052",
    journal = "Phys. Lett. B",
    volume = "662",
    pages = "53--61",
    year = "2008"
}

@article{Gaskins:2016cha,
    author = "Gaskins, Jennifer M.",
    title = "{A review of indirect searches for particle dark matter}",
    eprint = "1604.00014",
    archivePrefix = "arXiv",
    primaryClass = "astro-ph.HE",
    doi = "10.1080/00107514.2016.1175160",
    journal = "Contemp. Phys.",
    volume = "57",
    number = "4",
    pages = "496--525",
    year = "2016"
}

@article{Schumann:2019eaa,
    author = "Schumann, Marc",
    title = "{Direct Detection of WIMP Dark Matter: Concepts and Status}",
    eprint = "1903.03026",
    archivePrefix = "arXiv",
    primaryClass = "astro-ph.CO",
    doi = "10.1088/1361-6471/ab2ea5",
    journal = "J. Phys. G",
    volume = "46",
    number = "10",
    pages = "103003",
    year = "2019"
}

@article{Boveia:2018yeb,
    author = "Boveia, Antonio and Doglioni, Caterina",
    title = "{Dark Matter Searches at Colliders}",
    eprint = "1810.12238",
    archivePrefix = "arXiv",
    primaryClass = "hep-ex",
    doi = "10.1146/annurev-nucl-101917-021008",
    journal = "Ann. Rev. Nucl. Part. Sci.",
    volume = "68",
    pages = "429--459",
    year = "2018"
}

@article{Cirelli:2024ssz,
    author = "Cirelli, Marco and Strumia, Alessandro and Zupan, Jure",
    title = "{Dark Matter}",
    eprint = "2406.01705",
    archivePrefix = "arXiv",
    primaryClass = "hep-ph",
    month = "6",
    year = "2024"
}

@article{DiMauro:2024kml,
    author = "Di Mauro, Mattia and Fornengo, Nicolao and Jueid, Adil and de Austri, Roberto Ruiz and Bellini, Francesca",
    title = "{Nailing Down the Theoretical Uncertainties of D{\textasciimacron} Spectrum Produced from Dark Matter}",
    eprint = "2411.04815",
    archivePrefix = "arXiv",
    primaryClass = "astro-ph.HE",
    reportNumber = "CTPU-PTC-24-31, CERN-TH-2024-164",
    doi = "10.1103/w6n5-vs4d",
    journal = "Phys. Rev. Lett.",
    volume = "135",
    number = "13",
    pages = "131002",
    year = "2025"
}

@article{Genolini:2021doh,
    author = "G\'enolini, Yoann and Boudaud, Mathieu and Cirelli, Marco and Derome, Laurent and Lavalle, Julien and Maurin, David and Salati, Pierre and Weinrich, Nathanael",
    title = "{New minimal, median, and maximal propagation models for dark matter searches with Galactic cosmic rays}",
    eprint = "2103.04108",
    archivePrefix = "arXiv",
    primaryClass = "astro-ph.HE",
    reportNumber = "LAPTH-011/21, LUPM:21-002",
    month = "3",
    year = "2021"
}

@article{Fermi-LAT:2016uux,
    author = "Albert, A. and others",
    collaboration = "Fermi-LAT, DES",
    title = "{Searching for Dark Matter Annihilation in Recently Discovered Milky Way Satellites with Fermi-LAT}",
    eprint = "1611.03184",
    archivePrefix = "arXiv",
    primaryClass = "astro-ph.HE",
    reportNumber = "FERMILAB-PUB-16-073-AE",
    doi = "10.3847/1538-4357/834/2/110",
    journal = "Astrophys. J.",
    volume = "834",
    number = "2",
    pages = "110",
    year = "2017"
}

@article{DiMauro:2015jxa,
    author = "Di Mauro, Mattia and Donato, Fiorenza and Fornengo, Nicolao and Vittino, Andrea",
    title = "{Dark matter vs. astrophysics in the interpretation of AMS-02 electron and positron data}",
    eprint = "1507.07001",
    archivePrefix = "arXiv",
    primaryClass = "astro-ph.HE",
    doi = "10.1088/1475-7516/2016/05/031",
    journal = "JCAP",
    volume = "05",
    pages = "031",
    year = "2016"
}

@ARTICLE{1997ApJ...490..493N,
       author = {{Navarro}, Julio F. and {Frenk}, Carlos S. and {White}, Simon D.~M.},
        title = "{A Universal Density Profile from Hierarchical Clustering}",
      journal = {\apj},
         year = "1997",
        month = "Dec",
       volume = {490},
       number = {2},
        pages = {493-508},
          doi = {10.1086/304888},
archivePrefix = {arXiv},
       eprint = {astro-ph/9611107},
 primaryClass = {astro-ph},
       adsurl = {https://ui.adsabs.harvard.edu/abs/1997ApJ...490..493N}
}

@article{Arcadi:2019lka,
    author = "Arcadi, Giorgio and Djouadi, Abdelhak and Raidal, Martti",
    title = "{Dark Matter through the Higgs portal}",
    eprint = "1903.03616",
    archivePrefix = "arXiv",
    primaryClass = "hep-ph",
    reportNumber = "LAPTH-010/19",
    doi = "10.1016/j.physrep.2019.11.003",
    journal = "Phys. Rept.",
    volume = "842",
    pages = "1--180",
    year = "2020"
}

@article{Arcadi:2017kky,
    author = "Arcadi, Giorgio and Dutra, Ma\'\i{}ra and Ghosh, Pradipta and Lindner, Manfred and Mambrini, Yann and Pierre, Mathias and Profumo, Stefano and Queiroz, Farinaldo S.",
    title = "{The waning of the WIMP? A review of models, searches, and constraints}",
    eprint = "1703.07364",
    archivePrefix = "arXiv",
    primaryClass = "hep-ph",
    doi = "10.1140/epjc/s10052-018-5662-y",
    journal = "Eur. Phys. J. C",
    volume = "78",
    number = "3",
    pages = "203",
    year = "2018"
}

@ARTICLE{2019JCAP...10..037D,
       author = {{de Salas}, P.~F. and {Malhan}, K. and {Freese}, K. and {Hattori}, K. and
         {Valluri}, M.},
        title = "{On the estimation of the local dark matter density using the rotation curve of the Milky Way}",
      journal = {\jcap},
         year = 2019,
        month = oct,
       volume = {2019},
       number = {10},
          eid = {037},
        pages = {037},
          doi = {10.1088/1475-7516/2019/10/037},
archivePrefix = {arXiv},
       eprint = {1906.06133},
 primaryClass = {astro-ph.GA},
       adsurl = {https://ui.adsabs.harvard.edu/abs/2019JCAP...10..037D}
}

@article{Korsmeier:2017xzj,
    author = "Korsmeier, Michael and Donato, Fiorenza and Fornengo, Nicolao",
    title = "{Prospects to verify a possible dark matter hint in cosmic antiprotons with antideuterons and antihelium}",
    eprint = "1711.08465",
    archivePrefix = "arXiv",
    primaryClass = "astro-ph.HE",
    reportNumber = "TTK-17-42",
    doi = "10.1103/PhysRevD.97.103011",
    journal = "Phys. Rev. D",
    volume = "97",
    number = "10",
    pages = "103011",
    year = "2018"
}

@article{Orusa:2022pvp,
    author = "Orusa, Luca and Di Mauro, Mattia and Donato, Fiorenza and Korsmeier, Michael",
    title = "{New determination of the production cross section for secondary positrons and electrons in the Galaxy}",
    eprint = "2203.13143",
    archivePrefix = "arXiv",
    primaryClass = "astro-ph.HE",
    doi = "10.1103/PhysRevD.105.123021",
    journal = "Phys. Rev. D",
    volume = "105",
    number = "12",
    pages = "123021",
    year = "2022"
}

@article{Korsmeier:2018gcy,
    author = "Korsmeier, Michael and Donato, Fiorenza and Di Mauro, Mattia",
    title = "{Production cross sections of cosmic antiprotons in the light of new data from the NA61 and LHCb experiments}",
    eprint = "1802.03030",
    archivePrefix = "arXiv",
    primaryClass = "astro-ph.HE",
    reportNumber = "TTK-18-06",
    doi = "10.1103/PhysRevD.97.103019",
    journal = "Phys. Rev. D",
    volume = "97",
    number = "10",
    pages = "103019",
    year = "2018"
}

@article{Hoof:2018hyn,
    author = "Hoof, Sebastian and Geringer-Sameth, Alex and Trotta, Roberto",
    title = "{A Global Analysis of Dark Matter Signals from 27 Dwarf Spheroidal Galaxies using 11 Years of Fermi-LAT Observations}",
    eprint = "1812.06986",
    archivePrefix = "arXiv",
    primaryClass = "astro-ph.CO",
    doi = "10.1088/1475-7516/2020/02/012",
    journal = "JCAP",
    volume = "02",
    pages = "012",
    year = "2020"
}

@article{DiMauro:2015tfa,
    author = "Di Mauro, Mattia and Donato, Fiorenza",
    title = "{Composition of the Fermi-LAT isotropic gamma-ray background intensity: Emission from extragalactic point sources and dark matter annihilations}",
    eprint = "1501.05316",
    archivePrefix = "arXiv",
    primaryClass = "astro-ph.HE",
    doi = "10.1103/PhysRevD.91.123001",
    journal = "Phys. Rev. D",
    volume = "91",
    number = "12",
    pages = "123001",
    year = "2015"
}

@article{Planck:2018vyg,
    author = "Aghanim, N. and others",
    collaboration = "Planck",
    title = "{Planck 2018 results. VI. Cosmological parameters}",
    eprint = "1807.06209",
    archivePrefix = "arXiv",
    primaryClass = "astro-ph.CO",
    doi = "10.1051/0004-6361/201833910",
    journal = "Astron. Astrophys.",
    volume = "641",
    pages = "A6",
    year = "2020",
    note = "[Erratum: Astron.Astrophys. 652, C4 (2021)]"
}

@article{AGUILAR2020,
    author = "Aguilar, M. and others",
    collaboration = "AMS",
    title = "{The Alpha Magnetic Spectrometer (AMS) on the international space station: Part II \textemdash{} Results from the first seven years}",
    doi = "10.1016/j.physrep.2020.09.003",
    journal = "Phys. Rept.",
    volume = "894",
    pages = "1--116",
    year = "2021"
}

@article{Balan:2023lwg,
    author = "Balan, Sowmiya and Kahlhoefer, Felix and Korsmeier, Michael and Manconi, Silvia and Nippel, Kathrin",
    title = "{Fast and accurate AMS-02 antiproton likelihoods for global dark matter fits}",
    eprint = "2303.07362",
    archivePrefix = "arXiv",
    primaryClass = "hep-ph",
    reportNumber = "TTK-23-02, TTP-23-008, LAPTH-008/23",
    month = "3",
    year = "2023"
}

@article{Bierlich:2022pfr,
    author = "Bierlich, Christian and others",
    title = "{A comprehensive guide to the physics and usage of PYTHIA 8.3}",
    eprint = "2203.11601",
    archivePrefix = "arXiv",
    primaryClass = "hep-ph",
    reportNumber = "LU-TP 22-16, MCNET-22-04, FERMILAB-PUB-22-227-SCD",
    doi = "10.21468/SciPostPhysCodeb.8",
    month = "3",
    year = "2022"
}

@article{DiMauro:2022hue,
    author = "Di Mauro, Mattia and Stref, Martin and Calore, Francesca",
    title = "{Investigating the effect of Milky~Way dwarf spheroidal galaxies extension on dark matter searches with Fermi-LAT data}",
    eprint = "2212.06850",
    archivePrefix = "arXiv",
    primaryClass = "astro-ph.HE",
    doi = "10.1103/PhysRevD.106.123032",
    journal = "Phys. Rev. D",
    volume = "106",
    number = "12",
    pages = "123032",
    year = "2022"
}

@article{Arcadi:2024ukq,
    author = "Arcadi, Giorgio and Cabo-Almeida, David and Dutra, Ma\'\i{}ra and Ghosh, Pradipta and Lindner, Manfred and Mambrini, Yann and Neto, Jacinto P. and Pierre, Mathias and Profumo, Stefano and Queiroz, Farinaldo S.",
    title = "{The Waning of the WIMP: Endgame?}",
    eprint = "2403.15860",
    archivePrefix = "arXiv",
    primaryClass = "hep-ph",
    month = "3",
    year = "2024"
}

@article{diMauro:2014zea,
    author = "di Mauro, Mattia and Donato, Fiorenza and Goudelis, Andreas and Serpico, Pasquale Dario",
    title = "{New evaluation of the antiproton production cross section for cosmic ray studies}",
    eprint = "1408.0288",
    archivePrefix = "arXiv",
    primaryClass = "hep-ph",
    reportNumber = "LAPTH-051-14",
    doi = "10.1103/PhysRevD.90.085017",
    journal = "Phys. Rev. D",
    volume = "90",
    number = "8",
    pages = "085017",
    year = "2014",
    note = "[Erratum: Phys.Rev.D 98, 049901 (2018)]"
}

@article{Heisig:2020nse,
    author = "Heisig, Jan and Korsmeier, Michael and Winkler, Martin Wolfgang",
    title = "{Dark matter or correlated errors? Systematics of the AMS-02 antiproton excess}",
    eprint = "2005.04237",
    archivePrefix = "arXiv",
    primaryClass = "astro-ph.HE",
    reportNumber = "CP3-20-19, TTK-20-14",
    doi = "10.1103/PhysRevResearch.2.043017",
    journal = "Phys. Rev. Res.",
    volume = "2",
    number = "4",
    pages = "043017",
    year = "2020"
}

@article{Kappl:2014hha,
    author = "Kappl, Rolf and Winkler, Martin Wolfgang",
    title = "{The Cosmic Ray Antiproton Background for AMS-02}",
    eprint = "1408.0299",
    archivePrefix = "arXiv",
    primaryClass = "hep-ph",
    reportNumber = "DESY-14-138",
    doi = "10.1088/1475-7516/2014/09/051",
    journal = "JCAP",
    volume = "09",
    pages = "051",
    year = "2014"
}

@article{Weinrich:2020ftb,
    author = "Weinrich, N. and Boudaud, M. and Derome, L. and Genolini, Y. and Lavalle, J. and Maurin, D. and Salati, P. and Serpico, P. and Weymann-Despres, G.",
    title = "{Galactic halo size in the light of recent AMS-02 data}",
    eprint = "2004.00441",
    archivePrefix = "arXiv",
    primaryClass = "astro-ph.HE",
    reportNumber = "LAPTH-009/20, LUPM:20-019",
    doi = "10.1051/0004-6361/202038064",
    journal = "Astron. Astrophys.",
    volume = "639",
    pages = "A74",
    year = "2020"
}

@article{Maurin:2001sj,
    author = "Maurin, D. and Donato, F. and Taillet, R. and Salati, P.",
    title = "{Cosmic rays below z=30 in a diffusion model: new constraints on propagation parameters}",
    eprint = "astro-ph/0101231",
    archivePrefix = "arXiv",
    reportNumber = "PREPRINT-LAPTH-817-00",
    doi = "10.1086/321496",
    journal = "Astrophys. J.",
    volume = "555",
    pages = "585--596",
    year = "2001"
}

@article{Donato:2001ms,
    author = "Donato, F. and Maurin, D. and Salati, P. and Barrau, A. and Boudoul, G. and Taillet, R.",
    title = "{Anti-protons from spallations of cosmic rays on interstellar matter}",
    eprint = "astro-ph/0103150",
    archivePrefix = "arXiv",
    doi = "10.1086/323684",
    journal = "Astrophys. J.",
    volume = "563",
    pages = "172--184",
    year = "2001"
}

@article{Tan:1983de,
    author = "Tan, L.C. and Ng, L.K.",
    title = "{CALCULATION OF THE EQUILIBRIUM ANTI-PROTON SPECTRUM}",
    doi = "10.1088/0305-4616/9/2/015",
    journal = "J. Phys. G",
    volume = "9",
    pages = "227--242",
    year = "1983"
}

@article{Barrau:2001ev,
    author = "Barrau, Aurelien and Boudoul, Gaelle and Donato, Fiorenza and Maurin, David and Salati, Pierre and Taillet, Richard",
    title = "{Anti-protons from primordial black holes}",
    eprint = "astro-ph/0112486",
    archivePrefix = "arXiv",
    doi = "10.1051/0004-6361:20020313",
    journal = "Astron. Astrophys.",
    volume = "388",
    pages = "676",
    year = "2002"
}

@article{Boudaud:2019efq,
    author = "Boudaud, Mathieu and G\'enolini, Yoann and Derome, Laurent and Lavalle, Julien and Maurin, David and Salati, Pierre and Serpico, Pasquale D.",
    title = "{AMS-02 antiprotons' consistency with a secondary astrophysical origin}",
    eprint = "1906.07119",
    archivePrefix = "arXiv",
    primaryClass = "astro-ph.HE",
    doi = "10.1103/PhysRevResearch.2.023022",
    journal = "Phys. Rev. Res.",
    volume = "2",
    number = "2",
    pages = "023022",
    year = "2020"
}

@misc{Tingcern2016,
  author       = {S. Ting},
  title        = {The First Five Years of the Alpha Magnetic Spectrometer on the International Space Station: Unlocking the Secrets of the Cosmos},
  howpublished = {CERN},
  year         = {2016},
  url          = {https://indico.cern.ch/event/592392/attachments/1381599/2110332/AMS-CERN-Dec-2016.pdf}
}

@misc{Miapp2022DbarHebar,
  author       = {P. Zuccon},
  title        = {AMS-02 results \& upgrade},
  howpublished = {MIAPP},
  year         = {2022},
  url          = {https://indico.ph.tum.de/event/6990/contributions/4988/attachments/3947/4992/Zuccon_miapp.pdf}
}

@misc{Miapp2022Dbar,
  author       = {Senquan Lu},
  title        = {Cosmic Ray Antideuteron Search with Alpha Magnetic Spectrometer (AMS)},
  howpublished = {MIAPP},
  year         = {2022},
  url          = {https://indico.ph.tum.de/event/6990/contributions/4970/attachments/3924/4946/Lu_Slides.pdf}
}

@article{Winkler:2020ltd,
    author = "Winkler, Martin Wolfgang and Linden, Tim",
    title = "{Dark Matter Annihilation Can Produce a Detectable Antihelium Flux through $\bar{\Lambda}_b$ Decays}",
    eprint = "2006.16251",
    archivePrefix = "arXiv",
    primaryClass = "hep-ph",
    doi = "10.1103/PhysRevLett.126.101101",
    journal = "Phys. Rev. Lett.",
    volume = "126",
    number = "10",
    pages = "101101",
    year = "2021"
}

@article{Cirelli:2014qia,
    author = "Cirelli, Marco and Fornengo, Nicolao and Taoso, Marco and Vittino, Andrea",
    title = "{Anti-helium from Dark Matter annihilations}",
    eprint = "1401.4017",
    archivePrefix = "arXiv",
    primaryClass = "hep-ph",
    reportNumber = "SACLAY-T14-003",
    doi = "10.1007/JHEP08(2014)009",
    journal = "JHEP",
    volume = "08",
    pages = "009",
    year = "2014"
}

@INPROCEEDINGS{2008ICRC....4..765C,
       author = {{Choutko}, V. and {Giovacchini}, F.},
        title = "{Cosmic Rays Antideuteron Sensitivity for AMS-02 Experiment}",
    booktitle = {International Cosmic Ray Conference},
         year = 2008,
       series = {International Cosmic Ray Conference},
       volume = {4},
        month = jan,
        pages = {765-768},
       adsurl = {https://ui.adsabs.harvard.edu/abs/2008ICRC....4..765C}
}

@article{Carlson:2014ssa,
    author = "Carlson, Eric and Coogan, Adam and Linden, Tim and Profumo, Stefano and Ibarra, Alejandro and Wild, Sebastian",
    title = "{Antihelium from Dark Matter}",
    eprint = "1401.2461",
    archivePrefix = "arXiv",
    primaryClass = "hep-ph",
    doi = "10.1103/PhysRevD.89.076005",
    journal = "Phys. Rev. D",
    volume = "89",
    number = "7",
    pages = "076005",
    year = "2014"
}

@article{Fornengo:2013osa,
    author = "Fornengo, N. and Maccione, L. and Vittino, A.",
    title = "{Dark matter searches with cosmic antideuterons: status and perspectives}",
    eprint = "1306.4171",
    archivePrefix = "arXiv",
    primaryClass = "hep-ph",
    reportNumber = "LMU-ASC-41-13, MPP-2013-158",
    doi = "10.1088/1475-7516/2013/09/031",
    journal = "JCAP",
    volume = "09",
    pages = "031",
    year = "2013"
}

@article{Orusa:2024ewq,
    author = "Orusa, Luca and Manconi, Silvia and Donato, Fiorenza and Di Mauro, Mattia",
    title = "{Disclosing the catalog pulsars dominating the Galactic positron flux}",
    eprint = "2410.10951",
    archivePrefix = "arXiv",
    primaryClass = "astro-ph.HE",
    reportNumber = "LAPTH-051/24",
    doi = "10.1088/1475-7516/2025/02/029",
    journal = "JCAP",
    volume = "02",
    pages = "029",
    year = "2025"
}

@article{DiMauro:2023oqx,
    author = "Di Mauro, Mattia and Donato, Fiorenza and Korsmeier, Michael and Manconi, Silvia and Orusa, Luca",
    title = "{Novel prediction for secondary positrons and electrons in the Galaxy}",
    eprint = "2304.01261",
    archivePrefix = "arXiv",
    primaryClass = "astro-ph.HE",
    reportNumber = "LAPTH-014/23, TTK-23-07",
    doi = "10.1103/PhysRevD.108.063024",
    journal = "Phys. Rev. D",
    volume = "108",
    number = "6",
    pages = "063024",
    year = "2023"
}

@article{DiMauro:2023jgg,
    author = "Di Mauro, Mattia and Korsmeier, Michael and Cuoco, Alessandro",
    title = "{Data-driven constraints on cosmic-ray diffusion: Probing self-generated turbulence in the Milky~Way}",
    eprint = "2311.17150",
    archivePrefix = "arXiv",
    primaryClass = "astro-ph.HE",
    doi = "10.1103/PhysRevD.109.123003",
    journal = "Phys. Rev. D",
    volume = "109",
    number = "12",
    pages = "123003",
    year = "2024"
}

@article{PhysRevLett.132.131001,
  title = {Search for Antideuterons of Cosmic Origin Using the BESS-Polar II Magnetic-Rigidity Spectrometer},
  author = {Sakai, K. and Fuke, H. and Yoshimura, K. and Sasaki, M. and Abe, K. and Haino, S. and Hams, T. and Hasegawa, M. and Kim, K. C. and Lee, M. H. and Makida, Y. and Mitchell, J. W. and Nishimura, J. and Nozaki, M. and Orito, R. and Ormes, J. F. and Seo, E. S. and Streitmatter, R. E. and Thakur, N. and Yamamoto, A. and Yoshida, T.},
  collaboration = {BESS Collaboration},
  journal = {Phys. Rev. Lett.},
  volume = {132},
  issue = {13},
  pages = {131001},
  numpages = {6},
  year = {2024},
  month = {Mar},
  publisher = {American Physical Society},
  doi = {10.1103/PhysRevLett.132.131001},
  url = {https://link.aps.org/doi/10.1103/PhysRevLett.132.131001}
}

@article{vonDoetinchem:2020vbj,
    author = "von Doetinchem, P. and others",
    title = "{Cosmic-ray antinuclei as messengers of new physics:  status and outlook for the new decade}",
    eprint = "2002.04163",
    archivePrefix = "arXiv",
    primaryClass = "astro-ph.HE",
    doi = "10.1088/1475-7516/2020/08/035",
    journal = "JCAP",
    volume = "08",
    pages = "035",
    year = "2020"
}

@article{Aramaki:2015laa,
    author = "Aramaki, T. and Hailey, C. J. and Boggs, S. E. and von Doetinchem, P. and Fuke, H. and Mognet, S. I. and Ong, R. A. and Perez, K. and Zweerink, J.",
    collaboration = "GAPS",
    title = "{Antideuteron Sensitivity for the GAPS Experiment}",
    eprint = "1506.02513",
    archivePrefix = "arXiv",
    primaryClass = "astro-ph.HE",
    doi = "10.1016/j.astropartphys.2015.09.001",
    journal = "Astropart. Phys.",
    volume = "74",
    pages = "6--13",
    year = "2016"
}

@unpublished{Oliva:JENAA:2024,
  author       = {Oliva, A.},
  title        = {Latest Results of the {Alpha Magnetic Spectrometer} on the {International Space Station}},
  note         = {Invited talk at the JENAA Workshop @ CERN, 20 Aug 2024},
  organization = {INFN Bologna},
  address      = {CERN, Geneva, Switzerland},
  year         = {2024},
  month        = aug
}

@phdthesis{HabibyAlaoui:2016:PhD,
  author    = {Habiby Alaoui, Marion},
  title     = {Measurement of the Cosmic Ray Helium Flux with the AMS-02 Experiment},
  school    = {Universit{\'e} de Gen{\`e}ve, Facult{\'e} des Sciences, Section de Physique, D{\'e}partement de Physique Nucl{\'e}aire et Corpusculaire},
  address   = {Gen{\`e}ve, Switzerland},
  year      = {2016},
  type      = {Ph.D. thesis},
  number    = {Th{\`e}se No.~4912},
  note      = {Advisor: Prof.\ Martin Pohl},
  language  = {English}
}

@inproceedings{Spada:2008xk,
    author = "Spada, Francesca R.",
    collaboration = "AMS 02",
    title = "{Antimatter and Dark Matter search in space with AMS-02}",
    booktitle = "{34th International Conference on High Energy Physics}",
    eprint = "0810.3831",
    archivePrefix = "arXiv",
    primaryClass = "hep-ex",
    month = "10",
    year = "2008"
}

@article{GAPS:2020axg,
    author = "Saffold, N. and others",
    collaboration = "GAPS",
    title = "{Cosmic antihelium-3 nuclei sensitivity of the GAPS experiment}",
    eprint = "2012.05834",
    archivePrefix = "arXiv",
    primaryClass = "hep-ph",
    doi = "10.1016/j.astropartphys.2021.102580",
    journal = "Astropart. Phys.",
    volume = "130",
    pages = "102580",
    year = "2021"
}

@article{Ibarra:2012cc,
    author = "Ibarra, Alejandro and Wild, Sebastian",
    title = "{Prospects of antideuteron detection from dark matter annihilations or decays at AMS-02 and GAPS}",
    eprint = "1209.5539",
    archivePrefix = "arXiv",
    primaryClass = "hep-ph",
    doi = "10.1088/1475-7516/2013/02/021",
    journal = "JCAP",
    volume = "02",
    pages = "021",
    year = "2013"
}

@article{Herms:2016vop,
    author = "Herms, Johannes and Ibarra, Alejandro and Vittino, Andrea and Wild, Sebastian",
    title = "{Antideuterons in cosmic rays: sources and discovery potential}",
    eprint = "1610.00699",
    archivePrefix = "arXiv",
    primaryClass = "astro-ph.HE",
    reportNumber = "TUM-HEP-1063-16",
    doi = "10.1088/1475-7516/2017/02/018",
    journal = "JCAP",
    volume = "02",
    pages = "018",
    year = "2017"
}

@article{McDaniel:2023bju,
    author = "McDaniel, Alex and Ajello, Marco and Karwin, Christopher M. and Di Mauro, Mattia and Drlica-Wagner, Alex and S\'anchez-Conde, Miguel A.",
    title = "{Legacy analysis of dark matter annihilation from the Milky~Way dwarf spheroidal galaxies with 14~years of Fermi-LAT data}",
    eprint = "2311.04982",
    archivePrefix = "arXiv",
    primaryClass = "astro-ph.HE",
    reportNumber = "FERMILAB-PUB-23-686-PPD",
    doi = "10.1103/PhysRevD.109.063024",
    journal = "Phys. Rev. D",
    volume = "109",
    number = "6",
    pages = "063024",
    year = "2024"
}

@article{2006192,
title = {Deuteron and anti-deuteron production in e+e− collisions at the Z resonance},
journal = {Physics Letters B},
volume = {639},
number = {3},
pages = {192-201},
year = {2006},
issn = {0370-2693},
doi = {https://doi.org/10.1016/j.physletb.2006.06.043},
url = {https://www.sciencedirect.com/science/article/pii/S037026930600774X},
author = {S. Schael and R. Barate and R. Brunelière and others}
}

@article{Donato:1999gy,
      author         = "Donato, Fiorenza and Fornengo, Nicolao and Salati,
                        Pierre",
      title          = "{Anti-deuterons as a signature of supersymmetric dark
                        matter}",
      journal        = "Phys. Rev.",
      volume         = "D62",
      year           = "2000",
      pages          = "043003",
      doi            = "10.1103/PhysRevD.62.043003",
      eprint         = "hep-ph/9904481",
      archivePrefix  = "arXiv",
      primaryClass   = "hep-ph",
      reportNumber   = "IFIC-99-9, FTUV-99-9, LAPTH-722-99",
      SLACcitation   = "%%CITATION = HEP-PH/9904481;%%"
}

@article{Calore:2022stf,
    author = "Calore, Francesca and Cirelli, Marco and Derome, Laurent and Genolini, Yoann and Maurin, David and Salati, Pierre and Serpico, Pasquale Dario",
    title = "{AMS-02 antiprotons and dark matter: Trimmed hints and robust bounds}",
    eprint = "2202.03076",
    archivePrefix = "arXiv",
    primaryClass = "hep-ph",
    reportNumber = "LAPTH-003/22",
    doi = "10.21468/SciPostPhys.12.5.163",
    journal = "SciPost Phys.",
    volume = "12",
    number = "5",
    pages = "163",
    year = "2022"
}

@article{Strong:2007nh,
    author = "Strong, Andrew W. and Moskalenko, Igor V. and Ptuskin, Vladimir S.",
    title = "{Cosmic-ray propagation and interactions in the Galaxy}",
    eprint = "astro-ph/0701517",
    archivePrefix = "arXiv",
    reportNumber = "SLAC-PUB-12312",
    doi = "10.1146/annurev.nucl.57.090506.123011",
    journal = "Ann. Rev. Nucl. Part. Sci.",
    volume = "57",
    pages = "285--327",
    year = "2007"
}

@article{DiMauro:2021qcf,
    author = "Di Mauro, Mattia and Winkler, Martin Wolfgang",
    title = "{Multimessenger constraints on the dark matter interpretation of the Fermi-LAT Galactic center excess}",
    eprint = "2101.11027",
    archivePrefix = "arXiv",
    primaryClass = "astro-ph.HE",
    doi = "10.1103/PhysRevD.103.123005",
    journal = "Phys. Rev. D",
    volume = "103",
    number = "12",
    pages = "123005",
    year = "2021"
}

@article{Holdom:1985ag,
  author    = {Holdom, Bob},
  title     = {Two U(1)'s and Epsilon Charge Shifts},
  journal   = {Phys. Lett. B},
  volume    = {166},
  pages     = {196},
  year      = {1986},
  doi       = {10.1016/0370-2693(86)91377-8}
}

@article{Heisig:2024jkk,
    author = {Heisig, Jan and Korsmeier, Michael and Kr\"amer, Michael and Nippel, Kathrin and Rathmann, Lena},
    title = "{D̅arkRayNet: emulation of cosmic-ray antideuteron fluxes from dark matter}",
    eprint = "2406.18642",
    archivePrefix = "arXiv",
    primaryClass = "hep-ph",
    reportNumber = "TTK-24-25",
    doi = "10.1088/1475-7516/2024/11/017",
    journal = "JCAP",
    volume = "11",
    pages = "017",
    year = "2024"
}

@article{DeLaTorreLuque:2024htu,
    author = "De La Torre Luque, Pedro and Winkler, Martin Wolfgang and Linden, Tim",
    title = "{Cosmic-ray propagation models elucidate the prospects for antinuclei detection}",
    eprint = "2404.13114",
    archivePrefix = "arXiv",
    primaryClass = "astro-ph.HE",
    doi = "10.1088/1475-7516/2024/10/017",
    journal = "JCAP",
    volume = "10",
    pages = "017",
    year = "2024"
}

@article{PhysRevC.97.024615,
  title = {Production of deuterons, tritons, $^{3}\mathrm{He}$ nuclei, and their antinuclei in $pp$ collisions at $\sqrt{s}=0.9$, 2.76, and 7 TeV},
  author = {Acharya, S. and others},
  collaboration = {ALICE Collaboration},
  journal = {Phys. Rev. C},
  volume = {97},
  issue = {2},
  pages = {024615},
  numpages = {17},
  year = {2018},
  month = {Feb},
  publisher = {American Physical Society},
  doi = {10.1103/PhysRevC.97.024615},
  url = {https://link.aps.org/doi/10.1103/PhysRevC.97.024615}
}

@article{ALEPH:2006qoi,
    author = "Schael, S. and others",
    collaboration = "ALEPH",
    title = "{Deuteron and anti-deuteron production in e+ e- collisions at the Z resonance}",
    eprint = "hep-ex/0604023",
    archivePrefix = "arXiv",
    reportNumber = "CERN-PH-EP-2006-009",
    doi = "10.1016/j.physletb.2006.06.043",
    journal = "Phys. Lett. B",
    volume = "639",
    pages = "192--201",
    year = "2006"
}

@article{Wiringa:1994wb,
    author = "Wiringa, Robert B. and Stoks, V. G. J. and Schiavilla, R.",
    title = "{An Accurate nucleon-nucleon potential with charge independence breaking}",
    eprint = "nucl-th/9408016",
    archivePrefix = "arXiv",
    reportNumber = "PHY-7742-TH-94, CEBAF-TH-94-19",
    doi = "10.1103/PhysRevC.51.38",
    journal = "Phys. Rev. C",
    volume = "51",
    pages = "38--51",
    year = "1995"
}

@ARTICLE{GleesonAxford:1968,
       author = {{Gleeson}, L.~J. and {Axford}, W.~I.},
        title = "{Solar Modulation of Galactic Cosmic Rays}",
      journal = {\apj},
         year = 1968,
        month = dec,
       volume = {154},
        pages = {1011},
          doi = {10.1086/149822},
       adsurl = {https://ui.adsabs.harvard.edu/abs/1968ApJ...154.1011G}
}

@article{Smirnov:2019ngs,
    author = "Smirnov, Juri and Beacom, John F.",
    title = "{TeV-Scale Thermal WIMPs: Unitarity and its Consequences}",
    eprint = "1904.11503",
    archivePrefix = "arXiv",
    primaryClass = "hep-ph",
    doi = "10.1103/PhysRevD.100.043029",
    journal = "Phys. Rev. D",
    volume = "100",
    number = "4",
    pages = "043029",
    year = "2019"
}

@article{DiMauro:2025vxp,
    author = "Di Mauro, Mattia and Jueid, Adil and Koechler, Jordan and de Austri, Roberto Ruiz",
    title = "{Robust determination of antinuclei production from dark matter via weakly decaying beauty hadrons}",
    eprint = "2504.07172",
    archivePrefix = "arXiv",
    primaryClass = "hep-ph",
    doi = "10.1103/s6cm-45b4",
    journal = "Phys. Rev. D",
    volume = "112",
    number = "8",
    pages = "083017",
    year = "2025"
}

@article{deSalas:2020hbh,
    author = "de Salas, Pablo F. and Widmark, Axel",
    title = "{Dark matter local density determination: recent observations and future prospects}",
    eprint = "2012.11477",
    archivePrefix = "arXiv",
    primaryClass = "astro-ph.GA",
    doi = "10.1088/1361-6633/ac24e7",
    journal = "Rept. Prog. Phys.",
    volume = "84",
    number = "10",
    pages = "104901",
    year = "2021"
}

@ARTICLE{2025MNRAS.542.2987S,
       author = {{S{\"o}ding}, Laurin and {Bartel}, Ruben L. and {Mertsch}, Philipp},
        title = "{Local dark matter density from Gaia DR3 K-dwarfs using Gaussian processes}",
      journal = {\mnras},
         year = 2025,
        month = oct,
       volume = {542},
       number = {4},
        pages = {2987-2997},
          doi = {10.1093/mnras/staf1391},
archivePrefix = {arXiv},
       eprint = {2506.02956},
 primaryClass = {astro-ph.GA},
       adsurl = {https://ui.adsabs.harvard.edu/abs/2025MNRAS.542.2987S}
}

@ARTICLE{1978ApJ...223.1015G,
       author = {{Gunn}, J.~E. and {Lee}, B.~W. and {Lerche}, I. and {Schramm}, D.~N. and {Steigman}, G.},
        title = "{Some astrophysical consequences of the existence of a heavy stable neutral lepton.}",
      journal = {\apj},
         year = 1978,
        month = aug,
       volume = {223},
        pages = {1015-1031},
          doi = {10.1086/156335},
       adsurl = {https://ui.adsabs.harvard.edu/abs/1978ApJ...223.1015G}
}

@article{Lee:1977ua,
    author = "Lee, Benjamin W. and Weinberg, Steven",
    editor = "Srednicki, M. A.",
    title = "{Cosmological Lower Bound on Heavy Neutrino Masses}",
    reportNumber = "FERMILAB-PUB-77-041-T",
    doi = "10.1103/PhysRevLett.39.165",
    journal = "Phys. Rev. Lett.",
    volume = "39",
    pages = "165--168",
    year = "1977"
}

@article{Clowe:2006eq,
    author = "Clowe, Douglas and Bradac, Marusa and Gonzalez, Anthony H. and Markevitch, Maxim and Randall, Scott W. and Jones, Christine and Zaritsky, Dennis",
    title = "{A direct empirical proof of the existence of dark matter}",
    eprint = "astro-ph/0608407",
    archivePrefix = "arXiv",
    reportNumber = "SLAC-PUB-12078",
    doi = "10.1086/508162",
    journal = "Astrophys. J. Lett.",
    volume = "648",
    pages = "L109--L113",
    year = "2006"
}

@article{LZ:2023,
  author = "Aalbers, J. and others",
  title = "{First Dark Matter Search Results from the LUX-ZEPLIN (LZ) Experiment}",
  journal = "Phys. Rev. Lett.",
  volume = "131",
  pages = "041002",
  year = "2023",
  doi = "10.1103/PhysRevLett.131.041002",
  eprint = "2207.03764",
  archivePrefix = "arXiv",
  primaryClass = "hep-ex"
}

@article{Aprile:2023XENONnT,
  author = "Aprile, E. and others",
  title = "{First Dark Matter Search with Nuclear Recoils from the XENONnT Experiment}",
  journal = "Phys. Rev. Lett.",
  volume = "131",
  pages = "041003",
  year = "2023",
  doi = "10.1103/PhysRevLett.131.041003"
}

@article{Strassler:2006im,
author = {Strassler, Matthew J. and Zurek, Kathryn M.},
title = {Echoes of a hidden valley at hadron colliders},
journal = {Physics Letters B},
volume = {651},
pages = {374--379},
year = {2007},
doi = {10.1016/j.physletb.2007.06.055},
eprint = {hep-ph/0604261},
archivePrefix= {arXiv}
}

@article{Barron:2021SUEP,
author = {Barron, Joshua and others},
title = {Unsupervised hadronic SUEP at the LHC},
journal = {Journal of High Energy Physics},
volume = {2021},
number = {12},
pages = {129},
year = {2021},
doi = {10.1007/JHEP12(2021)129},
eprint = {2107.12379},
archivePrefix= {arXiv},
primaryClass = {hep-ph}
}

@article{DiMauro:2025jia,
    author = "Di Mauro, Mattia and Xie, Bohan",
    title = "{Dark matter Simplified models in the Resonance Region}",
    eprint = "2510.08677",
    archivePrefix = "arXiv",
    primaryClass = "hep-ph",
    month = "10",
    year = "2025"
}

@article{Evoli:2019iih,
    author = "Evoli, Carmelo and Morlino, Giovanni and Blasi, Pasquale and Aloisio, Roberto",
    title = "{AMS-02 beryllium data and its implication for cosmic ray transport}",
    eprint = "1910.04113",
    archivePrefix = "arXiv",
    primaryClass = "astro-ph.HE",
    doi = "10.1103/PhysRevD.101.023013",
    journal = "Phys. Rev. D",
    volume = "101",
    number = "2",
    pages = "023013",
    year = "2020"
}

@article{Heeck:2019ego,
    author = "Heeck, Julian and Rajaraman, Arvind",
    title = "{How to produce antinuclei from dark matter}",
    eprint = "1906.01667",
    archivePrefix = "arXiv",
    primaryClass = "hep-ph",
    reportNumber = "UCI-TR-2019-14",
    doi = "10.1088/1361-6471/ab9f03",
    journal = "J. Phys. G",
    volume = "47",
    number = "10",
    pages = "105202",
    year = "2020"
}

@article{Fedderke:2024hfy,
    author = "Fedderke, Michael A. and Kaplan, David E. and Mathur, Anubhav and Rajendran, Surjeet and Tanin, Erwin H.",
    title = "{Fireball antinucleosynthesis}",
    eprint = "2402.15581",
    archivePrefix = "arXiv",
    primaryClass = "hep-ph",
    reportNumber = "FERMILAB-PUB-24-0514-SQMS-V",
    doi = "10.1103/PhysRevD.109.123028",
    journal = "Phys. Rev. D",
    volume = "109",
    number = "12",
    pages = "123028",
    year = "2024"
}

@article{Knapen:2016hky,
    author = "Knapen, Simon and Pagan Griso, Simone and Papucci, Michele and Robinson, Dean J.",
    title = "{Triggering Soft Bombs at the LHC}",
    eprint = "1612.00850",
    archivePrefix = "arXiv",
    primaryClass = "hep-ph",
    doi = "10.1007/JHEP08(2017)076",
    journal = "JHEP",
    volume = "08",
    pages = "076",
    year = "2017"
}

@article{Strassler:2008bv,
    author = "Strassler, Matthew J.",
    title = "{Why Unparticle Models with Mass Gaps are Examples of Hidden Valleys}",
    eprint = "0801.0629",
    archivePrefix = "arXiv",
    primaryClass = "hep-ph",
    month = "1",
    year = "2008"
}

@article{Sjostrand:2014zea,
    author = {Sj{\"o}strand, Torbj{\"o}rn and Ask, Stefan and Christiansen, Jesper R. and Corke, Richard and Desai, Nishita and Ilten, Philip and Mrenna, Stephen and Prestel, Stefan and Rasmussen, Christine O. and Skands, Peter Z.},
    title = "{An introduction to PYTHIA 8.2}",
    eprint = "1410.3012",
    archivePrefix = "arXiv",
    primaryClass = "hep-ph",
    reportNumber = "LU-TP-14-36, MCNET-14-22, CERN-PH-TH-2014-190, FERMILAB-PUB-14-316-CD, DESY-14-178, SLAC-PUB-16122",
    doi = "10.1016/j.cpc.2015.01.024",
    journal = "Comput. Phys. Commun.",
    volume = "191",
    pages = "159--177",
    year = "2015"
}

@misc{suepcode,
  howpublished = {\url{https://gitlab.com/simonknapen/suep_generator}},
}

@article{Blanchard:2004du,
    author = "Blanchard, Philippe and Fortunato, Santo and Satz, Helmut",
    title = "{The Hagedorn temperature and partition thermodynamics}",
    eprint = "hep-ph/0401103",
    archivePrefix = "arXiv",
    reportNumber = "BI-TP-2004-01",
    doi = "10.1140/epjc/s2004-01673-0",
    journal = "Eur. Phys. J. C",
    volume = "34",
    pages = "361--366",
    year = "2004"
}

@article{Noronha-Hostler:2010nut,
    author = "Noronha-Hostler, J. and Noronha, Jorge and Greiner, Carsten",
    editor = "Fraga, Eduardo and Kodama, Takeshi and Padula, Sandra and Takahashi, Jun",
    title = "{Particle Ratios and the QCD Critical Temperature}",
    eprint = "1001.2610",
    archivePrefix = "arXiv",
    primaryClass = "nucl-th",
    doi = "10.1088/0954-3899/37/9/094062",
    journal = "J. Phys. G",
    volume = "37",
    pages = "094062",
    year = "2010"
}

@article{CMS:2024nca,
    author = "Hayrapetyan, Aram and others",
    collaboration = "CMS",
    title = "{Search for Soft Unclustered Energy Patterns in Proton-Proton Collisions at 13~TeV}",
    eprint = "2403.05311",
    archivePrefix = "arXiv",
    primaryClass = "hep-ex",
    reportNumber = "CMS-EXO-23-002, CERN-EP-2024-054",
    doi = "10.1103/PhysRevLett.133.191902",
    journal = "Phys. Rev. Lett.",
    volume = "133",
    number = "19",
    pages = "191902",
    year = "2024"
}

@article{Kribs:2018ilo,
    author = "Kribs, Graham D. and Martin, Adam and Ostdiek, Bryan and Tong, Tom",
    title = "{Dark Mesons at the LHC}",
    eprint = "1809.10184",
    archivePrefix = "arXiv",
    primaryClass = "hep-ph",
    doi = "10.1007/JHEP07(2019)133",
    journal = "JHEP",
    volume = "07",
    pages = "133",
    year = "2019"
}

@article{Fleming:2024flc,
    author = "Fleming, George T. and Kribs, Graham D. and Neil, Ethan T. and Schaich, David and Vranas, Pavlos M.",
    title = "{Hyperstealth dark matter and long-lived particles}",
    eprint = "2412.14540",
    archivePrefix = "arXiv",
    primaryClass = "hep-ph",
    reportNumber = "LLNL-JRNL-2001590, FERMILAB-PUB-24-0954-T",
    doi = "10.1103/9nv9-rqq7",
    journal = "Phys. Rev. D",
    volume = "112",
    number = "7",
    pages = "075004",
    year = "2025"
}

@article{Chacko:2005pe,
    author = "Chacko, Z. and Goh, Hock-Seng and Harnik, Roni",
    title = "{The Twin Higgs: Natural electroweak breaking from mirror symmetry}",
    eprint = "hep-ph/0506256",
    archivePrefix = "arXiv",
    doi = "10.1103/PhysRevLett.96.231802",
    journal = "Phys. Rev. Lett.",
    volume = "96",
    pages = "231802",
    year = "2006"
}

@article{Poulin:2018wzu,
    author = "Poulin, Vivian and Salati, Pierre and Cholis, Ilias and Kamionkowski, Marc and Silk, Joseph",
    title = "{Where do the AMS-02 antihelium events come from?}",
    eprint = "1808.08961",
    archivePrefix = "arXiv",
    primaryClass = "astro-ph.HE",
    doi = "10.1103/PhysRevD.99.023016",
    journal = "Phys. Rev. D",
    volume = "99",
    number = "2",
    pages = "023016",
    year = "2019"
}

@article{Bykov:2023nnr,
    author = "Bykov, Andrey and Postnov, Konstantin and Bondar, Alexander and Blinnikov, Serguey and Dolgov, Aleksander",
    title = "{Antistars as possible sources of antihelium cosmic rays}",
    eprint = "2304.04623",
    archivePrefix = "arXiv",
    primaryClass = "astro-ph.HE",
    doi = "10.1088/1475-7516/2023/08/027",
    journal = "JCAP",
    volume = "08",
    pages = "027",
    year = "2023"
}

@article{Hur:2007uz,
    author = "Hur, Taeil and Jung, Dong-Won and Ko, P. and Lee, Jae Yong",
    title = "{Electroweak symmetry breaking and cold dark matter from strongly interacting hidden sector}",
    eprint = "0709.1218",
    archivePrefix = "arXiv",
    primaryClass = "hep-ph",
    doi = "10.1016/j.physletb.2010.12.047",
    journal = "Phys. Lett. B",
    volume = "696",
    pages = "262--265",
    year = "2011"
}

@article{Kribs:2009fy,
    author = "Kribs, Graham D. and Roy, Tuhin S. and Terning, John and Zurek, Kathryn M.",
    title = "{Quirky Composite Dark Matter}",
    eprint = "0909.2034",
    archivePrefix = "arXiv",
    primaryClass = "hep-ph",
    reportNumber = "FERMILAB-PUB-09-425-T",
    doi = "10.1103/PhysRevD.81.095001",
    journal = "Phys. Rev. D",
    volume = "81",
    pages = "095001",
    year = "2010"
}

@article{Beauchesne:2018myj,
    author = "Beauchesne, Hugues and Bertuzzo, Enrico and Grilli Di Cortona, Giovanni",
    title = "{Dark matter in Hidden Valley models with stable and unstable light dark mesons}",
    eprint = "1809.10152",
    archivePrefix = "arXiv",
    primaryClass = "hep-ph",
    doi = "10.1007/JHEP04(2019)118",
    journal = "JHEP",
    volume = "04",
    pages = "118",
    year = "2019"
}

@article{Francis:2018xjd,
    author = "Francis, Anthony and Hudspith, Renwick J. and Lewis, Randy and Tulin, Sean",
    title = "{Dark Matter from Strong Dynamics: The Minimal Theory of Dark Baryons}",
    eprint = "1809.09117",
    archivePrefix = "arXiv",
    primaryClass = "hep-ph",
    reportNumber = "CERN-TH-2018-207",
    doi = "10.1007/JHEP12(2018)118",
    journal = "JHEP",
    volume = "12",
    pages = "118",
    year = "2018"
}

@article{Craig:2015pha,
    author = "Craig, Nathaniel and Katz, Andrey and Strassler, Matt and Sundrum, Raman",
    title = "{Naturalness in the Dark at the LHC}",
    eprint = "1501.05310",
    archivePrefix = "arXiv",
    primaryClass = "hep-ph",
    reportNumber = "UMD-PP-014-028, CERN-PH-TH-2014-263",
    doi = "10.1007/JHEP07(2015)105",
    journal = "JHEP",
    volume = "07",
    pages = "105",
    year = "2015"
}

@article{Burdman:2006tz,
    author = "Burdman, Gustavo and Chacko, Z. and Goh, Hock-Seng and Harnik, Roni",
    title = "{Folded supersymmetry and the LEP paradox}",
    eprint = "hep-ph/0609152",
    archivePrefix = "arXiv",
    reportNumber = "SLAC-PUB-12115",
    doi = "10.1088/1126-6708/2007/02/009",
    journal = "JHEP",
    volume = "02",
    pages = "009",
    year = "2007"
}

@article{Elor:2015bho,
    author = "Elor, Gilly and Rodd, Nicholas L. and Slatyer, Tracy R. and Xue, Wei",
    title = "{Model-Independent Indirect Detection Constraints on Hidden Sector Dark Matter}",
    eprint = "1511.08787",
    archivePrefix = "arXiv",
    primaryClass = "hep-ph",
    reportNumber = "MIT-CTP-4742",
    doi = "10.1088/1475-7516/2016/06/024",
    journal = "JCAP",
    volume = "06",
    pages = "024",
    year = "2016"
}

@article{Costantino:2020msc,
    author = "Costantino, Alexandria and Fichet, Sylvain and Tanedo, Philip",
    title = "{Effective Field Theory in AdS: Continuum Regime, Soft Bombs, and IR Emergence}",
    eprint = "2002.12335",
    archivePrefix = "arXiv",
    primaryClass = "hep-th",
    reportNumber = "UCR-TR-2020-FLIP-IG-11",
    doi = "10.1103/PhysRevD.102.115038",
    journal = "Phys. Rev. D",
    volume = "102",
    number = "11",
    pages = "115038",
    year = "2020"
}

@article{Cesarotti:2020uod,
    author = "Cesarotti, Cari and Reece, Matthew and Strassler, Matthew J.",
    title = "{Spheres To Jets: Tuning Event Shapes with 5d Simplified Models}",
    eprint = "2009.08981",
    archivePrefix = "arXiv",
    primaryClass = "hep-ph",
    doi = "10.1007/JHEP05(2021)096",
    journal = "JHEP",
    volume = "05",
    pages = "096",
    year = "2021"
}

@article{VanApeldoorn:1981gx,
    author = "Van Apeldoorn, G. W. and others",
    title = "{Thermal Emission of Pions in Anti-proton Proton Annihilation at 12-{GeV}/c}",
    doi = "10.1007/BF01436312",
    journal = "Z. Phys. C",
    volume = "7",
    pages = "235--239",
    year = "1981"
}

@article{Hatta:2008tn,
    author = "Hatta, Yoshitaka and Matsuo, Toshihiro",
    title = "{Jet fragmentation and gauge/string duality}",
    eprint = "0804.4733",
    archivePrefix = "arXiv",
    primaryClass = "hep-th",
    doi = "10.1016/j.physletb.2008.10.043",
    journal = "Phys. Lett. B",
    volume = "670",
    pages = "150--153",
    year = "2008"
}

@article{Hagedorn:1965st,
    author = "Hagedorn, R.",
    title = "{Statistical thermodynamics of strong interactions at high-energies}",
    reportNumber = "CERN-TH-520",
    journal = "Nuovo Cim. Suppl.",
    volume = "3",
    pages = "147--186",
    year = "1965"
}

@article{Bjorken:1969wi,
    author = "Bjorken, J. D. and Brodsky, Stanley J.",
    title = "{Statistical Model for electron-Positron Annihilation Into Hadrons}",
    reportNumber = "SLAC-PUB-0662",
    doi = "10.1103/PhysRevD.1.1416",
    journal = "Phys. Rev. D",
    volume = "1",
    pages = "1416--1420",
    year = "1970"
}

@article{Patt:2006fw,
    author = "Patt, Brian and Wilczek, Frank",
    title = "{Higgs-field portal into hidden sectors}",
    eprint = "hep-ph/0605188",
    archivePrefix = "arXiv",
    reportNumber = "MIT-CTP-3745",
    month = "5",
    year = "2006"
}

@article{Falkowski:2009yz,
    author = "Falkowski, Adam and Juknevich, Jose and Shelton, Jessie",
    title = "{Dark Matter Through the Neutrino Portal}",
    eprint = "0908.1790",
    archivePrefix = "arXiv",
    primaryClass = "hep-ph",
    reportNumber = "RU-NHETC-09-15",
    month = "8",
    year = "2009"
}

@article{Knapen:2021eip,
    author = "Knapen, Simon and Shelton, Jessie and Xu, Dong",
    title = "{Perturbative benchmark models for a dark shower search program}",
    eprint = "2103.01238",
    archivePrefix = "arXiv",
    primaryClass = "hep-ph",
    doi = "10.1103/PhysRevD.103.115013",
    journal = "Phys. Rev. D",
    volume = "103",
    number = "11",
    pages = "115013",
    year = "2021"
}

@article{Bai_2014,
   title={Scale of dark QCD},
   volume={89},
   ISSN={1550-2368},
   url={http://dx.doi.org/10.1103/PhysRevD.89.063522},
   DOI={10.1103/physrevd.89.063522},
   number={6},
   journal={Physical Review D},
   publisher={American Physical Society (APS)},
   author={Bai, Yang and Schwaller, Pedro},
   year={2014},
   month=mar }

@article{Kang:2008ea,
    author = "Kang, Junhai and Luty, Markus A.",
    title = "{Macroscopic Strings and 'Quirks' at Colliders}",
    eprint = "0805.4642",
    archivePrefix = "arXiv",
    primaryClass = "hep-ph",
    doi = "10.1088/1126-6708/2009/11/065",
    journal = "JHEP",
    volume = "11",
    pages = "065",
    year = "2009"
}

@article{Renner:2018fhh,
    author = "Renner, Sophie and Schwaller, Pedro",
    title = "{A flavoured dark sector}",
    eprint = "1803.08080",
    archivePrefix = "arXiv",
    primaryClass = "hep-ph",
    doi = "10.1007/JHEP08(2018)052",
    journal = "JHEP",
    volume = "08",
    pages = "052",
    year = "2018"
}

@article{Baumgart:2009tn,
    author = "Baumgart, Matthew and Cheung, Clifford and Ruderman, Joshua T. and Wang, Lian-Tao and Yavin, Itay",
    title = "{Non-Abelian Dark Sectors and Their Collider Signatures}",
    eprint = "0901.0283",
    archivePrefix = "arXiv",
    primaryClass = "hep-ph",
    doi = "10.1088/1126-6708/2009/04/014",
    journal = "JHEP",
    volume = "04",
    pages = "014",
    year = "2009"
}

@article{Bai:2010qg,
    author = "Bai, Yang and Hill, Richard J.",
    title = "{Weakly Interacting Stable Pions}",
    eprint = "1005.0008",
    archivePrefix = "arXiv",
    primaryClass = "hep-ph",
    reportNumber = "FERMILAB-PUB-10-001-T, EFI-PREPRINT-10-9",
    doi = "10.1103/PhysRevD.82.111701",
    journal = "Phys. Rev. D",
    volume = "82",
    pages = "111701",
    year = "2010"
}

@article{Buckley:2012ky,
    author = "Buckley, Matthew R. and Neil, Ethan T.",
    title = "{Thermal Dark Matter from a Confining Sector}",
    eprint = "1209.6054",
    archivePrefix = "arXiv",
    primaryClass = "hep-ph",
    reportNumber = "FERMILAB-PUB-12-533-A-T",
    doi = "10.1103/PhysRevD.87.043510",
    journal = "Phys. Rev. D",
    volume = "87",
    number = "4",
    pages = "043510",
    year = "2013"
}

@article{Antipin:2014qva,
    author = "Antipin, Oleg and Redi, Michele and Strumia, Alessandro",
    title = "{Dynamical generation of the weak and Dark Matter scales from strong interactions}",
    eprint = "1410.1817",
    archivePrefix = "arXiv",
    primaryClass = "hep-ph",
    doi = "10.1007/JHEP01(2015)157",
    journal = "JHEP",
    volume = "01",
    pages = "157",
    year = "2015"
}

@article{Appelquist:2015yfa,
    author = "Appelquist, Thomas and others",
    title = "{Stealth Dark Matter: Dark scalar baryons through the Higgs portal}",
    eprint = "1503.04203",
    archivePrefix = "arXiv",
    primaryClass = "hep-ph",
    reportNumber = "INT-PUB-15-004, LLNL-JRNL-667446",
    doi = "10.1103/PhysRevD.92.075030",
    journal = "Phys. Rev. D",
    volume = "92",
    number = "7",
    pages = "075030",
    year = "2015"
}

@article{Antipin:2015xia,
    author = "Antipin, Oleg and Redi, Michele and Strumia, Alessandro and Vigiani, Elena",
    title = "{Accidental Composite Dark Matter}",
    eprint = "1503.08749",
    archivePrefix = "arXiv",
    primaryClass = "hep-ph",
    reportNumber = "IFTP-TH-2015",
    doi = "10.1007/JHEP07(2015)039",
    journal = "JHEP",
    volume = "07",
    pages = "039",
    year = "2015"
}

@article{Mitridate:2017oky,
    author = "Mitridate, Andrea and Redi, Michele and Smirnov, Juri and Strumia, Alessandro",
    title = "{Dark Matter as a weakly coupled Dark Baryon}",
    eprint = "1707.05380",
    archivePrefix = "arXiv",
    primaryClass = "hep-ph",
    reportNumber = "CERN-TH-2017-151",
    doi = "10.1007/JHEP10(2017)210",
    journal = "JHEP",
    volume = "10",
    pages = "210",
    year = "2017"
}

@article{Contino:2020god,
    author = "Contino, Roberto and Podo, Alessandro and Revello, Filippo",
    title = "{Composite Dark Matter from Strongly-Interacting Chiral Dynamics}",
    eprint = "2008.10607",
    archivePrefix = "arXiv",
    primaryClass = "hep-ph",
    doi = "10.1007/JHEP02(2021)091",
    journal = "JHEP",
    volume = "02",
    pages = "091",
    year = "2021"
}

@article{Asadi:2021yml,
    author = "Asadi, Pouya and Kramer, Eric David and Kuflik, Eric and Ridgway, Gregory W. and Slatyer, Tracy R. and Smirnov, Juri",
    title = "{Accidentally Asymmetric Dark Matter}",
    eprint = "2103.09822",
    archivePrefix = "arXiv",
    primaryClass = "hep-ph",
    reportNumber = "MIT-CTP/5284",
    doi = "10.1103/PhysRevLett.127.211101",
    journal = "Phys. Rev. Lett.",
    volume = "127",
    number = "21",
    pages = "211101",
    year = "2021"
}

@article{Asadi:2021pwo,
    author = "Asadi, Pouya and Kramer, Eric David and Kuflik, Eric and Ridgway, Gregory W. and Slatyer, Tracy R. and Smirnov, Juri",
    title = "{Thermal squeezeout of dark matter}",
    eprint = "2103.09827",
    archivePrefix = "arXiv",
    primaryClass = "hep-ph",
    doi = "10.1103/PhysRevD.104.095013",
    journal = "Phys. Rev. D",
    volume = "104",
    number = "9",
    pages = "095013",
    year = "2021"
}

@article{Cline:2021itd,
    author = "Cline, James M.",
    title = "{Dark atoms and composite dark matter}",
    eprint = "2108.10314",
    archivePrefix = "arXiv",
    primaryClass = "hep-ph",
    doi = "10.21468/SciPostPhysLectNotes.52",
    journal = "SciPost Phys. Lect. Notes",
    volume = "52",
    pages = "1",
    year = "2022"
}

@article{Asadi:2024bbq,
    author = "Asadi, Pouya and Kribs, Graham D. and Mantel, Chester J. Hamilton",
    title = "{Direct detection of dark baryons naturally suppressed by H-parity}",
    eprint = "2410.23631",
    archivePrefix = "arXiv",
    primaryClass = "hep-ph",
    doi = "10.1103/PhysRevD.111.095030",
    journal = "Phys. Rev. D",
    volume = "111",
    number = "9",
    pages = "095030",
    year = "2025"
}

@article{Asadi:2024tpu,
    author = "Asadi, Pouya and Batz, Austin and Kribs, Graham D.",
    title = "{Noble dark matter: Surprising elusiveness of dark baryons}",
    eprint = "2412.14240",
    archivePrefix = "arXiv",
    primaryClass = "hep-ph",
    doi = "10.1103/PhysRevD.111.095025",
    journal = "Phys. Rev. D",
    volume = "111",
    number = "9",
    pages = "095025",
    year = "2025"
}

@article{Veneziano:1976wm,
    author = "Veneziano, G.",
    title = "{Some Aspects of a Unified Approach to Gauge, Dual and Gribov Theories}",
    reportNumber = "CERN-TH-2200",
    doi = "10.1016/0550-3213(76)90412-0",
    journal = "Nucl. Phys. B",
    volume = "117",
    pages = "519--545",
    year = "1976"
}

@article{Witten:1979vv,
    author = "Witten, Edward",
    title = "{Current Algebra Theorems for the U(1) Goldstone Boson}",
    reportNumber = "HUTP-79/A014",
    doi = "10.1016/0550-3213(79)90031-2",
    journal = "Nucl. Phys. B",
    volume = "156",
    pages = "269--283",
    year = "1979"
}

@article{Morrison:2020yeg,
    author = "Morrison, Logan and Profumo, Stefano and Robinson, Dean J.",
    title = "{Large $N$-ightmare Dark Matter}",
    eprint = "2010.03586",
    archivePrefix = "arXiv",
    primaryClass = "hep-ph",
    doi = "10.1088/1475-7516/2021/05/058",
    journal = "JCAP",
    volume = "05",
    pages = "058",
    year = "2021"
}

@article{Morningstar:1999rf,
    author = "Morningstar, Colin J. and Peardon, Mike J.",
    title = "{The Glueball spectrum from an anisotropic lattice study}",
    eprint = "hep-lat/9901004",
    archivePrefix = "arXiv",
    reportNumber = "UCSD-PTH-98-36, HLRZ-1998-62",
    doi = "10.1103/PhysRevD.60.034509",
    journal = "Phys. Rev. D",
    volume = "60",
    pages = "034509",
    year = "1999"
}

\end{document}